\documentclass[a4paper,11pt]{article}
\pdfoutput=1 
\usepackage{jheppub}
\usepackage{amssymb,amsmath,verbatim,mathtools,needspace,enumitem,graphicx,physics,mathrsfs}
\usepackage{gensymb}
\usepackage{appendix}
\usepackage{tensor}
\usepackage{array}
\usepackage[normalem]{ulem}
\usepackage[dvipsnames, usenames]{xcolor}
\usepackage{xr-hyper}
\definecolor{linkcolor}{rgb}{0.0,0.3,0.5}
\definecolor{myred}{RGB}{179, 27, 27}
\usepackage[utf8]{inputenc}
\usepackage{hyperref}
\hypersetup{colorlinks=true,citecolor=myred,linkcolor=myred,urlcolor=myred,breaklinks}
\usepackage[all]{hypcap}
\usepackage[T1]{fontenc}
\usepackage{orcidlink}

\newcommand{\tho}{\text{\th}}

\newcommand{\thop}{\text{\th}^\prime}

\newcommand{\C}{\mathbb{C}}
\newcommand{\J}{\mathbb{J}}

\title{Unifying the Regge--Wheeler--Zerilli and Bardeen--Press--Teukolsky formalisms on spherical backgrounds}

\author{David Pereñiguez\orcidlink{0000-0002-1007-4551}}

\affiliation{William H. Miller III Department of Physics and Astronomy, Johns Hopkins
University, 3400 North Charles Street, Baltimore, Maryland, 21218, USA }

\emailAdd{dpereni1@jhu.edu}

\abstract{We develop a formulation of perturbation theory on spherically symmetric backgrounds based on self-dual curvature equations combined with spherical harmonic expansions. The resulting framework unifies the Regge–Wheeler–Zerilli (RWZ) and Bardeen--Press--Teukolsky (BPT) formalisms and is designed to combine key advantages of both. The use of self-dual variables is crucial, and makes quasinormal mode isospectrality manifest, when present. We present the formalism first for a general energy–momentum tensor, and then specialize to vacuum General Relativity with matter sources to illustrate its practical advantages. A central result is that the RWZ and BPT equations arise directly as different components of a single tensorial curvature equation. We also show that, in the frequency domain, the metric can be reconstructed algebraically from any of the proposed master functions and their derivatives, and we comment on possible obstructions to such a reconstruction in the time domain. A Mathematica notebook, based on xAct, that implements the formalism and was used in our computations is released alongside this work.}

\begin{document} 
\maketitle
\flushbottom


\section{Introduction}\label{sec:Introduction}

Understanding the generation and propagation of gravitational waves in black hole spacetimes is a central problem in gravity theory. A precise grasp of this problem is essential for exploiting the full potential of gravitational-wave astronomy as a probe of General Relativity (GR) and its high-energy extensions \cite{Berti:2009kk,Barack:2018yly,LISAConsortiumWaveformWorkingGroup:2023arg,Berti:2025hly}.

Within black hole perturbation theory, gravitational waves are described as small fluctuations about a background black hole spacetime. The governing equations are, in general, a system of coupled linear equations whose structure makes them difficult to solve in full generality. A major simplification arises when the background is assumed to be spherically symmetric, in which case the equations become significantly more tractable. This setting already captures a wide range of physically relevant phenomena, allowing one to extract robust insights while avoiding some of the technical complications associated with more general backgrounds. Historically, many key results were first established in spherical symmetry and only later generalized, for instance, to rotating spacetimes. As a result, perturbation theory on spherical backgrounds continues to play a central role in current research.

Two complementary approaches have traditionally been employed. The first is the metric-based formalism introduced by Regge and Wheeler \cite{PhysRev.108.1063} and Zerilli \cite{PhysRevLett.24.737}, commonly referred to as the RWZ approach, which has been extensively developed in subsequent work (see, e.g., \cite{Martel:2005ir,Spiers:2023mor}). The second is the curvature-based approach, formulated in the Newman--Penrose (NP) \cite{Newman:1961qr} or Geroch--Held--Penrose (GHP) \cite{Geroch:1973am} frameworks and developed in seminal work by Bardeen and Press \cite{Bardeen:1973xb} and Teukolsky \cite{Teukolsky:1973ha} (BPT) (see, e.g., \cite{Chandrasekhar:1985kt}). Each approach has its own advantages and limitations and differs in methodology, but they should ultimately yield equivalent physical predictions.

On the one hand, the RWZ formalism addresses metric perturbations directly by decomposing them into standard harmonic modes, which further decouple into two parity sectors—even and odd—defined by their transformation under the spherical antipodal map. This approach is appealing due to its reliance on familiar spherical harmonic expansions and its relative technical simplicity. However, it does not fully exploit the geometric structure underlying GR, which leads to several drawbacks. First, the reduction of the equations of motion within each parity sector to decoupled master equations is not systematic, relying instead on seemingly arbitrary choices of combinations of variables and equations (see \cite{Lenzi:2021wpc} for a detailed discussion about this aspect). Second, although the even and odd sectors take markedly different forms, they have a hidden symmetry---isospectrality---when describing free perturbations such as quasinormal modes (QNMs) of black holes. This symmetry is not manifest within the RWZ framework and only emerges after applying Chandrasekhar \cite{Chandrasekhar:1985kt,Chandrasekhar_AlgSpecial} (or equivalently Darboux \cite{Glampedakis:2017rar}) transformations within each sector.

On the other hand, the BPT approach takes advantage of the Newman--Penrose (NP) or Geroch--Held--Penrose (GHP) formulations of GR, which encode the theory in terms of 2-spinors \cite{Witten:1959zza,Penrose:1960eq,Penrose:1985bww} and express the field equations using a null tetrad basis. Linearising these equations on a black hole background, Bardeen and Press \cite{Bardeen:1973xb} obtained quite straightforwardly decoupled equations for the perturbations of the extreme-weight Weyl scalars, $\delta\Psi_{0}$ and $\delta\Psi_{4}$. Remarkably, these equations remain decoupled even in the case of rotating black hole backgrounds, as shown by Teukolsky \cite{Teukolsky:1973ha}. When restricted to spherical backgrounds, however, the BPT formalism presents certain disadvantages relative to RWZ. Working with dynamical null frames instead of tensor perturbations introduces additional technical complexity, including an enlarged gauge redundancy that now incorporates null-frame Lorentz transformations alongside diffeomorphisms. Relatedly, NP or GHP scalars transform nontrivially under frame rotations, preventing a decomposition in ordinary spherical harmonics and instead requiring spin-weighted spherical harmonics, which are often more cumbersome to handle. For this reason, RWZ is typically preferred in the context of spherical backgrounds. A further limitation is that, unlike the RWZ master equations, the BPT equations involve a long-range potential. This complicates both analytical and numerical treatments, either by increasing computational cost or by necessitating additional nontrivial transformations of the variables \cite{Sasaki1981TheRE}.

These considerations motivate the search for a unified framework satisfying the following criteria: $(i)$ avoidance of dynamical null frames and spin-weighted spherical harmonics, $(ii)$ a simple and systematic derivation of both the RWZ and BPT equations, $(iii)$ a formulation in which the RWZ variables arise naturally with a clear geometric interpretation, and
$(iv)$ a manifest realization of symmetries such as isospectrality, when present. A first step in this direction was taken in \cite{Mukkamala:2024dxf}, where it was proposed to work with self-dual curvature variables and the non-perturbative wave equations they satisfy.\footnote{See also \cite{Gasperin:2026znw,Chaverra:2012bh} for other interesting perspectives.} 

The aim of this paper is to develop a formalism that unifies the RWZ and BPT approaches, satisfying the criteria $(i)$-$(iv)$ above. This work can be viewed as a completion of \cite{Mukkamala:2024dxf}, extending the framework so that not only the RWZ equations but also the BPT equations emerge naturally. In addition, we show that, for perturbations of Schwarzschild in GR, the metric in the frequency domain can be reconstructed from any of the proposed master variables using purely differential reconstruction maps, involving only the variables and their derivatives. We also discuss obstructions to such reconstructions in the time domain, recently identified in \cite{Poisson:2025oic}. For generality, the formalism is first presented in a setting with arbitrary matter content, encoded in the energy-momentum tensor and the self-dual Cotton tensor (or curvature current). Finally, we provide a \texttt{Mathematica} notebook, based on the \texttt{xAct} package suite, implementing the formalism for efficient symbolic computations \cite{pereniguez2026ptcwe}.

\subsection{Basic Idea}\label{sec:BasicIdea}

The basic idea is to exploit a set of features specific to four-dimensional GR. In this setting, vacuum curvature is described by a single irreducible component. This is most naturally expressed in the 2-spinor formalism \cite{Witten:1959zza,Penrose:1960eq,Penrose:1985bww}, where the full vacuum curvature content is encoded in the Weyl spinor $\Psi_{ABCD}$. This object represents the self-dual part of the Weyl tensor, a notion that exists only in four dimensions. Physically relevant solutions, such as the Kerr spacetime, are characterized by particularly simple Weyl spinors, which are classified according to their Petrov type. In vacuum, $\Psi_{ABCD}$ satisfies the nonlinear wave equation
\begin{equation}\label{eq:Psi_Eq}
    \square \Psi_{ABCD}-6\tensor{\Psi}{^E^F_{(AB}}\tensor{\Psi}{_{CD)EF}}=0\,,
\end{equation}
where $\square=\nabla_{AA'}\nabla^{AA'}$ denotes the spacetime d'Alembertian. This equation can be interpreted as governing the propagation of the gravitational degrees of freedom.

This simplicity underlies many key results in the study of perturbations of exact backgrounds. For instance, projecting Eq.~\eqref{eq:Psi_Eq} onto the extreme components of a spinor dyad, $o^{A}o^{B}o^{C}o^{D}$ and $\iota^{A}\iota^{B}\iota^{C}\iota^{D}$, yields exact nonlinear wave equations for the curvature scalars $\Psi_{0}$ and $\Psi_{4}$, expressed in the GHP formalism \cite{Fransen:2025cgv}.\footnote{Projecting onto other elements of the dyad basis yields nonlinear equations for $\Psi_{1,2,3}$, reproducing the results of \cite{Bini:2002jx,Bini:2003km}, who followed the approach of Stewart and Walker \cite{Stewart:1974uz}.} These equations were first obtained by Stewart and Walker through direct manipulation of the GHP system \cite{Stewart:1974uz}. Upon linearisation about a Petrov type $D$ background, they reduce directly to Teukolsky's equations (see also \cite{Ryan:1974nt} and \cite{Green:2026nlt}). The same strategy extends to other classes of backgrounds, such as type $N$ spacetimes \cite{Fransen:2025cgv}.

We now specialise to spherically symmetric backgrounds. These form a subclass of type $D$ spacetimes, independently of the field equations or additional symmetry assumptions such as time symmetry. In this setting, spherical harmonics provide a useful basis for describing fluctuations, since harmonic-mode decoupling is guaranteed, and therefore seem more natural than a dynamical null-tetrad approach; otherwise, one does not fully exploit the stronger assumption of spherical symmetry. Moreover, this choice avoids the additional gauge redundancies associated with local frame degrees of freedom and leads to a technically simpler formulation. Accordingly, rather than projecting onto null frames or spinor dyads, we expand the fluctuations of Eq.~\eqref{eq:Psi_Eq}, or equivalently its tensorial version derived below, in a basis of spherical harmonics. We will show that the RWZ and BPT equations arise naturally from different components of this equation, for some components of the self-dual Weyl tensor. Since the real and imaginary parts of the self-dual variables correspond respectively to purely even and purely odd metric perturbations (as first observed in \cite{Price_Despun}), isospectrality becomes manifest (whenever it holds). As discussed below, this framework can be extended straightforwardly to incorporate the effects of matter or other extensions of vacuum GR, such as higher-derivative corrections.

\subsection{Summary and Outline}

The formulation of perturbation theory in this approach begins with a set of exact, non-perturbative equations, referred to as the fundamental curvature equations, derived in Section \ref{sec:Sec1}. These follow directly from Einstein's equations and play the role of structure equations, analogous to those of the NP or GHP formalisms.

Section \ref{sec:Sec2} develops the description of linear fluctuations. The properties of self-dual curvature variables are analysed, showing that they render QNM isospectrality manifest whenever it is present. Contact with the GHP curvature scalars is established within this framework, and a two-dimensional version of the GHP formalism is introduced, which proves useful for the covariant projection of the linearised equations. Sections \ref{sec:Sec1} and \ref{sec:Sec2} are formulated in full generality, including arbitrary matter contributions encoded in the energy-momentum tensor and in a self-dual version of the Cotton tensor (or curvature current).

In Section \ref{sec:Sec3}, the formalism is applied to a well-studied setting: perturbations of Schwarzschild black holes in GR, including linear matter sources. The standard equations in the literature---the Regge--Wheeler \cite{PhysRev.108.1063}, Bardeen--Press \cite{Bardeen:1973xb} (or Teukolsky \cite{Teukolsky:1973ha}), and Zerilli \cite{PhysRevLett.24.737} equations---are shown to arise naturally within this framework. The relation between the variables introduced in \cite{Mukkamala:2024dxf} and earlier observations by Aksteiner and Andersson \cite{Aksteiner:2010rh} (see also \cite{Shah:2016juc}) is also clarified. It is further shown that the full metric perturbation in the frequency domain can be reconstructed from any of the proposed master variables using purely differential reconstruction maps, involving only the variables and their derivatives. Obstructions to such reconstructions in the time domain are identified, in agreement with the findings of \cite{Poisson:2025oic}. Section \ref{sec:Conclusions} concludes with a discussion of future directions.

\section{Self-dual Variables and Fundamental Curvature Equations}\label{sec:Sec1}

The aim of this section is to introduce the self-dual variables employed in this formulation and to derive the exact wave equations they satisfy. For generality, no assumptions are made on the spacetime dynamics, and Einstein's equation is written as
\begin{equation}\label{eq:Einstein}
    G_{\mu\nu}=T_{\mu\nu}\,,
\end{equation}
where $T_{\mu\nu}$ is left unspecified. Thus, all equations presented here are geometric identities whenever $T_{\mu\nu}$ is replaced by the Einstein tensor $G_{\mu\nu}$. In GR, $T_{\mu\nu}$ is identified with the usual energy-momentum tensor (modulo prefactors), while in more general theories it may be replaced by other tensors, such as curvature densities in higher-derivative theories. In what follows, $T_{\mu\nu}$ will nevertheless be referred to as the energy-momentum tensor. Although its precise nature is not specified (except in Section \ref{sec:Sec3}), consistency of \eqref{eq:Einstein} requires that $T_{\mu\nu}$ be conserved,
\begin{equation}\label{eq:ConsT}
    \nabla^{\mu}T_{\mu\nu}=0\,.
\end{equation}
The starting point is the \textit{self-dual Weyl tensor}, defined as\footnote{Sometimes it is customary to include a factor of $1/2$. Here it is more convenient to omit it.}
\begin{equation}\label{eq:C}
    \C_{\mu\nu\rho\sigma} \coloneqq C_{\mu\nu\rho\sigma}-i\frac{1}{2}\epsilon_{\mu\nu\alpha\beta}\tensor{C}{^\alpha^\beta_\rho_\sigma}\, ,
\end{equation}
where $C_{\mu\nu\rho\sigma}$ is the Weyl tensor. It inherits the algebraic symmetries of the Riemann tensor and, in addition, satisfies
\begin{equation}\label{eq:algC}
\begin{aligned}
     &\C^{\mu}_{\ \nu\mu\sigma}=0 \quad \text{\textit{(tracefree)}}\,,\\
     &\epsilon_{\mu\nu\alpha\beta}\tensor{\C}{_\rho_\sigma^\alpha^\beta}=\epsilon_{\rho\sigma\alpha\beta}\tensor{\C}{_\mu_\nu^\alpha^\beta} \quad \text{\textit{(equal left and right duals)}}\,,\\
     &\C_{\mu\nu\rho\sigma}+i \frac{1}{2}\tensor{\epsilon}{_\mu_\nu^\alpha^\beta}\tensor{\C}{_\alpha_\beta_\rho_\sigma}=0\, \quad \text{\textit{(selfdual)}}\,.
\end{aligned}
\end{equation}
The proof of these identities is given in Appendix \ref{sec:Ap1}. The first and second identities in \eqref{eq:algC} are also satisfied by the Weyl tensor, but in general not by the Riemann tensor (unless $T_{\mu\nu}=0$, in which case the Weyl and Riemann tensors coincide) \cite{Penrose:1985bww}.

The next step is to derive a wave equation for $\C$. In vacuum GR, $\C$ satisfies a homogeneous quadratic wave equation, as first observed by Penrose in the 2-spinor formalism \cite{Penrose:1960eq}. In the presence of sources, it acquires a contribution proportional to $T_{\mu\nu}$ and its derivatives, which is to be expressed in a self-dual form. To this end, the \textit{self-dual curvature current} is introduced as
\begin{equation}\label{eq:Jcurrent}
    \tensor{\J}{_\mu^\nu^\rho}\coloneqq \nabla^{[\nu}\tilde{T}^{\rho]}_{\ \mu}-i\frac{1}{2}\tensor{\epsilon}{_\alpha_\beta^\nu^\rho}\nabla^{[\alpha}\tilde{T}^{\beta]}_{\ \mu}\quad\text{where}\quad\tilde{T}_{\mu\nu}\coloneqq T_{\mu\nu}-\frac{1}{3}T^{\alpha}_{\ \alpha}g_{\mu\nu}\, .
\end{equation}
This quantity is natural because it satisfies
\begin{equation}\label{eq:difC}
    \nabla^{\mu}\tensor{\C}{_\mu_\nu^\rho^\sigma}=\tensor{\J}{_\nu^\rho^\sigma}\, ,
\end{equation}
so that it acts as the source for $\C$. It can be seen as a self-dual version of the Cotton tensor \cite{Penrose:1985bww}. It obeys the following algebraic symmetries, together with the identity $\tensor{\J}{_\mu^{(\rho}^{\sigma)}}=0$,
\begin{equation}\label{eq:algsdJ}
\begin{aligned}
    &\J_{\mu\nu\rho}+i\frac{1}{2}\tensor{\epsilon}{_\nu_\rho^\alpha^\beta}\J_{\mu\alpha\beta}=0 \quad \text{\textit{(selfdual)}}\, ,\\
    &\tensor{\J}{_\mu^\mu_\nu}=0 \quad \text{\textit{(tracefree)}}\,,\\
    &\J_{[\mu\nu\rho]}=0 \quad \text{\textit{(cyclic identity)}}\, .
\end{aligned}
\end{equation}
Applying a suitable derivative operator on \eqref{eq:difC} and performing straightforward, though lengthy, tensor manipulations yields the desired wave equation for $\C$,
\begin{equation}\label{eq:triC2}
\begin{aligned}
    \Box \tensor{\C}{_\mu_\nu^\rho^\sigma}+4\tensor{\C}{_{[\mu\vert}^{\lambda\gamma}^{[\rho}}\tensor{\C}{^{\sigma]}_{(\lambda\gamma)}_{\vert \nu]}}+\tensor{\C}{^\lambda^\gamma^\delta^{[\rho}}\delta^{\sigma]}_{[\mu}\tensor{\C}{_{\nu]}_\delta_\lambda_\gamma}&=2\nabla_{[\mu}\tensor{\J}{_{\nu]}^\rho^\sigma}-i\tensor{\epsilon}{_\mu_\nu^\alpha^\beta}\nabla_{\alpha}\tensor{\J}{_\beta^\rho^\sigma}+2 T^{\gamma}_{\lambda}\delta^{[\rho}_{[\mu\vert}\tensor{\C}{^{\sigma]}^\lambda_\gamma_{\vert\nu]}}\\
    &-T^{[\rho}_{\lambda}\tensor{\C}{^{\sigma]}^\lambda_\mu_\nu}-T^{\gamma}_{[\mu}\tensor{\C}{_{\nu]}_\gamma^\rho^\sigma}-T^{\alpha}_{\ \alpha}\tensor{\C}{_\mu_\nu^\rho^\sigma}\, ,
\end{aligned}
\end{equation}
where $\Box=\nabla^{\mu}\nabla_{\mu}$. We notice that the right hand side vanishes in the vacuum of GR, so this equation is a tensor analogue of Eq.~\eqref{eq:Psi_Eq}. The derivations of \eqref{eq:difC}, \eqref{eq:algsdJ}, and \eqref{eq:triC2} follow from standard tensor identities and are presented in detail in Appendix \ref{sec:Ap1}.

Equations \eqref{eq:difC} and \eqref{eq:triC2} form the basis for the analysis of fluctuations in the next section. Before proceeding, it is useful to highlight a geometric interpretation that facilitates the derivation of \eqref{eq:triC2}. Following \cite{Bini:2002jx,Mukkamala:2024dxf} (using the notation of \cite{Mukkamala:2024dxf}), $\tensor{\J}{_\mu^\alpha^\beta}$ and $\tensor{\C}{_\mu_\nu^\alpha^\beta}$ can be viewed as tensor-valued one- and two-forms, respectively. In this language, equations \eqref{eq:difC} and \eqref{eq:triC2} take the compact form
\begin{equation}
    \star D\star \C=\J\, ,\quad \triangle \C=\left(D-i\star D\right)\J\, , 
\end{equation}
where $D$ and $\star$ denote the exterior covariant derivative and Hodge dual acting on tensor-valued differential forms, and $\triangle=\star D\star D+D\star D\star$ is the associated harmonic operator (see \cite{Mukkamala:2024dxf} and Appendix \ref{sec:Ap1} for details). In this formulation, $\J$ quantifies the failure of $\C$ to be a harmonic curvature two-form, i.e. $\C\in\text{Ker}{\triangle}$. The analogy with electromagnetism becomes manifest upon the replacement $\C\mapsto\mathbb{F}$, where $\mathbb{F}_{\mu\nu}=F_{\mu\nu}-i\star F_{\mu\nu}$ is the self-dual Maxwell field strength \cite{Mukkamala:2024dxf}. It would be of interest to explore the relation of this perspective with the weighted de Rham operators introduced in \cite{Edgar:2005zr}.

\section{Perturbation Theory}\label{sec:Sec2}

In this section, linear fluctuations of the metric and the energy-momentum tensor on a spherically symmetric background spacetime are considered. After describing the background geometry in Section \ref{sec:Sec2.1}, the treatment of fluctuations is presented in Sections \ref{sec:Sec2.2}. In Section \ref{sec:Sec2.2.2}, the properties of the self-dual curvature variables are analysed in detail, focusing on the even-odd splitting of their real and imaginary parts and their relation to the NP curvature scalars. The fundamental equations governing the fluctuations are then presented in Section \ref{sec:Sec2.3}, obtained by linearising the nonperturbative curvature equations derived above. The section concludes in Section \ref{sec:Sec2.4} with the presentation of a two-dimensional version of the GHP formalism, which is useful for projecting the linearised fundamental equations.

\subsection{Background Spacetime and Conventions}\label{sec:Sec2.1}

Following the covariant approach of \cite{Moncrief:1974am,PhysRevD.12.1526,Gerlach:1979rw,Gerlach:1980tx,Gundlach:1999bt,Martin-Garcia:2000cgm} (further developed in \cite{Martel:2005ir,Spiers:2023mor,Pereniguez:2023wxf}), the background metric and energy-momentum tensor are given by
\begin{align}\label{linel}
ds^{2}&=g_{ab}(y)dy^{a}dy^{b}+r^{2}(y)\Omega_{AB}(z)dz^{A}dz^{B}\,, \\ \notag \\ \label{T}
T&=T_{ab}(y)dy^{a}dy^{b}+r^{2}(y)\mathcal{P}(y)\Omega_{AB}dz^{A}dz^{B}\, .
\end{align}
These quantities are defined on a manifold with structure $M=\mathcal{N}^{2}\times S^{2}$. The fields $g_{ab}(y)$ and $r^{2}(y)$ are a Lorentzian metric and a positive function on the two-dimensional manifold $\mathcal{N}^{2}$, parametrised by coordinates $y^{a}$ (with $a=1,2$). Similarly, $T_{ab}(y)$ and $\mathcal{P}(y)$ are a symmetric tensor and a scalar function on $\mathcal{N}^{2}$. The coordinates $z^{A}$ (with $A=3,4$) parametrise the unit round two-sphere $S^{2}$ with metric $\Omega_{AB}(z)$. General covariance is retained separately on $\mathcal{N}^{2}$ and $S^{2}$, so $y^{a}$ and $z^{A}$ remain arbitrary coordinates. Indices on each submanifold are raised and lowered with the corresponding metrics $g_{ab}$ and $\Omega_{AB}$. Their associated covariant derivatives and curvature tensors are denoted by $\nabla$ (or ``$:$'') and $R^{a}_{\ bcd}$ on $\mathcal{N}^{2}$, and by $D$ and $\mathcal{R}^{A}_{\ BCD}$ on $S^{2}$. It is also convenient to introduce the 1-form $r_{a}\equiv \nabla_{a}r$. (The symbol $\nabla$ was used for the four-dimensional covariant derivative in Section \ref{sec:Sec1}, but from now on it denotes the covariant derivative on $\mathcal{N}^{2}$.) All geometric quantities of the full background spacetime, such as the Riemann, Ricci, and Einstein tensors, can be expressed in terms of these submanifold structures. In particular, the background Einstein equations take the form
\begin{equation}\label{eq:bgrules}
\begin{aligned}
    \nabla_{a}r_{b}&=\frac{1-r^{a}r_{a}}{2 r}g_{ab}-\frac{r}{2}\left(T_{ab}-T^{c}_{\ c} g_{ab}\right)\, ,\\
     R&=T^{a}_{\ a}-2 \mathcal{P} +2 \frac{1-r^{a}r_{a}}{r^{2}}\, ,
\end{aligned}
\quad (\text{\textit{background Einstein equations}})
\end{equation}
where $R$ denotes the Ricci scalar of $g_{ab}$. \textit{In this work Eqs.~\eqref{eq:bgrules} will be used automatically to eliminate $R$ and derivatives of $r_{a}$}. Besides simplifying the equations, this prescription allows for a straightforward evaluation in the vacuum of GR. 

Next, expansions of several geometric objects are presented. While standard quantities such as connection components or the Riemann tensor can be found in e.g. \cite{Martel:2005ir,Pereniguez:2023wxf}, the expansions of $\C$ and $\J$ are less commonly discussed. These expressions will be required for the linearisation of the curvature equations introduced in Section \ref{sec:Sec1}.
\begin{itemize}
    \item \textbf{Volume form}: the total volume form $\epsilon^{(4)}=(1/4!)\epsilon^{(4)}_{\mu\nu\rho\sigma}dx^{\mu}\wedge dx^{\nu}\wedge dx^{\rho}\wedge dx^{\sigma}$ reads
    \begin{equation}\label{eq:volformdec}
        \epsilon^{(4)}=r^{2}\varepsilon \wedge \epsilon=\frac{r^{2}}{4}\varepsilon_{ab}\epsilon_{AB}dy^{a}\wedge dy^{b}\wedge dz^{A}\wedge dz^{B}\,,
    \end{equation}
    where $\varepsilon_{ab}$ and $\epsilon_{AB}$ denote the volume forms on $\mathcal{N}^{2}$ and $S^{2}$, respectively.
    
    \item \textbf{Selfdual Weyl tensor}: using \eqref{eq:volformdec}, one finds 
    \begin{equation}
        \begin{aligned}\label{eq:sdWdec}
            &\C_{abcd}=-\mathcal{C}\varepsilon_{ab}\varepsilon_{cd}\, , \\  
            &\C_{abAB}=-ir^{2}\mathcal{C}\varepsilon_{ab}\epsilon_{AB}\, , \\ 
            &\C_{aAbB}=-\frac{r^{2}}{2}\mathcal{C}\left(g_{ab}\Omega_{AB}+i\varepsilon_{ab}\epsilon_{AB}\right)\, , \\
            &\C_{ABCD}=r^{4}\mathcal{C}\epsilon_{AB}\epsilon_{CD}\, , \\ 
        \end{aligned}
    \end{equation}
    where the function $\mathcal{C}= \frac{1-r^{a}r_{a}}{r^{2}}+\frac{1}{3}T^{a}_{\ a}-\frac{1}{3}\mathcal{P}$ has been introduced, and the remaining components either vanish or are related to \eqref{eq:sdWdec} by algebraic symmetries. In particular, the structure of $\C$ is strongly constrained by spherical symmetry, with a single independent component given by $\mathcal{C}$. It follows that the background is always of Petrov type $D$, without additional assumptions such as time symmetry.

    \item \textbf{Selfdual curvature current (or Cotton tensor)}: finally, the background value of $\J$ is
    \begin{equation}
        \begin{aligned}\label{eq:sdJdec}
            &\J_{abc}=\mathcal{Q}_{a}\varepsilon_{bc}\, , \\ 
            &\J_{aAB}=i r^{2}\mathcal{Q}_{a}\epsilon_{AB}\, , \\ 
            &\J_{ABa}=\mathcal{P}_{a}\Omega_{AB}-i \varepsilon_{ab}\mathcal{P}^{b}\epsilon_{AB}\, ,
        \end{aligned}
    \end{equation}
    where the quantities
    \begin{equation}
    \begin{aligned}\label{eq:sdJdecCurrents}
        \mathcal{Q}_{a}&\equiv \frac{1}{2}\varepsilon^{de} \nabla_{e}T_{ad}- \frac{1}{6}\left(\nabla^{d}T^{e}{}_{e} +  2 \nabla^{d}\mathcal{P}\right)\varepsilon_{ad}  \, ,\\
        \mathcal{P}_{a}&\equiv \frac{r^2}{6}\nabla_{a} \left(T^{b}{}_{b}-\mathcal{P}\right)+\frac{r r^{b}}{2}\left(T_{ab}-\mathcal{P} g_{ab}\right)\, .
    \end{aligned}
    \end{equation}
    have been introduced. As before, the remaining components either vanish or are related to \eqref{eq:sdJdec} by algebraic symmetries.
\end{itemize}
All background expansions presented here can be reproduced using the associated \texttt{mathematica} notebook \cite{pereniguez2026ptcwe}.

\subsection{First-order Fluctuations}\label{sec:Sec2.2}

The traditional RWZ approach to perturbation theory on spherically symmetric backgrounds consists in expanding the fluctuations of the metric $\delta g_{\mu\nu}$ (denoted as usual by $h_{\mu\nu}$), the energy-momentum tensor $\delta T_{\mu\nu}$, and the linearised Einstein equations in spherical tensor harmonics. In contrast, a key feature of the present approach is to treat the fluctuations $\delta\C_{\mu\nu\rho\sigma}$ and $\delta\J_{\mu\nu\rho}$ formally as independent quantities, although they are entirely determined by $h_{\mu\nu}$ and $\delta T_{\mu\nu}$. This choice is motivated by the fact that the components of $\delta\C_{\mu\nu\rho\sigma}$ and $\delta\J_{\mu\nu\rho}$ provide natural variables for the linearised equations, while their explicit expressions in terms of $h_{\mu\nu}$ and $\delta T_{\mu\nu}$ are comparatively complicated and lack a transparent structure.

This section presents the expansions of $h_{\mu\nu},\delta T_{\mu\nu},\delta \C_{\mu\nu\rho\sigma}$, and $\delta \J_{\mu\nu\rho}$ in terms of spherical tensor harmonics. Complex-valued spherical harmonics $Y^{l m}$ are employed, satisfying
\begin{equation}
    \left(D^{A}D_{A}+\Lambda^{2}\right)Y^{l m}=0\,, \quad\text{with}\quad \Lambda^{2}=l(l+1)\,,\quad   l\in Z_{\geq0}\, ,
\end{equation}
from which the even $\{Y^{l m},Z^{l m}_{A},U^{l m}_{AB},V^{l m}_{AB}\}$ and odd $\{X^{l m}_{A},W^{l m}_{AB}\}$ tensor harmonics are constructed as\footnote{The conventions of \cite{Pereniguez:2023wxf} are followed, where $X_{A},W_{AB}$ differ by a sign from those of \cite{Martel:2005ir,Spiers:2023mor}.}
\begin{equation}\label{eq:HarmEv}
     Z^{l m}_{A}=D_{A}Y^{l m}\, ,\quad U^{l m}_{AB}=Y^{l m}\Omega_{AB}\, , \quad V^{l m}_{AB}=D_{A}D_{B}Y^{l m}+\frac{1}{2}l(l+1)U^{l m}_{AB}\,,
\end{equation}
and
\begin{equation}\label{eq:HarmOd}
     X^{l m}_{A}=\epsilon_{AB}D^{B}Y^{l m}\, ,\quad W^{l m}_{AB}=D_{(A}X^{l m}_{B)}\, .
\end{equation}
The distinct parity properties of these two sets of harmonics and more general ones are discussed in Section \ref{sec:Sec2.2.2}.

\begin{itemize}
    \item \textbf{Metric}: the metric perturbation $h=h_{\mu\nu}dx^{\mu}dx^{\nu}$ is written as
    \begin{equation}
    \begin{aligned}
              h=\sum_{l,m}\Bigl\{&h^{l m+}_{ab} Y^{l m}dy^{a}dy^{b}+2\left(h^{l m+}_{a}Z^{l m}_{A}+h^{l m-}_{a}X^{l m}_{A}\right)dy^{a}dz^{A}\\ & +\left(h^{l m+}U^{l m}_{AB}+h^{l m \oplus}V^{l m}_{AB}+h^{l m-}W^{l m}_{AB}\right)dz^{A}dz^{B}\Bigr\}\, .
    \end{aligned}
    \end{equation}
    The harmonic components $h^{l m \pm},h^{l m \oplus},h^{l m \pm}_{a},h^{l m +}_{ab}$ are (symmetric) tensors on $\mathcal{N}^{2}$. The superscripts $\pm$ and $\oplus$ (partly following \cite{Spiers:2023mor}) indicate whether the components multiply even ($+$ and $\oplus$) or odd ($-$) tensor harmonics. To simplify notation, the harmonic sum $\sum_{l,m}$ and the labels $l,m$ will be omitted henceforth, yielding
    \begin{equation}
    \begin{aligned}\label{eq:hPert}
        h=&h^{+}_{ab} Y dy^{a}dy^{b}+2\left(h^{+}_{a}Z_{A}+h^{-}_{a}X_{A}\right)dy^{a}dz^{A}\\
        +&\left(h^{+}U_{AB}+h^{\oplus}V_{AB}+h^{-}W_{AB}\right)dz^{A}dz^{B}\, .
    \end{aligned}
    \end{equation}

    \item \textbf{Energy-momentum tensor}: in the conventions above, $\delta T=\delta T_{\mu\nu}dx^{\mu}dx^{\nu}$ is written as
    \begin{equation}
    \begin{aligned}\label{eq:TPert}
        \delta T=&\Phi^{+}_{ab} Y dy^{a}dy^{b}+2\left(\Phi^{+}_{a}Z_{A}+\Phi^{-}_{a}X_{A}\right)dy^{a}dz^{A}\\
        +&\left(\Phi^{+}U_{AB}+\Phi^{\oplus}V_{AB}+\Phi^{-}W_{AB}\right)dz^{A}dz^{B}\, ,
    \end{aligned}
    \end{equation}
    where $\Phi^{\pm},\Phi^{\oplus},\Phi^{\pm}_{a},\Phi^{+}_{ab}$ are (symmetric) tensors on $\mathcal{N}^{2}$.
    
    \item \textbf{Selfdual Weyl tensor}: since $\C_{\mu\nu\rho\sigma}$ has the algebraic symmetries of the Riemann tensor, the expansion of $\delta \C_{\mu\nu\rho\sigma}$ in tensor spherical harmonics takes the form
    \begin{equation}\label{eq:sdCPert}
        \begin{aligned}
            \delta \C_{abcd} &= \Psi^{+} Y  {\varepsilon}_{ab} {\varepsilon}_{cd},\\
            \delta \C_{Abcd} &= \left(\Psi^{+}_{b} Z_A + \Psi^{-}_{b} X_A \right){\varepsilon}_{cd},\\
            \delta \C_{AaBb} &= \frac{1}{2}\Psi^{-}\varepsilon_{ab}Y {\epsilon}_{AB} + \Psi^{+}_{ab} U_{AB} + \Psi^{\oplus}{}_{ab} V_{AB} + \Psi^{-}_{ab} W_{AB},\\
            \delta \C_{abAB} &= \Psi^{-} Y {\varepsilon}_{ab}  {\epsilon}_{AB},\\
            \delta \C_{ABDb} &=\left( \Xi^{-}_{b} Z_D + \Xi^{+}_{b} X_D\right) {\epsilon}_{AB},\\
            \delta \C_{ABCD} &= \Xi^{+} Y {\epsilon}_{AB} {\epsilon}_{CD}.
        \end{aligned}
    \end{equation}
    Here, $\Psi^{\pm},\Psi^{\pm}_{a},\Psi^{\pm}_{ab},\Psi^{\oplus}{}_{ab}$ and $\Xi^{\pm},\Xi^{\pm}_{a}$ are (symmetric) tensors on $\mathcal{N}^{2}$, and the remaining components of $\delta\C_{\mu\nu\rho\sigma}$ follow from the algebraic Riemann symmetries. In \eqref{eq:sdCPert}, only the symmetries $\delta \C_{\mu\nu(\rho\sigma)}=0$ and $\delta \C_{\mu[\nu\rho\sigma]}=0$ have been imposed, while tracelessness and selfduality (see \eqref{eq:algC}) are not yet enforced. These conditions yield additional algebraic relations among the components of $\delta \C_{\mu\nu\rho\sigma}$ and those of $h_{\mu\nu}$, and will be included in Section \ref{sec:Sec2.3} as part of the fundamental equations. The sign superscripts of $\Xi^{+}$ and $\Xi^{\pm}_{a}$ account for the parity properties of the corresponding tensor harmonics: $Y\epsilon_{AB}\epsilon_{CD}$ is even, while $Z_{A}\epsilon_{BC}$ and $X_{A}\epsilon_{BC}$ define rank-3 tensor harmonics of odd and even parity, respectively (see Section \ref{sec:Sec2.2.2}).
    
    \item \textbf{Selfdual curvature current (or Cotton tensor)}: using $\delta \J_{\mu(\nu\rho)}=0$ and $\delta \J_{[\mu\nu\rho]}=0$, the harmonic expansion of $\delta\J_{\mu\nu\rho}$ takes the form
        \begin{equation}
        \begin{aligned}\label{eq:sdJPert}
            \delta \J_{abc} &= \Pi^{+}_{a} Y  {\varepsilon}_{bc} ,\\
            \delta \J_{abA} &= \left(\Pi^{+} \varepsilon_{ab}+\Pi^{+}_{ab}\right)Z_{A}+\left(\Pi^{-}       \varepsilon_{ab}+\Pi^{-}_{ab}\right)X_{A},\\
            \delta \J_{aAB} &= -2 \Pi^{-}_{a}Y \epsilon_{AB},\\
            \delta \J_{Aab} &=-2\left( \Pi^{+} Z_{A}+\Pi^{-}  X_{A}\right)\varepsilon_{ab} ,\\
            \delta \J_{ABa} &=\Pi^{-}_{a}Y\epsilon_{AB}+\Gamma^{+}_{a} U_{AB}+\Gamma^{-}_{a}W_{AB}+ \Pi^{\oplus}_{a}V_{AB}\,,\\
            \delta \J_{ABC} &= \left(\Gamma^{-} Z_{A}+\Gamma^{+} X_{A}\right)\epsilon_{BC},
        \end{aligned}
        \end{equation}
    where $\Pi^{\pm},\Pi^{\pm}_{a},\Pi^{\oplus}_{a},\Pi^{\pm}_{ab}$ and $\Gamma^{\pm},\Gamma^{\pm}_{a}$ are (symmetric) tensors on $\mathcal{N}^{2}$. Tracefreeness and selfduality of $\J_{\mu\nu\rho}$ (see \eqref{eq:algsdJ}) have not been imposed in \eqref{eq:sdJPert}; these conditions provide additional relations among the components of $\delta\J_{\mu\nu\rho}$ and those of $h_{\mu\nu}$.
\end{itemize}

\subsubsection{Gauge Considerations}\label{sec:Sec2.2.1}

Diffeomorphisms constitute the gauge symmetry of the theory, and are generated by vector fields $\xi_{\mu}$. If $\delta A$ denotes the fluctuation of a tensor field $A$ of (a not shown) arbitrary index structure, the action of a gauge transformation on $\delta A$ is given by
\begin{equation}\label{eq:gaugeaction}
    \delta A\mapsto \delta A-\pounds_{\xi}A\, ,
\end{equation}
where $\pounds_{\xi}A$ denotes the Lie derivative of the background value of $A$ along $\xi_{\mu}$. Both sides of \eqref{eq:gaugeaction} represent physically equivalent configurations. The fluctuation expansions introduced above were written in an arbitrary gauge. In order to simplify the equations, it is customary to adopt the RW gauge, defined by $h^{+}_{a}=h^{\oplus}=h^{-}=0$, which can always be imposed for multipoles with $l\geq2$, as considered in this work. Instead, a closely related approach will be adopted, which is formally equivalent but conceptually preferable, as it is based on gauge-invariant variables. The discussion that follows is standard; only the main steps are summarised here, and the reader is referred to \cite{Martel:2005ir,Pereniguez:2023wxf} for details.

Expanding the gauge parameter $\xi_{\mu}$ in spherical harmonics,
\begin{equation}\label{eq:xi_gauge}
\xi=\xi^{+}_{a}Ydy^{a}+\left[\xi^{+} Z_{A}+\xi^{-}  X_{A}\right]dz^{A}\, ,
\end{equation}
one finds that the fluctuation-dependent vector field 
\begin{equation}
    \eta[h]=\eta^{+}_{a}[h] dy^{a}+\left(\eta^{+}[h]Z_{A}+\eta^{-}[h]X_{A}\right)dz^{A}\,,
\end{equation}
with 
\begin{align}\label{vec1}
\eta^{+}_{a}[h]= -h^{+}_{a}+\frac{r^{2}}{2}\nabla_{a}\left(\frac{h^{\oplus}}{r^{2}}\right)\, ,\quad \eta^{+}[h]= -\frac{h^{\oplus}}{2}\, ,\quad \eta^{-}[h]= -\frac{h^{-}}{2}\, ,
\end{align}
transforms under a gauge transformation according to
\begin{equation}
\eta_{\mu}[h] \mapsto \eta_{\mu}[h]+\xi_{\mu}\, ,
\end{equation}
for multipoles with $l\geq2$, excluding the lower-order modes $l=0,1$. It follows that the harmonic components of order $l\geq2$ of the quantity
\begin{equation}\label{eq:gaugecomp}
   \tilde{\delta A} \equiv \delta A+\pounds_{\eta[h]}A\, ,
\end{equation}
are automatically gauge-invariant. \textit{Henceforth, the gauge-invariant perturbations defined in \eqref{eq:gaugecomp} will be employed, and the tilde will be omitted for notational simplicity. This procedure is formally equivalent to working in the RW gauge where $h^{+}_{a}=h^{\oplus}=h^{-}=0$}, as discussed in \cite{Martel:2005ir,Pereniguez:2023wxf}, and both viewpoints are compatible with the results presented here.

The harmonic expansion of the gauge transformations $\pounds_{\xi}A$ for the tensors considered here, namely $A=\{g_{\mu\nu},T_{\mu\nu},\C_{\mu\nu\rho\sigma}, \J_{\mu\nu\rho}\}$, can be obtained straightforwardly using the associated \texttt{mathematica} notebook \cite{pereniguez2026ptcwe}. Only those expressions relevant for the present analysis will be discussed explicitly in Section \ref{sec:Sec3.2}.

\subsection{Structure of Self-dual Variables}\label{sec:Sec2.2.2}

A central feature of this approach is that the perturbations are described in terms of the quantities $\delta\C$ and $\delta\J$, which are genuine complex-valued tensor fields. When expressed in terms of the metric and stress-energy perturbations, their real and imaginary parts depend exclusively on the even or odd components of $h$ and $\delta T$, respectively. This separation is a characteristic property of self-dual variables, reminiscent of an earlier observation by Price \cite{Price_Despun}, and it will play an important role for isospectrality (see Section \ref{sec:Sec2.2.3}). The purpose of this section is to explain this even/odd splitting for the self-dual variables. Along the way, the standard harmonic and parity decomposition of perturbations is briefly reviewed, and well-known results such as the decoupling of even and odd gravitational perturbations in GR are recovered from symmetry arguments.

The discussion proceeds in an abstract manner, omitting tensor indices. Let $A$ denote a tensor field of arbitrary rank satisfying
\begin{equation}\label{eq:eq0}
    E[A]=0\, ,
\end{equation}
where $E[A]$ is another tensor constructed locally from $A$ (e.g.~$G[g]$, where $G$ is the Einstein tensor and $g$ the spacetime metric). It is assumed that $E[A]$ is \textit{generally covariant} in the sense of \cite{Wald:1993nt}, namely that under any diffeomorphism $f$ one has
$f_{*}\!\left(E[A]\right)=E[f_{*}A]$, where $f_{*}$ denotes the pushforward under $f$. Expanding about a background, $A\to A+\delta A$, yields $E[A]\to E[A]+\delta E[\delta A]$. To leading order in $\delta A$, Eq.~\eqref{eq:eq0} becomes
\begin{equation}\label{eq:eq1}
    \delta E[\delta A]=0\, ,
\end{equation}
where $\delta E$ is regarded as a linear operator acting on $\delta A$. If the background admits a Killing vector $K$, such that $\pounds_{K}A=0$, general covariance implies
\begin{equation}\label{eq:KillingCom}
    [\pounds_{K},\delta E]=0\, .
\end{equation}
For a spherically symmetric background, there exist Killing vectors $K_{x},K_{y},K_{z}$ generating rotations about the $x$, $y$, and $z$ axes and spanning the $SO(3)$ algebra. From \eqref{eq:KillingCom}, the Casimir operator
$L^{2}\equiv \pounds_{K_{x}}^{2}+\pounds_{K_{y}}^{2}+\pounds_{K_{z}}^{2}$ satisfies
\begin{equation}\label{eq:Casimir}
    [L^{2},\delta E]=0\, .
\end{equation}
It follows that solutions to \eqref{eq:eq1} can be sought in eigenfunctions of $L^{2}$. The spherical tensor harmonics form a complete set of such eigenfunctions on the two-sphere, so any smooth tensor field admits a decomposition of the form
\begin{equation}
    \delta A=\sum_{l,m}\delta A^{l m}\,, \qquad L^{2}\delta A^{l m}=l(l+1)\delta A^{l m}\,, \qquad \pounds_{K_{z}}A^{l m}=imA^{l m}\,,\qquad l\geq\lvert m\rvert\geq 0\, .
\end{equation}
The tensor harmonics introduced in \eqref{eq:HarmEv} and \eqref{eq:HarmOd}, as well as the higher-rank harmonics appearing in \eqref{eq:sdCPert} and \eqref{eq:sdJPert} (e.g.\ $Z_{D}\epsilon_{AB}$ or $X_{D}\epsilon_{AB}$), are eigenfunctions of $L^{2}$ with eigenvalues $l(l+1)$. This follows from the relations
\begin{equation}
    \pounds_{K_{x^{i}}}\Omega=0\, ,\quad \pounds_{K_{x^{i}}}\epsilon=0\,, \quad [\pounds_{K_{x^{i}}},D]=0\,, \quad L^{2}\phi=-D^{A}D_{A}\phi\, ,
\end{equation}
where $K_{x^{i}}$ denotes any rotational Killing vector and $\phi$ is any scalar function. The usefulness of this harmonic expansion lies in the fact that, in equations such as \eqref{eq:eq1}, different harmonic levels decouple and no mixing between distinct $(l,m)$ modes occurs. Indeed, expanding $\delta E[\delta A]=\sum_{l,m}\delta E^{l m}[\delta A]$, Eqs.~\eqref{eq:Casimir} and \eqref{eq:KillingCom} imply that $l(l+1)\delta E^{l m}[\delta A]=\delta E^{l m}[L^{2}\delta A]$ and $i m\delta E^{l m}[\delta A]=\delta E^{l m}[\pounds_{K_{z}}\delta A]$ for arbitrary $\delta A$. Evaluating these relations on a single harmonic mode $\delta A=\delta A^{l'm'}$ yields
\begin{equation}
    \delta E^{l m}[\delta A^{l'm'}]=\delta^{l l'}\delta^{m m'} \delta E^{l m}[\delta A^{l m}] \quad .
\end{equation}
All equations considered in this work are of the form $E[A]=0$, where $E$ is generally covariant in the sense described above, when regarded as a function of the dynamical fields $g,T,\C,\J$. It is therefore sufficient to analyse perturbations at a fixed harmonic level $(l,m)$ (the labels will be suppressed in what follows unless needed).

A spherically symmetric background also admits the \textit{antipodal map} $\mathcal{A}$, defined in coordinates by $(\theta,\phi)\mapsto(\pi-\theta,\phi+\pi)$. This map is an isometry that reverses the spacetime orientation. Since $\mathcal{A}$ is not generated by the flow of a Killing vector, its action on tensor fields is defined through the pushforward rather than a Lie derivative. The parity operator is defined as
\begin{equation}
    P\equiv (-1)^{l}\mathcal{A}_{*}\, ,
\end{equation}
where $\mathcal{A}_{*}$ denotes the pushforward under $\mathcal{A}$. It satisfies
\begin{equation}\label{eq:Prels}
    P\Omega=(-1)^{l}\Omega\,, \qquad P\epsilon=-(-1)^{l}\epsilon\,, \qquad P^{2}=1\,,\qquad [L^{2},P]=0\, ,\qquad [\pounds_{K_{z}},P]=0\, ,
\end{equation}
reflecting the fact that $\mathcal{A}$ reverses orientation, squares to the identity, and is compatible with $L^{2}$ and $\pounds_{K_{z}}$. Any tensor can be decomposed uniquely into even and odd components,
\begin{equation}
    \delta A=\delta A^{+}+\delta A^{-}\,, \qquad
    \delta A^{\pm}\equiv\frac{1}{2}\left(\delta A\pm P\delta A\right)\, ,
\end{equation}
which satisfy
\begin{equation}
    P\delta A^{\pm}=\pm \delta A^{\pm}\, .
\end{equation}
Since $[L^{2},P]=[\pounds_{K_{z}},P]=0$, the fields $\delta A^{\pm}$ are eigenstates of $L^{2}$ and $\pounds_{K_{z}}$ separately. In particular, the scalar spherical harmonics satisfy
\begin{equation}\label{eq:PY}
    PY=Y\, ,
\end{equation}
so they have even parity. Using \eqref{eq:Prels}, \eqref{eq:PY}, and the standard properties of the pushforward, the parity of all tensor harmonics appearing in \eqref{eq:hPert}, \eqref{eq:TPert}, \eqref{eq:sdCPert}, and \eqref{eq:sdJPert} can be determined. 

The conditions under which equations such as \eqref{eq:eq1} decouple into even and odd sectors can now be identified. Let $\mathcal{E}[\delta A]$ and $\mathcal{O}[\delta A]$ be linear operators that commute and anti-commute with parity, respectively, i.e.\ $P\mathcal{E}=\mathcal{E}P$ and $P\mathcal{O}=-\mathcal{O}P$. Then
\begin{equation}\label{eq:Parity}
    \mathcal{E}^{\pm}[\delta A^{\mp}]=0\,, \qquad \mathcal{O}^{\pm}[\delta A^{\pm}]=0\, .
\end{equation}
Thus, the even and odd components of $\mathcal{E}$ depend only on the even and odd parts of $\delta A$, respectively, while the opposite holds for $\mathcal{O}$. As an example, consider the Einstein tensor $G[g]$. By general covariance it satisfies $PG[g]=G[Pg]$, and at linear order $P\delta G[h]=\delta G[Ph]$. Therefore, the linearised Einstein tensor commutes with $P$, and by \eqref{eq:Parity} the even and odd sectors of $h$ satisfy decoupled equations in vacuum GR, reproducing the standard result of \cite{PhysRev.108.1063}.

We now analyse the real and imaginary parts of $\delta\C$ and $\delta\J$. The real part of $\delta\C$ is given by
\begin{equation}
    \delta\C+\delta\bar{\C}=2\delta C(h)\, ,
\end{equation}
where $\delta C(h)$ denotes the linearised Weyl tensor. By general covariance it commutes with parity, $P(\delta C(h))=\delta C(Ph)$. Therefore, the even and odd components of $\delta\C+\delta\bar{\C}$ depend only on the even and odd parts of $h$, respectively.

For the imaginary part,
\begin{equation}
    \delta\C-\delta\bar{\C}=-i\star\left(h^{\mu}{}_{\mu}C+2\delta C(h)\right)\, ,
\end{equation}
and the presence of the volume form in the Hodge dual induces a parity flip. Explicitly,
\begin{equation}
    P\!\left[-i\star\left(h^{\mu}{}_{\mu}C+2\delta C(h)\right)\right]
    =+i\star\left((Ph^{\mu}{}_{\mu})C+2\delta C(Ph)\right)\, .
\end{equation}
Thus $\delta\C-\delta\bar{\C}$, viewed as an operator acting on $h$, anti-commutes with $P$, so its even and odd components depend only on the odd and even parts of $h$, respectively.

This can be verified by explicit computations. To take the real and imaginary part of an harmonic component, one should recall that our basis of harmonics is itself complex-valued and proceed as follows. For a generic perturbation $\delta A$, reinstating the harmonic labels we have
\begin{equation}\label{eq:gen}
    \delta A=\sum_{l,m}\delta A^{l m +}Y^{l m +}+\delta A^{l m-}Y^{l m-}\, ,
\end{equation}
where $Y^{l m\pm}$ denote even ($+$) and odd ($-$) tensor spherical harmonics, and the coefficients $\delta A^{l m\pm}$ are independent of the sphere coordinates. Real and imaginary parts are obtained from $\delta A\pm\delta\bar{A}$. Using $\bar{Y}^{l m\pm}=(-1)^{m}Y^{l,-m,\pm}$, one finds
\begin{equation}
    \delta A\pm \delta\bar{A}
    =\sum_{l,m}\delta A^{l m +}_{\pm}Y^{l m+}+\delta A^{l m -}_{\pm}Y^{l m-}\, ,
\end{equation}
where
\begin{equation}\label{eq:pm}
    \delta A^{l m\bullet}_{\pm}
    =\delta A^{l m\bullet}\pm(-1)^{m}\delta \bar{A}^{l,-m,\bullet}\,, \qquad \bullet=+,-\, .
\end{equation}
If $\delta A$ is real, then $\delta A^{l m\pm}_{+}=2\delta A^{l m\pm}$ and $\delta A^{l m\pm}_{-}=0$; for complex $\delta A$ both combinations are generally non-vanishing. As an example, consider the components $\Psi^{\oplus}{}_{ab}$ and $\Psi^{-}_{ab}$ of $\delta\C$ in \eqref{eq:sdCPert}. In terms of $h$,
\begin{equation}
    \Psi^{\oplus}{}_{ab}=- \frac{1}{2} h^{+}{}_{ab} + \frac{1}{4} g_{ab} h^{+}{}^{c}{}_{c}
    -\frac{1}{2}i\left(h^{-}{}^{c}{}_{:(b}\varepsilon_{a)c} + h^{-}{}_{(b}{}^{:c}\varepsilon_{a)c}\right)\, ,
\end{equation}
and
\begin{equation}
    \Psi^{-}{}_{ab}= h^{-}{}_{(a}{}_{:b)}-\frac{1}{2}g_{ab}h^{-}{}^{c}{}_{:c}
    -\frac{1}{2}i\varepsilon_{(a}{}^{c} h^{+}{}_{b)c}\, .
\end{equation}
Taking real parts using \eqref{eq:pm},
\begin{equation}
    \Psi^{\oplus}{}_{+ab}=-  h^{+}{}_{ab} + \frac{1}{2} g_{ab} h^{+}{}^{c}{}_{c}\, ,\qquad
    \Psi^{-}{}_{+ab}= 2h^{-}{}_{(a}{}_{:b)}-g_{ab}h^{-}{}^{c}{}_{:c}\, ,
\end{equation}
so $\Psi^{\oplus}{}_{+ab}$ and $\Psi^{-}{}_{+ab}$ depend only on the even and odd parts of $h$, respectively. For the imaginary parts,
\begin{equation}
    \Psi^{\oplus}{}_{-ab}=-i\left(h^{-}{}^{c}{}_{:(b}\varepsilon_{a)c} + h^{-}{}_{(b}{}^{:c}\varepsilon_{a)c}\right)\, ,\qquad
    \Psi^{-}{}_{-ab}= -i\varepsilon_{(a}{}^{c} h^{+}{}_{b)c}\, ,
\end{equation}
so $\Psi^{\oplus}{}_{-ab}$ and $\Psi^{-}{}_{-ab}$ depend only on the odd and even parts of $h$, respectively. The same reasoning applies to $\delta\J$, leading to analogous conclusions for its real and imaginary parts.

Next, we discuss briefly how this property is useful to analyse the isospectrality of the QNM spectrum.

\subsubsection{Consequences for Black Hole Isospectrality}\label{sec:Sec2.2.3}

QNMs are free fluctuations of black holes with definite frequency $\omega$, so their behaviour in time is $\sim e^{-i\omega t}$ for a suitable time coordinate $t$. In the case of asymptotically flat black holes, the boundary conditions defining them are universal: smoothness at the future event horizon and purely outgoing behaviour at future null infinity. Such boundary conditions over-determine the problem, which only admits solutions for a discrete set of $\omega$, known as quasi normal frequencies. Isospectrality is the statement that the quasinormal frequencies of the even and odd sectors of a fluctuation coincide.

In general, the equations governing these two sectors take very different forms. It is therefore nontrivial that they are nevertheless isospectral in GR. This property has often been attributed to the Chandrasekhar transformations \cite{Chandrasekhar:1985kt}, or more generally to Darboux transformations \cite{Glampedakis:2017rar}, which consider suitable transformations of variables \textit{within} each parity sector \cite{Lenzi:2021njy,Solomon:2023ltn,Jaramillo:2024qjz,Lenzi:2025kqs,Lenzi:2025man,DeLuca:2025zqr}. In our view, the need to invoke such transformations reflects a less natural choice of variables-- at least for describing the QNM spectrum of perturbations. A natural choice would be one where isospectrality (when present) is manifest. Here we argue that self-dual variables provide such a notion. 

Suppose that the perturbation equations can be reduced to a system involving only self-dual curvature variables, schematically of the form
\begin{equation}\label{eq:formaleq}
\mathcal{O}\left(\Psi\right)=0\, ,
\end{equation}
where $\mathcal{O}$ is a collection of second-order differential operators acting on self-dual variables, collectively denoted by $\Psi$. From the analysis above, the real and imaginary parts of $\Psi$, denoted $\Psi_{+}$ and $\Psi_{-}$, depend only on the even and odd components of the metric perturbation, respectively. Since the even and odd sectors satisfy decoupled equations, it follows that $\Psi_{\pm}$ must satisfy Eq.~\eqref{eq:formaleq} independently,
\begin{equation}
\mathcal{O}\left(\Psi_{\pm}\right)=0\, .
\end{equation}
Thus, the even and odd sectors are governed by the exact same differential operators, and therefore share the same quasi-normal spectrum.

In vacuum GR, the perturbation equations can in fact be reduced to systems of the form \eqref{eq:formaleq} in several different ways, as will be shown below. Hence, in this formulation, isospectrality follows directly without the need to invoke Chandrasekhar or Darboux transformations. 

\subsubsection{Relation to the Newman--Penrose Curvature Scalars}\label{sec:Sec2.5}

The self-dual Weyl tensor is closely related to the NP curvature scalars $(\Psi_{0},\Psi_{1},\Psi_{2},\Psi_{3},\Psi_{4})$: these are precisely the dyad components of the Weyl spinor mentioned in Section \ref{sec:BasicIdea} \cite{Penrose:1960eq,Penrose:1985bww}. Consequently, the components of our curvature variables must be related in a simple way to those scalars. Establishing this relation explicitly is straightforward, but technically involved. It requires introducing null frames and their perturbations, and relating the standard spherical harmonics used here to spin-weighted spherical harmonics. We will not present the details of this computation here; they are fully documented in the accompanying \texttt{mathematica} notebook \cite{pereniguez2026ptcwe}. Instead, we report the final result. 

Because the background spacetime is of type $D$ (recall this follows from \eqref{eq:sdWdec} and holds for any matter content), the perturbations of the maximal-weight scalars $\delta \Psi_{0}$ and $\delta \Psi_{4}$ receive no contributions from the perturbation of the null frame. Restoring explicitly the sum over harmonic labels, and following the conventions of \cite{Spiers:2023mor} for the spin-weighted spherical harmonics $\tensor[_s]{Y}{^l^m}$ and those of \cite{Stephani:2003tm} for the NP curvature scalars, we obtain
\begin{equation}
    \begin{aligned}\label{eq:Psi_o_ToWeyl}
    \delta \Psi_{0}&=\frac{1}{2r^{2}}\sum_{l,m}\sqrt{(l-1)l(l+1)(l+2)}\times \ell^{a}\ell^{b}\Psi^{lm\oplus}{}_{ab}(y)\times\tensor[_2]{Y}{^l^m}\, ,\, \\
    \delta \Psi_{4}&=\frac{1}{2r^{2}}\sum_{l,m}\sqrt{(l-1)l(l+1)(l+2)}\times n^{a}n^{b}\Psi^{lm\oplus}{}_{ab}(y)\times\tensor[_{-2}]{Y}{^l^m}\, ,
\end{aligned}
\end{equation}
where $\ell^{a}$ and $n^{a}$ denote any null frame of the \textit{background} two-dimensional Lorentzian manifold $\left(\mathcal{N}^{2},g_{ab}\right)$. The remaining scalars do receive contributions from perturbations of the four-dimensional null frame. However, $\delta\Psi_{2}$ is invariant under first-order Lorentz frame rotations \cite{Price_Despun,Aksteiner:2010rh}. Its gauge transformation therefore coincides with that of a scalar,
\begin{equation}
    \delta\Psi_{2}\mapsto \delta\Psi_{2}-\pounds_{\xi}\Psi_{2}\, .
\end{equation}
Two consequences follow. First, the contribution arising from the frame perturbation is itself invariant under first-order Lorentz frame rotations. Second, a gauge-invariant quantity can be associated with $\delta\Psi_{2}$, by applying the discussion in Section \ref{sec:Sec2.2.1}. Recalling that we make no notational difference between a quantity and its gauge-invariant version, we find (again restoring the sum over $lm$) 
\begin{equation}\label{eq:Psi2}
    \delta\Psi_{2}=\frac{1}{2}\sum_{l,m}\left\{\Psi^{lm+}+\mathcal{C}\tensor{h}{^l^m^+^a_a}\right\}Y^{lm}\, ,
\end{equation}
where $\mathcal{C}$ denotes the unique component of the background self-dual Weyl tensor, given below \eqref{eq:sdWdec}.

We will show that, in the case of GR perturbations of Schwarzschild, the linearised curvature wave equation for $\C$, Eq.~\eqref{eq:triC2}, directly yields the BPT equations for $r^{-2}\ell^{a}\ell^{b}\Psi^{\oplus}{}_{ab}$ and $r^{-2}n^{a}n^{b}\Psi^{\oplus}{}_{ab}$. In contrast, one obtains the Regge--Wheeler equation for $r^{3}\Psi^{+}$. The latter observation was already made in \cite{Mukkamala:2024dxf}, where it was also noted that in this setting the Regge--Wheeler equation governs the even sector of metric perturbations as well: as discussed above, the real and imaginary parts of $\Psi^{+}$ respectively contain the even and odd components of the metric perturbation.

Our computation \eqref{eq:Psi2} shows that this variable is directly related to $\delta\Psi_{2}$ (more precisely, to its gauge-invariant version following Section \ref{sec:Sec2.2.1}). A related observation was first made in \cite{Aksteiner:2010rh} (see also \cite{Shah:2016juc}), where it was shown that the perturbation of $\Psi_{2}$ satisfies a Regge--Wheeler equation with a gauge-dependent source term. Equation \eqref{eq:Psi2} expresses this result in a gauge-invariant form and extends it to include contributions from matter, coming both from the background and the perturbations. We will also show that $\delta\Psi_{2}$ is closely related to the curvature component $\Xi^{+}$ (introduced in \eqref{eq:sdCPert}). After a suitable redefinition, the real part of this variable satisfies the Zerilli equation, following from the curvature equation for $\delta\C$.

We will not discuss the variables $\delta\Psi_{1}$ and $\delta\Psi_{3}$. In contrast with $\delta\Psi_{2}$, the contributions arising from frame perturbations in these quantities are not invariant under first-order Lorentz frame rotations. The explicit expressions for $\delta\Psi_{1,3}$ in terms of the components of $\delta \C$ and the frame perturbations are provided in the \texttt{mathematica} notebook \cite{pereniguez2026ptcwe}, but they will not play a role in the present work.

\subsection{The Fundamental Equations}\label{sec:Sec2.3}

Having introduced the variables that will be used throughout the analysis, we now derive the fundamental equations governing their dynamics. These consist of the linearisation of the Einstein equation \eqref{eq:Einstein}, energy–momentum conservation \eqref{eq:ConsT}, and the algebraic, first-order, and second-order equations satisfied by the self-dual Weyl tensor, Eqs.~\eqref{eq:algC}, \eqref{eq:difC}, and \eqref{eq:triC2}, respectively.

These equations are expressed in terms of the fluctuations of the metric \eqref{eq:hPert}, the energy–momentum tensor \eqref{eq:TPert}, the self-dual Weyl tensor \eqref{eq:sdCPert}, and the self-dual curvature current \eqref{eq:sdJPert}. In addition, the background relations \eqref{eq:bgrules} are imposed to eliminate the Ricci scalar $R$ of the background Lorentzian 2-dimensional metric $g_{ab}$, as well as derivatives of $r_a$. As discussed above, it is essential to work in terms of the self-dual variables $\delta \C$ and $\delta \J$. Accordingly, we do not substitute their expressions in terms of the metric and energy–momentum perturbations $h$ and $\delta T$ --- these relations are provided separately. 

For the purpose of describing the dynamics, the linearised Einstein equations are, of course, sufficient. The remaining relations instead play the role of structure equations, which prove extremely useful in practice. This situation is formally analogous to that encountered in the NP or GHP formalisms, where the structure equations complement Einstein’s equations and are essential for the analysis. In particular, the linearisation of the curvature wave equation \eqref{eq:triC2} automatically has the form
\begin{equation}
    \nabla_{a}\nabla^{a} \Psi=f(\nabla\Psi,\Psi,...) \qquad \text{with $\Psi$ the components of $\delta \C$}\, ,
\end{equation}
so it yields wave equations for the $\Psi$'s with well-behaved hyperbolic principal parts. In the case of vacuum fluctuations of GR, we will see that some of these reduce to decoupled master equations. 

The linearisation of the fundamental equations, their expansion in submanifold products, and the projection onto spherical tensor harmonics is a computationally demanding task. However, this can be performed efficiently using the \texttt{xAct} packages, on which the provided \texttt{Mathematica} notebook is based \cite{pereniguez2026ptcwe}. The full set of equations together with the routines used to construct them can be found there. In the absence of simplifying assumptions the resulting expressions are rather lengthy. To illustrate their structure, in Appendix \ref{sec:Ap3} we present them explicitly for perturbations about a Schwarzschild black hole in GR, including the presence of matter sources. These are the equations employed in the present work. 

\subsection{A Two-dimensional GHP Formalism}\label{sec:Sec2.4}

Some of the fundamental equations presented above are tensor equations on the two-dimensional Lorentzian manifold $(\mathcal{N}^{2},g_{ab})$. At times it is convenient to project them onto a frame basis in this \textit{background} manifold (recall that we do not take the frames as dynamical quantities; we simply use the background frames to project and manipulate the linearised equations). The two-dimensional nature of the manifold makes null frames particularly convenient. An expansion in null frames can be performed covariantly, closely following the traditional GHP formalism \cite{Geroch:1973am}, but restricted here to two-dimensional Lorentzian spaces. On the one hand, this keeps the equations compact by collecting various terms into frame-covariant derivatives. On the other hand, it makes frame invariance manifest and allows immediate consistency checks through the counting of weights of each quantity (now given by a single real number, rather than two as in the standard GHP formalism). In this section we introduce such a two-dimensional version of the GHP formalism.

Choose any null frame $\ell^{a},n^{a}$ defined by the condition
\begin{equation}\label{eq:null_frame}
    \ell^{a}\ell_{a}=n^{a}n_{a}=0 \qquad \text{and}\qquad \ell^{a}n_{a}=-1\, .
\end{equation}
The null-frame condition \eqref{eq:null_frame} is invariant under the frame ``boost''
\begin{equation}\label{eq:frame_trafo}
    \left(\ell_{a},n_{a}\right)\mapsto\left(\lambda \ell_{a},\lambda^{-1} n_{a}\right)\, ,
\end{equation}
where $\lambda$ is a function from the manifold $\mathcal{N}^{2}$ into the positive real numbers. Since the null frame can be chosen arbitrarily, any expansion of the equations in a frame must be invariant (or covariant) under \eqref{eq:frame_trafo}, that is, this transformation should be a symmetry of the equations. However, this invariance may not be manifest in a naive expansion. The goal is therefore to write the equations so that invariance under \eqref{eq:frame_trafo} is \textit{manifest}. We say that a quantity $\eta$ has GHP weight $p$ if under \eqref{eq:frame_trafo} it transforms as
\begin{equation}
    \eta\mapsto\lambda^{p}\eta\, ,
\end{equation}
and we write $\eta\overset{\circ}{=}p$. Any tensor fluctuation introduced in Section \ref{sec:Sec2.2} can be expanded in null-frame components, each of which has a definite GHP weight. For example, $h^{+}_{ab}$ has components $h^{+}_{\ell\ell}$, $h^{+}_{\ell n}$ and $h^{+}_{nn}$ (where we use the notation $h^{+}_{\ell\ell}=h^{+}_{ab}\ell^{a}\ell^{b}$, $h^{+}_{\ell n}=h^{+}_{ab}\ell^{a}n^{b}$ and $h^{+}_{nn}=h^{+}_{ab}n^{a}n^{b}$), which have weights
\begin{equation}
    h^{+}_{\ell\ell}\overset{\circ}{=}2\,,\quad  h^{+}_{\ell n}\overset{\circ}{=}0\,,\quad  h^{+}_{n n}\overset{\circ}{=}-2\,. 
\end{equation}
Other tensor perturbations are expanded similarly. The sum of two GHP quantities with the same weight again has that weight. More generally, the space of GHP quantities of fixed weight can be regarded as a vector bundle associated with the symmetry \eqref{eq:frame_trafo}. We next introduce a differential operator that is covariant with respect to \eqref{eq:frame_trafo}. 

The standard covariant derivative of any two-dimensional null frame can be written without loss of generality as
\begin{equation}
    \begin{aligned}\label{eq:Deriv_Frames}
    \nabla_{a}\ell_{b}&=-2 \varepsilon \ n_{a}\ell_{b}+2\varepsilon'\ \ell_{a}\ell_{b}\, ,\\ \\
    \nabla_{a}n_{b}&=-2 \varepsilon' \ \ell_{a}n_{b}+2\varepsilon\ n_{a}n_{b}\, ,
\end{aligned}
\end{equation}
where we introduced
\begin{equation}
    \varepsilon=\frac{1}{2}\nabla_{a}\ell^{a}\, ,\quad \varepsilon'=\frac{1}{2}\nabla_{a}n^{a}\, .
\end{equation}
From $\ell^{a}\nabla_{a}\ell_{b}= 2\varepsilon \ell_{b}$ and $n^{a}\nabla_{a}n_{b}= 2 \varepsilon' n_{b}$ it follows that both vector fields are geodesic. We can then introduce the (2-dimensional) GHP connection one-form
\begin{equation}\label{eq:omega}
    \omega_{a}=2\varepsilon'\ell_{a}-2\varepsilon n_{a}\, .
\end{equation}
Under \eqref{eq:frame_trafo} it transforms as
\begin{equation}\label{eq:omega_trafo}
    \omega_{a}\mapsto\omega_{a}+\lambda^{-1}\nabla_{a}\lambda\, ,
\end{equation}
which allows us to define a GHP covariant derivative $\Theta_{a}$. Acting on a GHP quantity $\eta$ of weight $p$, it produces another quantity $\Theta_{a}\eta$ of the same weight,
\begin{equation}\label{eq:Theta_Deriv}
    \Theta_{a}\eta\equiv\nabla_{a}\eta-p\omega_{a}\eta\, .
\end{equation}
A direct computation shows that under \eqref{eq:frame_trafo} one has $\Theta_{a}\eta\mapsto \lambda^{p}\Theta_{a}\eta$, owing to the inhomogeneous transformation law \eqref{eq:omega_trafo}. The covariant derivative $\Theta_{a}$ yields a frame-covariant notion of directional derivative along $\ell_{a}$ and $n_{a}$, defined as
\begin{equation}\label{eq:tho_thop}
    \tho\eta\equiv\ell^{a}\Theta_{a}\eta=\left(\ell^{a}\nabla_{a}-2p\varepsilon \right)\eta\, ,\qquad \tho'\eta\equiv n^{a}\Theta_{a}\eta=\left(n^{a}\nabla_{a}+2p\varepsilon' \right)\eta\, .
\end{equation}
Note these map a quantity $\eta$ with weight $\eta\overset{\circ}{=}p$ to quantities $\tho\eta$ and $\thop \eta$ with weights
\begin{equation}
    \tho\eta\overset{\circ}{=}p+1\,,\qquad \tho'\eta\overset{\circ}{=}p-1\, .
\end{equation}
To complete the expansion in the null-frame basis, we write the background warp factor $r_{a}$ as
\begin{equation}
    r_{a}=r \rho' \ell_{a}+r \rho n_{a}\, ,\qquad \text{where} \qquad \rho=-\frac{\ell^{a}r_{a}}{r}\, ,\quad  \rho'=-\frac{n^{a}r_{a}}{r}\, .
\end{equation}
Below we will see that some of the fundamental equations take particularly simple forms when expanded in a null frame and written in terms of the GHP covariant derivative $\Theta_{a}$.

Finally, it is worth noting that the equations are not, in general, invariant under the formal replacement
\begin{equation}\label{eq:l_To_n}
    \ell_{a}\longleftrightarrow n_{a}\, ,
\end{equation}
contrary to what one might naively expect. In particular, \eqref{eq:l_To_n} does not preserve the orientation of the spacetime $\left(\mathcal{N}^{2},g_{ab}\right)$, since it sends $\varepsilon_{ab}\mapsto-\varepsilon_{ab}$. Thus, invariance under \eqref{eq:l_To_n} does not in general hold. In the standard, four-dimensional GHP formalism the equations are invariant under \eqref{eq:l_To_n} together with
\begin{equation}\label{eq:m_To_barm}
    m_{A}\longleftrightarrow \bar{m}_{A}\, ,
\end{equation}
where in our context $ m_{A},\bar{m}_{A}$ are a tetrad on the round two-sphere. The combined transformations \eqref{eq:l_To_n} and \eqref{eq:m_To_barm} are known as the \textit{priming operation}, and since they preserve the total spacetime orientation (as well as, of course, the null-frame orthogonality conditions) they are a symmetry of the GHP equations. In our two-dimensional GHP formalism, invariance under the priming operation is inherited as invariance of our equations under \eqref{eq:l_To_n}, together with the possible complex conjugation of some terms, as an effect descending from \eqref{eq:m_To_barm}.

\section{Application to Vacuum General Relativity}\label{sec:Sec3}

In this section we consider perturbations about Schwarzschild's black hole in GR, allowing for the presence of matter sources. Hence, we impose the following background relations,
\begin{equation}
    r^{a}r_{a}=1-\frac{2M}{r}\equiv f(r)\, ,\quad T_{ab}=\mathcal{P}=0\, ,
\end{equation}
and keep general the fluctuations of the energy-momentum tensor. The fundamental equations in this sitaution are explicitly presented in Appendix \ref{sec:Ap3}. We will use the linearised Einstein's equations \eqref{eq:Einstein_Trace}, \eqref{eq:Einstein_Div} and \eqref{eq:Einstein_Lorentz},  to fix automatically the trace $\tensor{h}{^+^a_a}$ and the divergences $\nabla^{a}\tensor{h}{^-_a}$ and $\nabla^{b}\tensor{h}{^+_b_a}$ in terms of the matter sources and $h^{+}$ as
\begin{equation}\label{eq:Einstein_Rep}
   \tensor{h}{^+^a_a}=- 2 \Phi^{\oplus}{}\, ,\quad  \nabla^{b}\tensor{h}{^-_b}=\Phi^{-}{}\, ,\quad \nabla^{b}\tensor{h}{^+_b_a}=\nabla_{a}\left(h^{+}/r^{2}\right)+2 \Phi^{+}{}_{a}-2r\nabla_{a}\left(\Phi^{\oplus}{}/r\right)\, .
\end{equation}
One can similarly use the Einstein equations in Appendix \ref{sec:EinstEq} to replace the second derivatives of $h^{+}$ by an expression containing only first derivatives:
\begin{equation}\label{eq:DDh}
    \begin{aligned}
        \nabla_{a}\nabla_{b}\mathit{h}^{+}&=\frac{2}{r}r_{(a}\nabla_{b)}h^{+}-\frac{r^{c}}{r}\nabla_{c}h^{+}g_{ab} + \left( \left(- \frac{2 M}{r^3}  + \
\frac{2+\Lambda^2}{2 r^2}\right)g_{ab} -  \frac{2 r_{a} r_{b}}{r^2}\right)h^{+}\\ 
&+ r r^{c}\left(2 \mathit{h}^{+}{}_{c(a}{}_{:b)} - \mathit{h}^{+}{}_{ab}{}_{:c} \right)+\left(\frac{\Lambda^2}{2} + \frac{2 M}{r}\right) \mathit{h}^{+}{}_{ab}+\mathit{h}^{+}{}_{cd}r^{c}r^{d}g_{ab}\\ 
&- r^2 \left(\Phi^{+}{}_{ab}-\Phi^{+}{}^{c}{}_{c}g_{ab}\right)\, .
    \end{aligned}
\end{equation}
Unlike \eqref{eq:Einstein_Rep}, however, this relation is not applied automatically; second derivatives of $\mathit{h}^{+}$ are left explicit unless indicated otherwise.

To simplify the presentation of the equations, the analysis will be carried out with the matter fluctuations set to zero. The source terms will be restored only at the end, in the final result.

\subsection{The Regge--Wheeler Equation}\label{sec:Sec3.1}

Consider the scalar curvature components of $\delta \C$, namely $\Psi^{\pm},\Xi^{+}$. The fundamental equations are already written in the form of wave equations for these variables, with principal part given by $\square=\nabla^{a}\nabla_{a}$; see e.g. Eqs.~\eqref{eq:CWE_Psi_p}, \eqref{eq:CWE_Psi_m}, and \eqref{eq:CWE_Xi_p}. The goal is to obtain from them a decoupled equation.

The algebraic relations of the self-dual Weyl tensor, presented in Appendix \ref{app:algebraic_C}, immediately imply the following relations among the curvature scalars,
\begin{equation}\label{eq:Psi_m_Relations}
   \Psi^{-}{} = - \frac{2i M \mathit{h}^{+}{}}{r^3} + i r^2 \Psi^{+}{}\, ,\quad \Psi^{-}{}=\frac{2i M \mathit{h}^{+}{}}{r^3} -  \frac{i \Xi^{+}{}}{r^2}\, ,
\end{equation}
where \eqref{eq:Einstein_Rep} has been used and the matter sources have been set to zero (they will be restored at the end). These relations suggest the following strategy. Starting from the curvature equation for $\Psi^{-}$, Eq.~\eqref{eq:CWE_Psi_m}, a sequence of straightforward eliminations is performed, after which the first relation in \eqref{eq:Psi_m_Relations} is used to obtain a decoupled equation for $\Psi^{+}$, namely the Regge--Wheeler equation. An analogous procedure, based on the second relation in \eqref{eq:Psi_m_Relations}, will be used in Section \ref{sec:Sec3.2} to derive Zerilli's equation for a suitable redefinition of $\Xi^{+}{}$.

The curvature wave equation for $\Psi^{-}$, Eq.~\eqref{eq:CWE_Psi_m}, after using \eqref{eq:Einstein_Rep} and setting the matter sources to zero reads
\begin{equation}\label{eq:Eq_Psi_m_Vac1}
    \begin{aligned}
        \square \Psi^{-}{}{} &=\frac{2 \
 }{r}r{}^{a}\nabla_{a}\Psi^{-}+\left(\frac{\Lambda^2+4}{r^2}- \
\frac{10 M}{r^3}\right) \Psi^{-}{}\\
&-  \frac{2i M \
}{r^3} \square h^{+} + \
\frac{4i M }{r^4}r^{a}\nabla_{a}h^{+}+\left(\frac{4i M^2}{r^6} -  \frac{4i M}{r^5} + \frac{2i \Lambda^2 M}{r^5}\right) \
\mathit{h}^{+}{}   \\
&-  \frac{6 \Lambda^2 M }{r^4}\varepsilon_{ab} \
\mathit{h}^{-}{}^{a} r{}^{b} -  \frac{2 \Lambda^2 }{r^3}\varepsilon_{ab} \Xi^{-}{}^{a} r{}^{b}-  \frac{6i M }{r^3}\Psi^{+}{}^{a}{}_{a} \
+ \frac{2 \Lambda^2 }{r}\Psi^{-}{}^{a} r{}_{a}\, ,
    \end{aligned}
\end{equation}
where $\Lambda^{2}=l(l+1)$.

The curvature variables appearing in the last line, $\tensor{\Psi}{^+^a_a}$, $\Psi^{-a}$, and $\Xi^{-a}$, can be expressed in terms of $h^{+}$, $h^{-}_{a}$, and $\Psi^{-}$ using the lower-order equations for the self-dual Weyl tensor. In particular, the algebraic equations Eq.~\eqref{eq:SD_aAbB_U} and Eq.~\eqref{eq:Trace_aA_X}, together with the first-derivative equation Eq.~\eqref{eq:FirstOrder_aAB_epsY}, after using \eqref{eq:Einstein_Rep} and setting the matter sources to zero, yield
\begin{equation}
\begin{aligned}\label{eq:Rep_1}
    \tensor{\Psi}{^+^a_a}&=-i\Psi^{-}\, ,\\ \\
    \Psi^{-}{}_{a}&=r^{-2}\varepsilon_{ab}\left(\frac{M}{r}\mathit{h}^{-}{}^{b} + \Xi^{-}{}^{b}\right)\, , \\  \\
    \varepsilon_{ab} \Xi^{-}{}^{b}&= - \frac{2 M \varepsilon_{ab} \mathit{h}^{-}{}^{b}}{r}-r\Lambda^{-2}\nabla_{a}\left(r \Psi^{-}\right) +iM\Lambda^{-2}r\nabla_{a}\left(h^{+}/r^{2}\right)\, .
\end{aligned}
\end{equation}
Substituting these relations into \eqref{eq:Eq_Psi_m_Vac1} gives
\begin{equation}\label{eq:Eq_Psi_m_Vac2}
    \begin{aligned}
\square \Psi^{-}&=-  \frac{2 \
}{r}r^{a}\nabla_{a}\Psi^{-} + \left( \
\frac{\Lambda^2}{r^2}- \frac{8 M}{r^3}\right) \Psi^{-}{} \\
&-2iM r\nabla_{a}\left(r^{-4}\nabla^{a}\mathit{h}^{+}{}\right) +\frac{2iM}{r^{5}}\left(\frac{10 M}{r}+ \Lambda^2 - 6 \right) \mathit{h}^{+}{}\, ,
    \end{aligned}
\end{equation}
where the terms involving $h^{-}_{a}$ cancel. Using the first relation in \eqref{eq:Psi_m_Relations} then leads directly to a decoupled equation for $\Psi^{+}$, since all terms involving $h^{+}$ cancel exactly,
\begin{equation}
   \square\Psi^{+}+ \frac{6 }{r}r^{a}\nabla_{a}\Psi^{+}+ \left(\frac{6}{r^2} -  \frac{\Lambda^2}{r^2}\right) \Psi^{+}{} =0\, . 
\end{equation}
Introducing the rescaled variable
\begin{equation}\label{eq:psi_p}
    \psi^{+}\equiv r^{3}\Psi^{+}\, ,
\end{equation}
one obtains the Regge--Wheeler equation,
\begin{equation}\label{eq:RW_Sourced}
    \square\psi^{+}+\left(\frac{6M}{r^{3}}-\frac{\Lambda^{2}}{r^{2}}\right)\psi^{+}=S_{\psi^{+}}\, ,
\end{equation}
where the matter source terms have been restored on the right-hand side, yielding
\begin{equation}\label{eq:source_RW}
\begin{aligned}
     S_{\psi^{+}}&=-\frac{2 i}{r}\varepsilon^{ab}\nabla_{a}\left(r^{2}\Pi^{-}_{b}\right)+2 r^{a}\Gamma^{+}_{a}-\frac{\Lambda^{2}}{r}\Gamma^{+}+M \Phi^{+}{}^{a}{}_{a}+ \frac{12 M }{r}\Phi^{+}{}^{a} r_{a}\\
    &  + \frac{2 M \
}{r^2}\Phi^{+}+ 4 M \
\square\Phi^{\oplus} -  \frac{6 M}{r}r^{a}\nabla_{a}\Phi^{\oplus} - \frac{4M}{r^{2}}\left(\frac{3M}{r}+\Lambda^{2}-3\right) \Phi^{\oplus}{}\, .
\end{aligned}
\end{equation}
The advantages of deriving \eqref{eq:RW_Sourced} within this framework are now apparent. The derivation follows directly from the fundamental equations, and the resulting variable $\Psi^{+}$ arises naturally as a scalar component of $\delta\C$. In contrast, in the traditional approach based on metric perturbations, the master variables are constructed as nontrivial combinations of the metric components, and obtaining decoupled equations requires substantially more involved manipulations.

Moreover, the present formulation treats the even- and odd-parity sectors simultaneously through the self-dual curvature variables. As discussed in Section \ref{sec:Sec2.2.2}, the real and imaginary parts of $\psi^{+}$ encode exclusively the even- and odd-parity components of the metric perturbation, respectively (explicit expressions will be given in Section \ref{sec:Sec3.4}). Taking the real and imaginary parts of \eqref{eq:RW_Sourced} therefore yields the master equations for the two parity sectors. In the absence of sources, $S_{\psi^{+}}=0$, the master equation is identical for both sectors so, in particular, QNM isospectrality appears as a manifest symmetry.

The presence of sources generically breaks that symmetry. Indeed, taking the real and imaginary parts of $S_{\psi^{+}}$ does not, in general, yield identical expressions. This can be seen explicitly, for example, by expressing the source term \eqref{eq:source_RW} entirely in terms of components of $\delta T$, using the relations collected in Appendix \ref{app:dJ_dT}.

Finally, in Section \ref{sec:Sec3.4}, the variable $\psi^{+}$ will be related explicitly to standard variables in the literature, and both the equation and the source term in \eqref{eq:RW_Sourced} will be shown to agree with other known results.

\subsection{The Bardeen--Press--Teukolsky Equations}\label{sec:Sec3.3}

Another set of decoupled equations follows by considering the tensorial component $\Psi^{\oplus}{}_{ab}$ of $\delta \C$. As in the scalar case, the strategy is to start from the curvature wave equation for $\Psi^{\oplus}{}_{ab}$, Eq.~\eqref{eq:CWE_Psi_o}, and eliminate the remaining curvature components using the lower-order equations of $\delta \C$.

Setting the matter sources to zero, Eq.~\eqref{eq:CWE_Psi_o} becomes
\begin{equation}
    \begin{aligned}\label{eq:Teuk_1}
        \square\Psi^{\oplus}{}_{ab}&=\frac{2r^{c}}{r}\nabla_{c}\Psi^{\oplus}{}_{ab}+\left(\frac{10 M}{r^{3}}+\frac{\Lambda^{2}-4}{r^{2}}\right)\Psi^{\oplus}{}_{ab}\\
        &+\frac{6iM}{r^{3}}\tensor{\varepsilon}{_c_{(a}}\tensor{\Psi}{^-_{b)}^c}-\frac{4}{r^{3}}r_{(a}\Xi^{+}{}_{b)}+\frac{4r^{c}}{r}\Psi^{+}{}_{(a}\varepsilon_{b)c}+\frac{12iM}{r^{4}}h^{-c}r_{(a}\varepsilon_{b)c}\,,
    \end{aligned}
\end{equation}
where we also used $\tensor{\Psi}{^\oplus^c_c}=0$, which follows from the algebraic equation \eqref{eq:Trace_AB_V}. The curvature components appearing in the second line, namely $\Psi^{-}{}_{ab}$, $\Psi^{+}{}_{a}$, and $\Xi^{+}{}_b$, can be eliminated in favor of $\Psi^{\oplus}{}_{ab}$ and $h^{-}{}_{a}$ by means of the lower-order equations of $\delta\C$. In particular, with matter sources set to zero, the algebraic Eqs.~\eqref{eq:SD_aAbB_W},\eqref{eq:Trace_aA_Z}, together with the first-order Eq.~\eqref{eq:FirstOrder_ABa_V}, read
\begin{equation}\label{eq:Eliminating_Teuk}
    \begin{aligned}
        \Psi^{-}{}_{ab}&= i \varepsilon_{a}{}^{c} \Psi^{\oplus}{}_{bc}\,, \\ \\ 
        \Psi^{+}{}_{a}&=\frac{3i M }{r^3}\mathit{h}^{-}{}_{a} -  \frac{1}{r^2}\varepsilon_{ab} \
\Xi^{+}{}^{b} \, , \\ \\ 
        \Xi^{+}{}_{a}&=\frac{3i M }{r}\varepsilon_{ab} \mathit{h}^{-}{}^{b} \
+ r^2 \Psi^{\oplus}{}_{a}{}^{b}{}_{:b} -  r \Psi^{\oplus}{}_{ab}\
r{}^{b}\, .
    \end{aligned}
\end{equation}
Substituting \eqref{eq:Eliminating_Teuk} into \eqref{eq:Teuk_1}, the terms proportional to $h^{-}{}_{a}$ cancel identically, and one obtains a decoupled equation for $\Psi^{\oplus}{}_{ab}$. This equation can be written as
\begin{equation}
    \begin{aligned}\label{eq:Teuk_Psi_o}
        \square\Psi^{\oplus}{}_{ab}&=-8r_{(a}\nabla^{c}r^{-1}\Psi^{\oplus}{}_{b)c} +\frac{2r{}^{c}}{r} \nabla_{c}\Psi^{\oplus}{}_{ab}+4 r^{c}\nabla^{d}\left(r^{-1}\Psi^{\oplus}{}_{dc}\right)g_{ab} \\ 
&+\left(\frac{16 M}{r^3} + \frac{\Lambda^2-4}{r^2}\right) \
\Psi^{\oplus}{}_{ab}\, .
    \end{aligned}
\end{equation}
This is still a tensorial equation. To derive decoupled equations for scalar quantities, we expand $\Psi^{\oplus}{}_{ab}$ in its null-frame components,
\begin{equation}\label{eq:Psi_oplus_exp}
    \Psi^{\oplus}{}_{ab}=\Psi^{\oplus}{}_{\ell\ell}n_{a}n_{b}+\Psi^{\oplus}{}_{n n}\ell_{a}\ell_{b}\, ,
\end{equation}
where we used $\Psi^{\oplus}{}_{\ell n}=\Psi^{\oplus}{}_{n\ell}=0$, since $\Psi^{\oplus}{}_{\ell n}=-(1/2)\Psi^{\oplus}{}^{a}_{\ a}=0$ by the algebraic equation \eqref{eq:Trace_AB_V}. Substituting this expansion into \eqref{eq:Teuk_Psi_o}, using Eq.~\eqref{eq:Deriv_Frames} to replace the frame derivatives, and contracting the resulting equation with $\ell^{a}\ell^{b}$ and $n^{a}n^{b}$ yields, respectively,
\begin{equation}
    \begin{aligned}\label{eq:Teuk_ll_NonCov}
\square\Psi^{\oplus}{}_{\ell \ell}&=  - \left[(8  \varepsilon  + 6 \rho)n^{a}-\left(8\varepsilon ' + 2 \rho '\right)\ell^{a} \right] \nabla_{a}\Psi^{\oplus}{}_{\ell \ell} \\
       &- \left(- \frac{4}{r^2} -  \
\frac{\Lambda^2}{r^2} + 32 \varepsilon ' \varepsilon + 8 \varepsilon \rho ' + \
24 \varepsilon ' \rho - 8 \rho ' \rho - 4 \ell^{a} \nabla_{a}\varepsilon ' + 4 \
n^{a} \nabla_{a}\varepsilon \right)\Psi^{\oplus}{}_{\ell \ell} \,,
    \end{aligned}
\end{equation}
and
\begin{equation}
    \begin{aligned}\label{eq:Teuk_nn_NonCov}
\square\Psi^{\oplus}{}_{n n}&=  - \left[(8  \varepsilon'  + 6 \rho')\ell^{a}-\left(8\varepsilon  + 2 \rho \right)n^{a} \right] \nabla_{a}\Psi^{\oplus}{}_{n n} \\
       &- \left(- \frac{4}{r^2} -  \
\frac{\Lambda^2}{r^2} + 32 \varepsilon ' \varepsilon + 8 \varepsilon' \rho  + \
24 \varepsilon  \rho' - 8 \rho ' \rho - 4 n^{a} \nabla_{a}\varepsilon  + 4 \
\ell^{a} \nabla_{a}\varepsilon' \right)\Psi^{\oplus}{}_{n n} \,,
    \end{aligned}
\end{equation}
thus yielding decoupled equations for $\Psi^{\oplus}{}_{\ell \ell}$ and $\Psi^{\oplus}{}_{n n}$. No assumption has been made about the null frame $\ell_{a},n_{a}$, and the equations must therefore be covariant under \eqref{eq:frame_trafo}. In the form of Eqs.~\eqref{eq:Teuk_ll_NonCov} and \eqref{eq:Teuk_nn_NonCov}, however, this symmetry is not manifest. In the two-dimensional GHP formalism introduced in Section \ref{sec:Sec2.4}, the symmetry becomes explicit and the equations simplify considerably. Restoring the matter source terms, the GHP-covariant forms of Eqs.~\eqref{eq:Teuk_ll_NonCov} and \eqref{eq:Teuk_nn_NonCov} are
\begin{equation}\label{eq:Teuk_ll_Cov}
    \Theta_{a}\Theta^{a}\Psi^{\oplus}{}_{\ell \ell} +\left(6\rho n^{a}-2\rho'\ell^{a}\right)\Theta_{a}\Psi^{\oplus}{}_{\ell \ell}-\left(\frac{\Lambda^{2}+4}{r^2}+8 \rho ' \rho\right)\Psi^{\oplus}{}_{\ell \ell}=S_{\Psi^{\oplus}_{\ell \ell}}\, ,
\end{equation}
and
\begin{equation}\label{eq:Teuk_nn_Cov}
    \Theta_{a}\Theta^{a}\Psi^{\oplus}{}_{nn} +\left(6\rho' \ell^{a}-2\rho n^{a}\right)\Theta_{a}\Psi^{\oplus}{}_{nn}-\left(\frac{\Lambda^{2}+4}{r^2}+8 \rho ' \rho\right)\Psi^{\oplus}{}_{nn}=S_{\Psi^{\oplus}_{nn}}\, ,
\end{equation}
where $\Theta_{a}$ is the two-dimensional GHP covariant derivative introduced in Eq.~\eqref{eq:Theta_Deriv}, and the source terms are
\begin{equation}
    S_{\Psi^{\oplus}_{\ell \ell}}=i \ell^{a}\Theta_{a}{}(\Gamma^{-}{}_{\ell}{}) -  \ell^{a}\Theta_{a}{}(\Pi^{\oplus}{}_{\ell}{}) + i \Pi^{-}{}_{\ell \ell}{} 
-  \Pi^{+}{}_{\ell \ell}{} + i \rho \Gamma^{-}{}_{\ell}{}  + 7 
\rho\Pi^{\oplus}{}_{\ell}{} \, ,
\end{equation}
and
\begin{equation}
    S_{\Psi^{\oplus}_{n n}}=-i n^{a}\Theta_{a}{}(\Gamma^{-}{}_{\mathit{n}}{}) -  n^{a}\Theta_{a}{}(\Pi^{\oplus}{}_{\mathit{n}}{}) - i \Pi^{-}{}_{\mathit{n} \mathit{n}}{} -  \Pi^{+}{}_{n n}{} - i \rho ' \Gamma^{-}{}_{\mathit{n}}{}  + 7   \rho ' \Pi^{\oplus}{}_{\mathit{n}}{}\, .
\end{equation}
In this form, the coefficients $\varepsilon,\varepsilon'$ are absorbed into the connection 1-form \eqref{eq:omega} associated with the covariant derivative $\Theta_{a}$. This both simplifies the equations and source terms and makes invariance under \eqref{eq:frame_trafo} manifest. From \eqref{eq:Psi_o_ToWeyl}, the variables $\Psi^{\oplus}{}_{\ell\ell}$ and $\Psi^{\oplus}{}_{nn}$ are closely related to the perturbations of the maximal-weight Weyl scalars $\delta\Psi_{0}$ and $\delta\Psi_{4}$. Hence, Eqs.~\eqref{eq:Teuk_ll_Cov} and \eqref{eq:Teuk_nn_Cov} are a version of BPT's equations, obtained in a direct way using a formalism based solely on the tools of spherical symmetry. As for any other selfdual curvature component, the real and imaginary parts of $\Psi^{\oplus}{}_{\ell\ell}$ and $\Psi^{\oplus}{}_{n n}$ contain only even and odd metric perturbations, respectively. Their explicit expressions are given in Section \ref{sec:Sec3.4}. Hence, in the absence of sources, QNM isospectrality again follows automatically in this approach, whereas a non-zero source will in general break the symmetry between parity sectors.

\subsection{The Zerilli Equation}\label{sec:Sec3.2}

Obtaining Zerilli's equation in this framework is less direct than for the Regge--Wheeler and BPT equations. As discussed above, inserting the first of \eqref{eq:Psi_m_Relations} into \eqref{eq:Eq_Psi_m_Vac2} immediately yields a decoupled equation for $\Psi^{+}$. By contrast, using the second relation in \eqref{eq:Psi_m_Relations} leads to an equation in which $\Xi^{+}$ remains coupled to $h^{+}$, indicating that $\Xi^{+}$ is not as natural a variable as $\Psi^{+}$ for this purpose.

This feature can be traced back to the construction of gauge-invariant variables. As described in Section \ref{sec:Sec2.2.1}, gauge invariance is achieved by introducing a compensating term. While this procedure systematically produces gauge-invariant quantities, it does not ensure that the resulting variables yield the simplest form of the equations. The residual coupling between $\Xi^{+}$ and $h^{+}$ therefore motivates a reconsideration of how to construct a suitable gauge-invariant variable from $\Xi^{+}$.

To this end, the convention of automatically working with gauge-invariant variables is temporarily suspended. Examining the gauge transformations of $\Xi^{+}$, $h^{+}$, and $h^{\oplus}$ under a gauge parameter \eqref{eq:xi_gauge}, one finds that the combination
\begin{equation}\label{eq:hatXi_To_Xi}
    \hat{\Xi}^{+}\equiv \frac{1}{r}\Xi^{+}-\frac{M}{r^{2}}h^{+}+\frac{3M\Lambda^{2}}{2r^{2}}h^{\oplus}\,,
\end{equation}
is gauge-invariant, and differs from the gauge-invariant variable associated to $\Xi^{+}$ following Section \ref{sec:Sec2.2.1}. This motivates seeking an equation for $\hat{\Xi}^{+}$. Returning to the convention of working with gauge-invariant variables as constructed in Section \ref{sec:Sec2.2.1}, $\hat{\Xi}^{+}$ and $\Xi^{+}$ are related by \eqref{eq:hatXi_To_Xi} with $h^{\oplus}=0$, which is formally equivalent to working in the Regge--Wheeler gauge.

The relation between the real part of $\hat{\Xi}^{+}$ (denoted $\hat{\Xi}^{+}_{+}$, following Section \ref{sec:Sec2.2.2}) and the metric perturbation is obtained using the expressions in Appendix \ref{app:sdC_To_h}, together with \eqref{eq:Einstein_Rep} and \eqref{eq:DDh},
\begin{equation}
\hat{\Xi}^{+}_{+}=-2r^{a}\nabla_{a}h^{+}+\left(\frac{2+\Lambda^2}{r}- \frac{2 M}{r^2}\right) \mathit{h}^{+}{} + 2 r \
\mathit{h}^{+}{}_{ab} r{}^{a} r{}^{b}\, ,
\end{equation}
and, computing $r^{a}\nabla_{a}\hat{\Xi}^{+}_{+}$ from this expression (again using \eqref{eq:Einstein_Rep} and \eqref{eq:DDh}), one finds
\begin{align}\label{eq:hp_To_hatXip}
    h^{+}&=\frac{r^2}{6 M + (\Lambda^2-2 ) r}\left(\hat{\Xi}^{+}_{+}+\frac{2r}{\Lambda^{2}}r^{a}\nabla_{a}\hat{\Xi}^{+}_{+}\right)\, .
\end{align}
With this relation, inserting the second of \eqref{eq:Psi_m_Relations} into \eqref{eq:Eq_Psi_m_Vac2}, replacing $\Xi^{+}$ by $\hat{\Xi}^{+}$ using \eqref{eq:hatXi_To_Xi}, and taking the real part yields a second-order, decoupled equation for $\hat{\Xi}^{+}_{+}$. Introducing the rescaled variable $\theta^{+}$,
\begin{equation}\label{eq:theta_p}
    \theta^{+}\equiv \left( \Lambda^2-2 + \frac{6 M}{r}\right)^{-1}\hat{\Xi}^{+}_{+}\, ,
\end{equation}
the equation reduces to
\begin{equation}\label{eq:Zer}
    \square \theta^{+}+V_{\theta^{+}}\  \theta^{+}=S_{\theta^{+}}\, ,
\end{equation}
where the potential is
\begin{equation}\label{eq:ZerPot}
    V_{\theta^{+}}=- \frac{\bigl(72 M^3 + 36 (\Lambda^2-2 ) M^2 r + 6 (\Lambda^2-2 )^2 M r^2 + \Lambda^2 (\Lambda^2-2 )^2 r^3\bigr) \
}{r^3 \bigl(6 M + (\Lambda^2-2) r\bigr)^2}\, ,
\end{equation}
and the source term decomposes as
\begin{equation}
S_{\theta^{+}}=S_{(\theta^{+},\Phi^{\oplus})}+S_{(\theta^{+},\Phi^{+})}+S_{(\theta^{+},\Phi^{+}_{a})}+S_{(\theta^{+},\Phi^{+}_{ab})}\, .
\end{equation}
Explicitly,
\begin{align}
    S_{(\theta^{+},\Phi^{\oplus})}&=- \frac{\Lambda^2 }{r}\Phi^{\oplus}\, ,\\ \notag \\ \notag
    S_{(\theta^{+},\Phi^{+})}&=\bigl(6 M + (\Lambda^2-2 ) r\bigr)^{-2}\Biggl\{\left(  \frac{4-2\Lambda^{2}}{3} r^3 - 4 M \
r^2   \right)\square\Phi^{+}{}\\\notag
&+\left( \frac{8-4\Lambda^{2}}{3} r^2  - 16 M r \right)r^{a}\nabla_{a}\Phi^{+}\\
&+\left( \frac{104}{3} M  -  \frac{28}{3} \Lambda^2 M  - \
 \frac{40 M^2 }{r}    +\left( \frac{2}{3} \Lambda^4 + \frac{8}{3} \
\Lambda^2 - 8  \right)r \right)\Phi^{+}\Biggr\}\,, \\ \notag \\ 
    S_{(\theta^{+},\Phi^{+}_{a})}&=\frac{4 \Lambda^2 r }{6 M + (\Lambda^2-2 ) r}\Phi^{+}{}_{a} r{}^{a}\,, \\ \notag \\ \notag
   S_{(\theta^{+},\Phi^{+}_{ab})}&=\bigl(6 M + (\Lambda^2-2 ) r\bigr)^{-2}\Biggl\{\left( \left(\frac{2}{3} \Lambda^2  -  \frac{4}{3} \right)r^5+ 4 M r^4 \right)\square \Phi^{+}{}^{a}{}_{a}\\ \notag
   &+\left(20 M r^3 + \left(2 \Lambda^2 - 4 \right)r^4 \right)r{}^{a}\nabla_{a}\Phi^{+}{}^{b}{}_{b}{} \\ \notag
   &+\left( 12 M^2 r  - 4 M r^2  + 8 
\Lambda^2 M r^2  -  \frac{2}{3} \Lambda^2 r^3 
 + \frac{1}{3} \Lambda^4 r^3  \right)\Phi^{+}{}^{a}{}_{a}\\
 &+ 24 M r^2 \Phi^{+}{}_{ab} \
r{}^{a} r{}^{b}\Biggr\}
 \, .
\end{align}
The left-hand side of \eqref{eq:Zer} defines Zerilli's operator. As will be shown in Section \ref{sec:Sec3.4}, $\theta^{+}$ is closely related to the standard Zerilli variable, up to matter source terms and constant prefactors. The source terms have been simplified using the conservation of the energy-momentum tensor (see Appendix \ref{app:Tcons}) to eliminate the divergences $\nabla^{a}\Phi^{+}_{a}$ and $\nabla^{a}\Phi^{+}_{ab}$.

Equation \eqref{eq:Zer} is qualitatively distinct from the Regge--Wheeler and BPT equations derived above. In particular, $\theta^{+}$ corresponds only to the real part of a self-dual variable, and therefore the arguments leading to isospectrality in Sections \ref{sec:Sec3.1} and \ref{sec:Sec3.3} do not apply. This further supports the conclusion that Zerilli's variable is less natural in this framework, as also reflected in the more involved derivation.

\subsection{Relation to Known Master Variables}\label{sec:Sec3.4}

To conclude, explicit contact with variables commonly used in the literature is established. Consider first $\psi^{+}$, $\Psi^{\oplus}_{\ell\ell}$, and $\Psi^{\oplus}_{nn}$, defined from the harmonic components of $\delta\C$ in \eqref{eq:sdCPert} as (cf.~\eqref{eq:psi_p} and \eqref{eq:Psi_oplus_exp})
\begin{align}
    \psi^{+}=r^{3}\Psi^{+}\, ,\qquad \Psi^{\oplus}_{\ell\ell}=\ell^{a}\ell^{b}\Psi^{\oplus}_{ab}\, ,\qquad \Psi^{\oplus}_{n n}=n^{a}n^{b}\Psi^{\oplus}_{ab}\, ,
\end{align}
where $\ell_{a},n_{a}$ is any null frame of the 2$D$ Lorentzian background spacetime $(\mathcal{N}^2,g_{ab})$. For perturbations of Schwarzschild black holes in GR, including matter sources, these variables satisfy the decoupled equations \eqref{eq:RW_Sourced}, \eqref{eq:Teuk_ll_Cov}, and \eqref{eq:Teuk_nn_Cov}, respectively. Using the general expressions in Section \ref{sec:Sec2.5}, these variables can be related directly to the extremal- and zero-weight Weyl scalar perturbations. Expanding the latter in spin-weighted spherical harmonics $\tensor[_{s}]{Y}{^l^m}(z^{A})$,
\begin{align}\label{eq:Psi_to_sdC}
    \delta\Psi_{0}&=\sum_{l,m}\delta\Psi^{l m}_{0}(y^{a})\ \tensor[_{2}]{Y}{^l^m}(z^{A})\, ,\\
    \delta\Psi_{2}&=\sum_{l,m}\delta\Psi^{l m}_{2}(y^{a})\ \tensor{Y}{^l^m}(z^{A})\, ,\\
    \delta\Psi_{4}&=\sum_{l,m}\delta\Psi^{l m}_{4}(y^{a})\ \tensor[_{-2}]{Y}{^l^m}(z^{A})\, , 
\end{align}
one finds, after restoring the harmonic labels $l,m$,
\begin{align}
    \Psi^{lm\oplus}_{\ell\ell}(y^{a})&=\frac{2r^{2}}{\sqrt{(l-1)l(l+1)(l+2)}}\delta\Psi^{l m}_{0}(y^{a})\, ,\\\notag \\
    \psi^{lm+}(y^{a})&=2r^{3}\delta\Psi^{l m}_{2}(y^{a})+4M\tensor{\Phi}{^l^m^\oplus}(y^{a})\, , \\\notag\\
     \Psi^{lm\oplus}_{nn}(y^{a})&=\frac{2r^{2}}{\sqrt{(l-1)l(l+1)(l+2)}}\delta\Psi^{l m}_{4}(y^{a})\, .
\end{align}
When expressed in terms of the metric perturbation and its null frame components using the relations in Appendix \ref{app:sdC_To_h}, $\Psi^{\oplus}_{\ell\ell}$ and $\Psi^{\oplus}_{nn}$ take the form
\begin{equation}
    \Psi^{\oplus}_{\ell\ell}=-\frac{1}{2}h^{+}_{\ell\ell}-i \tho h^{-}_{\ell}\, ,\qquad \Psi^{\oplus}_{nn}=-\frac{1}{2}h^{+}_{nn}+i \tho' h^{-}_{n}\, ,
\end{equation}
where $\tho$ and $\tho'$ are the GHP directional derivatives introduced in \eqref{eq:tho_thop}. Their real and imaginary parts therefore involve only the even and odd metric perturbations, as expected.

To connect with previously used variables, the real and imaginary parts of $\psi^{+}$ are considered. Using Appendix \ref{app:sdC_To_h}, these are given by
\begin{equation}
    \begin{aligned}\label{eq:Psi_m_ToMet}
        \psi^{+}_{-}&=\frac{1}{2}i (l-1) l (l+1) (l+2) \Psi_{\text{odd}}\\
        &=-r^{3}i l (l+1)\nabla_{a}\left(r^{-2}\epsilon^{ab}\mathit{h}^{-}{}_{b}\right) \,, 
        \end{aligned}
\end{equation}
and
\begin{equation}
    \begin{aligned}\label{eq:Psi_p_ToMet}
        \psi^{+}_{+}&=-  \Psi_{\text{MP}}- \frac{2}{3} r^3 \Phi^{+}{}^{a}{}_{a} + 8 M \Phi^{\oplus}{} + \frac{2}{3} r \Phi^{+}{}\\
        &=\left(\frac{8 M}{r^2} -  \frac{\Lambda^{2}+2}{r} \right) \
\mathit{h}^{+}{}+ 2 \
r^{a}\nabla_{a}\mathit{h}^{+}{}{} - 2 r \mathit{h}^{+}{}_{ab} r{}^{a} \
r^{b} \\
&-  \frac{2}{3} r^3 \Phi^{+}{}^{a}{}_{a} + 8 M \
\Phi^{\oplus}{} + \frac{2}{3} r \Phi^{+}{}  \,.
    \end{aligned}
\end{equation}
Here $\Psi_{\text{odd}}$ denotes the odd master variable employed in \cite{Martel:2005ir,Spiers:2023mor}, and $\Psi_{\text{MP}}$ denotes the even master variable used in \cite{Mukkamala:2024dxf,Poisson:2025oic}. In each case, the second equality expresses the result explicitly in terms of the metric perturbation.

Finally, the even variable $\theta^{+}$ introduced in \eqref{eq:theta_p} is considered. Using Appendix \ref{app:sdC_To_h}, it satisfies
\begin{equation}\label{eq:theta_p_ToMet}
\begin{aligned}
    \theta^{+}&=\frac{\Lambda^{2}}{2}\Psi_{\text{even}}+\frac{2}{3}\left( r(\Lambda^2-2) + 6 M\right)^{-1}\left( r^4 \Phi^{+}{}^{a}{}_{a} -  r^2 \Phi^{+}{}\right)\\
    &=\left( r(\Lambda^2-2) + 6 M\right)^{-1}\Biggl\{\left(2 +\Lambda^{2} -  \frac{2M}{r}\right) \
\mathit{h}^{+}{} - 2 r r{}^{a}\nabla_{a}\mathit{h}^{+}{}{} + 2 r^2 \
\mathit{h}^{+}{}_{ab} r{}^{a} r{}^{b}\\
    &+\frac{2}{3}\left( r^4 \Phi^{+}{}^{a}{}_{a} -  r^2 \Phi^{+}{}\right)\Biggr\}
\end{aligned}
\end{equation}
where $\Psi_{\text{even}}$ denotes the even master variable used in \cite{Martel:2005ir,Spiers:2023mor}. In deriving this expression, conservation of the energy-momentum tensor (see Appendix \ref{app:Tcons}) has been used to simplify the dependence on the $\Phi$ components. The second equality again expresses the result in terms of the metric perturbation.

These relations provide a non-trivial consistency check of the formalism. In particular, the source terms obtained for the master variables must reproduce those of previously known variables upon using the relations above. A direct computation, implemented in the accompanying \texttt{mathematica} notebook \cite{pereniguez2026ptcwe}, confirms this expectation: the source terms of $\Psi^{\oplus}_{\ell\ell}$, $\Psi^{\oplus}_{nn}$, $\psi^{+}_{-}$, and $\theta^{+}$ agree with those in \cite{Martel:2005ir,Spiers:2023mor}, while the source term of $\psi^{+}_{+}$ agrees with that of \cite{Poisson:2025oic} (note that $\psi^{+}_{+}$ was not considered in \cite{Martel:2005ir,Spiers:2023mor}, and was included in \cite{Mukkamala:2024dxf} but without sources). In establishing this correspondence, energy-momentum conservation must be imposed, as required by consistency of the linearised Einstein equations.

\subsection{Comments on Metric Reconstruction and (Time-Domain) Darboux Transformations}\label{sec:Darboux_And_Recons}

In the previous sections, decoupled wave equations were obtained as a consequence of identities in Lorentzian geometry together with Einstein's equation. To complete the analysis, the metric perturbation must be reconstructed, in a chosen gauge, from the decoupled master variables.

The reconstruction of the metric in the Regge--Wheeler gauge from $\Psi_{\text{odd}}$ in the odd sector (and hence $\psi^{+}_{-}$, see \eqref{eq:Psi_m_ToMet}) and from $\Psi_{\text{even}}$ in the even sector (and hence $\theta^{+}$, see \eqref{eq:theta_p_ToMet}) was given in \cite{Martel:2003jj,Martel2003}. Similarly, from either $\delta\Psi_{0,4}$ (and hence $\Psi^{\oplus}_{\ell\ell}$ and $\Psi^{\oplus}_{nn}$, see \eqref{eq:Psi_to_sdC}) one can reconstruct the metric perturbation in ingoing or outgoing radiation gauges \cite{Chandrasekhar:1985kt,Kegeles:1979,Wald:1978vm,Green:2019nam,Toomani:2021jlo,Hollands:2024iqp,Aksteiner:2016pjt,Wardell:2024yoi}. 

The reconstruction of the metric from the remaining even-sector variable $\psi^{+}_{+}$ is now considered. This problem was first analysed in \cite{Poisson:2025oic}, where it was shown that the even metric in the time domain and Regge--Wheeler gauge can be reconstructed from $\psi^{+}_{+}$ at the cost of an additional radial integral. While this remains true in the time domain, the situation simplifies for mode solutions (equivalently, in the frequency domain), where the metric can be reconstructed from $\psi^{+}_{+}$ and its radial derivatives --- in fact, this has been recently noticed in \cite{Pitre:2026msx}. In addition, the appearance of the algebraically special frequencies identified in \cite{Poisson:2025oic} is clarified. To simplify the expressions, the matter sources are set to zero in the following discussion.

Combining Eqs.~\eqref{eq:Psi_p_ToMet}, \eqref{eq:theta_p_ToMet}, \eqref{eq:hp_To_hatXip}, and \eqref{eq:theta_p} yields
\begin{equation}\label{eq:Darboux_Rel}
    \frac{\Lambda^{2}}{12 M}\psi^{+}_{+}=r^{a}\nabla_{a}\theta^{+}+\frac{144 M^3 - 72 M^2 r - 6 \Lambda^2 (-2 + \Lambda^2) M r^2 - \
 \Lambda^2 (-2 + \Lambda^2)^2 r^3 }{12 M r^2 \bigl(6 \
M + (-2 + \Lambda^2) r\bigr)}\theta^{+}\, .
\end{equation}
This relation shows that $\frac{\Lambda^{2}}{12 M}\psi^{+}_{+}$ and $\theta^{+}$ are connected by a \textit{time-domain version} of a Darboux transformation,\footnote{See e.g. \cite{Glampedakis:2017rar} for a discussion of frequency-domain Darboux transformations.} with Darboux potential
\begin{equation}\label{eq:W_RW}
    W(r)\equiv \frac{144 M^3 - 72 M^2 r - 6 \Lambda^2 (-2 + \Lambda^2) M r^2 - \
 \Lambda^2 (-2 + \Lambda^2)^2 r^3 }{12 M r^2 \bigl(6 \
M + (-2 + \Lambda^2) r\bigr)}\, ,
\end{equation}
which maps the Zerilli equation to the Regge--Wheeler equation, both in the time domain. Darboux transformations arise naturally in this framework: since isospectrality is manifest, the master variables must be related by a constrained class of transformations. All Darboux generators $W(r)$ satisfy a Riccati equation, which can be solved in terms of solutions of the equations they relate. In the present case, \eqref{eq:W_RW} is constructed from an algebraically special solution of the Regge--Wheeler equation \cite{Glampedakis:2017rar}, explaining the appearance of the corresponding algebraically special frequencies noted in \cite{Poisson:2025oic}.

If \eqref{eq:Darboux_Rel} could be inverted in the time domain and $\theta^{+}$ expressed in terms of $\psi^{+}_{+}$ and its derivatives, the metric could be reconstructed from $\psi^{+}_{+}$ via the reconstruction map for $\theta^{+}$. However, Darboux transformations such as \eqref{eq:Darboux_Rel} are not, in general, invertible in the time domain. This accounts for the obstruction to a purely differential reconstruction map in time domain, in agreement with the observation in \cite{Poisson:2025oic}.

However, for mode solutions with definite frequency,
\begin{equation}
    \theta^{+}(y^{a})=e^{-i\omega t}\theta^{+}_{\omega}(r)\,,\qquad \psi^{+}_{+}(y^{a})=e^{-i\omega t}\psi^{+}_{+\omega}(r)\,,
\end{equation}
the master equations reduce to second-order ordinary differential equations, and the Darboux transformation becomes invertible except in special cases. In terms of the tortoise coordinate $x=\int^{x}dr/f(r)$, $\theta^{+}_{\omega}$ satisfies
\begin{equation}\label{eq:theta_p_FreqDom}
    \frac{d^{2}}{dx^{2}}\theta^{+}_{\omega}+\left(\omega^{2}+fV_{\theta^{+}}\right)\theta^{+}_{\omega}=0\, ,
\end{equation}
where $V_{\theta^{+}}$ is given in \eqref{eq:ZerPot}, and \eqref{eq:Darboux_Rel} becomes
\begin{equation}\label{eq:rels_Theta_Psi}
    \frac{\Lambda^{2}}{12 M}\psi^{+}_{+\omega}=\frac{d}{dx}\theta^{+}_{\omega}+W \theta^{+}_{\omega}\, .
\end{equation}
The Riccati equation satisfied by $W$ is 
\begin{equation}\label{eq:Riccati}
    \frac{d}{dx}W(x)-W^{2}(x)-fV_{\theta^{+}}=\omega_{\text{Alg}}^{2}\, ,
\end{equation}
where
\begin{equation}
    \omega_{\text{Alg}}=i\frac{(l-1)l(l+1)(l+2)}{12M}
\end{equation}
is the algebraically special frequency. Differentiating \eqref{eq:rels_Theta_Psi} and using \eqref{eq:theta_p_FreqDom} and \eqref{eq:Riccati} yields
\begin{equation}
    \theta^{+}_{\omega}=\frac{l(l+1)}{12M\left(\omega_{\text{Alg}}^{2}-\omega^{2}\right)}\left(\frac{d}{dx}\psi^{+}_{+\omega}-W \psi^{+}_{+\omega}\right)\, .
\end{equation}
Thus, unless $\omega^{2}=\omega_{\text{Alg}}^{2}$, the Darboux transformation is invertible, and a purely differential reconstruction map exists for mode solutions.

\section{Conclusions and Outlook}\label{sec:Conclusions}

We have presented a framework that unifies the RWZ and BPT formalisms for the study of gravitational fluctuations on spherical backgrounds. The formalism has been developed allowing for a general energy-momentum tensor, and illustrated in the case of vacuum GR with sources. A number of nontrivial results---including the RWZ and BPT equations, as well as QNM isospectrality---are recovered in a straightforward manner within a unified setup. It is further shown how the full solution to the perturbative equations can be reconstructed easily from any of the master variables in the frequency domain, while also commenting on some technical complications that can arise in the time domain. Several directions for future work remain to be explored, where this approach may prove useful.

QNM isospectrality is in general broken when GR is extended, either through the inclusion of matter fields or higher-curvature corrections \cite{Li:2023ulk,Silva:2024ffz,Silva:2026jih}. However, for certain well-motivated completions, such as supergravity \cite{Pope:2024ncb,Pope:2025jgz} or type II string theory \cite{Cano:2024wzo}, it is preserved. The present framework may provide a simple setting in which to further investigate the mechanisms underlying isospectrality breaking and preservation.

Another natural direction concerns the role of gravitational nonlinearities in the ringdown phase. These effects have recently been observed in numerical simulations \cite{Cheung:2022rbm,Mitman:2022qdl,Dyer:2025hdt}, and there are prospects for their detection with future \cite{Yi:2024elj,Lagos:2024ekd} and, perhaps, even current detectors \cite{Wang:2026rev}. Since the analysis presented here allows one to work with a single propagator---the Regge--Wheeler one---for both parity sectors, the formalism may simplify the computation of nonlinear solutions \cite{Brizuela:2006ne,Brizuela:2007zza,Brizuela:2009qd,Ioka:2007ak,Bucciotti:2023ets,Redondo-Yuste:2023seq,Redondo-Yuste:2023ipg,May:2024rrg,Bucciotti:2024zyp,Bucciotti:2024jrv,BenAchour:2024skv,Ma:2024qcv,Bourg:2024jme,Singh:2025xzd,Cardoso:2026llh}.

It would also be interesting to explore whether the treatment of matter, encoded in the components of self-dual curvature currents (or the Cotton tensor), leads to simplifications when considering fluid models \cite{Speeney:2024mas,Redondo-Yuste:2024vdb} or other compact objects. In this context, variables closely related to $\psi^{+}_{\pm}$ have recently been employed to study instabilities in gravastar models \cite{Pitre:2026msx}.

Finally, following \cite{Franchini:2023xhd}, an important extension would be to incorporate corrections associated with the spin of the background, particularly in the presence of matter fields. This is especially relevant for astrophysical systems such as neutron stars, which are expected to be well described in the slowly rotating regime \cite{Berti:2004ny,Gerosa:2013laa}.  

\acknowledgments
I thank Emanuele Berti, Bogdan Ganchev, Gerardo García--Moreno, Loris del Grosso, Leah Jenks, Takuya Katagiri, Paolo Pani, Eric Poisson and Nick Speeney for valuable discussions related to this work. I especially thank Ibrahima Bah for useful conversations in relation to aspects in Section \ref{sec:Sec2.2.2}, Valerio de Luca for illuminating discussions in relation to \ref{sec:Darboux_And_Recons}, and Rishi Mukkamala for his collaboration on many topics that are important for this work. I also thank Nicola Franchini for discussions and feedback on previous versions of this manuscript. The computations in this work were performed using the xAct package suite for Mathematica \cite{xAct}. I am supported by NSF Grants No. AST-2307146, PHY-2513337, PHY090003, and PHY-20043, by NASA Grant No. 21-ATP21-0010, by John Templeton Foundation Grant No. 62840, by the Simons Foundation, and by Italian Ministry of Foreign Affairs and
International Cooperation Grant No. PGR01167. 
\bibliographystyle{JHEP}
\bibliography{ref}

\providecommand{\href}[2]{#2}\begingroup\raggedright\begin{thebibliography}{10}

\bibitem{Berti:2009kk}
E.~Berti, V.~Cardoso and A.O.~Starinets, \emph{{Quasinormal modes of black
  holes and black branes}},
  \href{https://doi.org/10.1088/0264-9381/26/16/163001}{\emph{Class. Quant.
  Grav.} {\bfseries 26} (2009) 163001}
  [\href{https://arxiv.org/abs/0905.2975}{{\ttfamily 0905.2975}}].

\bibitem{Barack:2018yly}
L.~Barack et~al., \emph{{Black holes, gravitational waves and fundamental
  physics: a roadmap}},
  \href{https://doi.org/10.1088/1361-6382/ab0587}{\emph{Class. Quant. Grav.}
  {\bfseries 36} (2019) 143001}
  [\href{https://arxiv.org/abs/1806.05195}{{\ttfamily 1806.05195}}].

\bibitem{LISAConsortiumWaveformWorkingGroup:2023arg}
{\scshape LISA Consortium Waveform Working Group} collaboration,
  \emph{{Waveform modelling for the Laser Interferometer Space Antenna}},
  \href{https://doi.org/10.1007/s41114-025-00056-1}{\emph{Living Rev. Rel.}
  {\bfseries 28} (2025) 9} [\href{https://arxiv.org/abs/2311.01300}{{\ttfamily
  2311.01300}}].

\bibitem{Berti:2025hly}
J.~Abedi et~al., \emph{{Black hole spectroscopy: from theory to experiment}},
  \href{https://arxiv.org/abs/2505.23895}{{\ttfamily 2505.23895}}.

\bibitem{PhysRev.108.1063}
T.~Regge and J.A.~Wheeler, \emph{Stability of a schwarzschild singularity},
  \href{https://doi.org/10.1103/PhysRev.108.1063}{\emph{Phys. Rev.} {\bfseries
  108} (1957) 1063}.

\bibitem{PhysRevLett.24.737}
F.J.~Zerilli, \emph{Effective potential for even-parity regge-wheeler
  gravitational perturbation equations},
  \href{https://doi.org/10.1103/PhysRevLett.24.737}{\emph{Phys. Rev. Lett.}
  {\bfseries 24} (1970) 737}.

\bibitem{Martel:2005ir}
K.~Martel and E.~Poisson, \emph{{Gravitational perturbations of the
  Schwarzschild spacetime: A Practical covariant and gauge-invariant
  formalism}}, \href{https://doi.org/10.1103/PhysRevD.71.104003}{\emph{Phys.
  Rev. D} {\bfseries 71} (2005) 104003}
  [\href{https://arxiv.org/abs/gr-qc/0502028}{{\ttfamily gr-qc/0502028}}].

\bibitem{Spiers:2023mor}
A.~Spiers, A.~Pound and B.~Wardell, \emph{{Second-order perturbations of the
  Schwarzschild spacetime: Practical, covariant, and gauge-invariant
  formalisms}}, \href{https://doi.org/10.1103/PhysRevD.110.064030}{\emph{Phys.
  Rev. D} {\bfseries 110} (2024) 064030}
  [\href{https://arxiv.org/abs/2306.17847}{{\ttfamily 2306.17847}}].

\bibitem{Newman:1961qr}
E.~Newman and R.~Penrose, \emph{{An Approach to gravitational radiation by a
  method of spin coefficients}},
  \href{https://doi.org/10.1063/1.1724257}{\emph{J. Math. Phys.} {\bfseries 3}
  (1962) 566}.

\bibitem{Geroch:1973am}
R.P.~Geroch, A.~Held and R.~Penrose, \emph{{A space-time calculus based on
  pairs of null directions}}, \href{https://doi.org/10.1063/1.1666410}{\emph{J.
  Math. Phys.} {\bfseries 14} (1973) 874}.

\bibitem{Bardeen:1973xb}
J.M.~Bardeen and W.H.~Press, \emph{{Radiation fields in the schwarzschild
  background}}, \href{https://doi.org/10.1063/1.1666175}{\emph{J. Math. Phys.}
  {\bfseries 14} (1973) 7}.

\bibitem{Teukolsky:1973ha}
S.A.~Teukolsky, \emph{{Perturbations of a rotating black hole. 1. Fundamental
  equations for gravitational electromagnetic and neutrino field
  perturbations}}, \href{https://doi.org/10.1086/152444}{\emph{Astrophys. J.}
  {\bfseries 185} (1973) 635}.

\bibitem{Chandrasekhar:1985kt}
S.~Chandrasekhar, \emph{The Mathematical Theory of Black Holes}, Oxford
  University Press (1985).

\bibitem{Lenzi:2021wpc}
M.~Lenzi and C.F.~Sopuerta, \emph{{Master functions and equations for
  perturbations of vacuum spherically symmetric spacetimes}},
  \href{https://doi.org/10.1103/PhysRevD.104.084053}{\emph{Phys. Rev. D}
  {\bfseries 104} (2021) 084053}
  [\href{https://arxiv.org/abs/2108.08668}{{\ttfamily 2108.08668}}].

\bibitem{Chandrasekhar_AlgSpecial}
S.~{Chandrasekhar}, \emph{{On Algebraically Special Perturbations of Black
  Holes}}, \href{https://doi.org/10.1098/rspa.1984.0021}{\emph{Proceedings of
  the Royal Society of London Series A} {\bfseries 392} (1984) 1}.

\bibitem{Glampedakis:2017rar}
K.~Glampedakis, A.D.~Johnson and D.~Kennefick, \emph{{Darboux transformation in
  black hole perturbation theory}},
  \href{https://doi.org/10.1103/PhysRevD.96.024036}{\emph{Phys. Rev. D}
  {\bfseries 96} (2017) 024036}
  [\href{https://arxiv.org/abs/1702.06459}{{\ttfamily 1702.06459}}].

\bibitem{Witten:1959zza}
L.~Witten, \emph{{Invariants of General Relativity and the Classification of
  Spaces}}, \href{https://doi.org/10.1103/PhysRev.113.357}{\emph{Phys. Rev.}
  {\bfseries 113} (1959) 357}.

\bibitem{Penrose:1960eq}
R.~Penrose, \emph{{A Spinor approach to general relativity}},
  \href{https://doi.org/10.1016/0003-4916(60)90021-X}{\emph{Annals Phys.}
  {\bfseries 10} (1960) 171}.

\bibitem{Penrose:1985bww}
R.~Penrose and W.~Rindler, \emph{{Spinors and Space-Time}}, Cambridge
  Monographs on Mathematical Physics, Cambridge Univ. Press, Cambridge, UK (4,
  2011),
  \href{https://doi.org/10.1017/CBO9780511564048}{10.1017/CBO9780511564048}.

\bibitem{Sasaki1981TheRE}
M.~Sasaki and T.~Nakamura, \emph{The regge-wheeler equation with sources for
  both even and odd parity perturbations of the schwarzschild geometry},
  {\emph{Physics Letters A} {\bfseries 87} (1981) 85}.

\bibitem{Mukkamala:2024dxf}
G.R.~Mukkamala and D.~Pere\~niguez, \emph{{Decoupled gravitational wave
  equations in spherical symmetry from curvature wave equations}},
  \href{https://doi.org/10.1088/1475-7516/2025/01/122}{\emph{JCAP} {\bfseries
  01} (2025) 122} [\href{https://arxiv.org/abs/2408.13557}{{\ttfamily
  2408.13557}}].

\bibitem{Gasperin:2026znw}
E.~Gasperin, R.~Panosso~Macedo and J.~Feng, \emph{{The linearised conformal
  Einstein field equations around a Petrov-type{\textasciitilde}D spacetime:
  the conformal Teukolsky equation}},
  \href{https://arxiv.org/abs/2602.19245}{{\ttfamily 2602.19245}}.

\bibitem{Chaverra:2012bh}
E.~Chaverra, N.~Ortiz and O.~Sarbach, \emph{{Linear perturbations of
  self-gravitating spherically symmetric configurations}},
  \href{https://doi.org/10.1103/PhysRevD.87.044015}{\emph{Phys. Rev. D}
  {\bfseries 87} (2013) 044015}
  [\href{https://arxiv.org/abs/1209.3731}{{\ttfamily 1209.3731}}].

\bibitem{Poisson:2025oic}
E.~Poisson, \emph{{Mukkamala-Pere{\~n}iguez master function for even-parity
  perturbations of the Schwarzschild spacetime}},
  \href{https://doi.org/10.1007/s10714-025-03384-3}{\emph{Gen. Rel. Grav.}
  {\bfseries 57} (2025) 51} [\href{https://arxiv.org/abs/2501.12377}{{\ttfamily
  2501.12377}}].

\bibitem{pereniguez2026ptcwe}
D.~Pereñiguez, ``{PT-from-CWE}.''
  \url{https://github.com/DavidPereniguez/PT-from-CWE}, 2026.

\bibitem{Fransen:2025cgv}
K.~Fransen, D.~Pere{\~n}iguez and J.~Redondo-Yuste, \emph{{Perturbations of
  plane waves and quadratic quasinormal modes on the lightring}},
  \href{https://doi.org/10.1007/JHEP12(2025)148}{\emph{JHEP} {\bfseries 12}
  (2025) 148} [\href{https://arxiv.org/abs/2509.03598}{{\ttfamily
  2509.03598}}].

\bibitem{Bini:2002jx}
D.~Bini, C.~Cherubini, R.T.~Jantzen and R.J.~Ruffini, \emph{{Teukolsky master
  equation: De Rham wave equation for the gravitational and electromagnetic
  fields in vacuum}}, \href{https://doi.org/10.1143/PTP.107.967}{\emph{Prog.
  Theor. Phys.} {\bfseries 107} (2002) 967}
  [\href{https://arxiv.org/abs/gr-qc/0203069}{{\ttfamily gr-qc/0203069}}].

\bibitem{Bini:2003km}
D.~Bini, C.~Cherubini, R.T.~Jantzen and R.~Ruffini, \emph{{De Rham wave
  equation for tensor valued p-forms}},
  \href{https://doi.org/10.1142/S0218271803003785}{\emph{Int. J. Mod. Phys. D}
  {\bfseries 12} (2003) 1363}.

\bibitem{Stewart:1974uz}
J.M.~Stewart and M.~Walker, \emph{{Perturbations of spacetimes in general
  relativity}}, \href{https://doi.org/10.1098/rspa.1974.0172}{\emph{Proc. Roy.
  Soc. Lond. A} {\bfseries 341} (1974) 49}.

\bibitem{Ryan:1974nt}
M.P.~Ryan, \emph{{Teukolsky equation and Penrose wave equation}},
  \href{https://doi.org/10.1103/PhysRevD.10.1736}{\emph{Phys. Rev. D}
  {\bfseries 10} (1974) 1736}.

\bibitem{Green:2026nlt}
S.R.~Green, K.~Krasnov and A.~Shaw, \emph{{Kahler decoupling for Kerr
  perturbations}},  \href{https://arxiv.org/abs/2604.22424}{{\ttfamily
  2604.22424}}.

\bibitem{Price_Despun}
R.H.~{Price}, \emph{{Nonspherical Perturbations of Relativistic Gravitational
  Collapse. II. Integer-Spin, Zero-Rest-Mass Fields}},
  \href{https://doi.org/10.1103/PhysRevD.5.2439}{\emph{PRD} {\bfseries 5}
  (1972) 2439}.

\bibitem{Aksteiner:2010rh}
S.~Aksteiner and L.~Andersson, \emph{{Linearized gravity and gauge
  conditions}},
  \href{https://doi.org/10.1088/0264-9381/28/6/065001}{\emph{Class. Quant.
  Grav.} {\bfseries 28} (2011) 065001}
  [\href{https://arxiv.org/abs/1009.5647}{{\ttfamily 1009.5647}}].

\bibitem{Shah:2016juc}
A.G.~Shah, B.F.~Whiting, S.~Aksteiner, L.~Andersson and T.~Backd{\"a}hl,
  \emph{{Gauge-invariant perturbations of Schwarzschild spacetime}},
  \href{https://arxiv.org/abs/1611.08291}{{\ttfamily 1611.08291}}.

\bibitem{Edgar:2005zr}
S.B.~Edgar and J.M.M.~Senovilla, \emph{{A Weighted de Rham operator acting on
  arbitrary tensor fields and their local potentials}},
  \href{https://doi.org/10.1016/j.geomphys.2005.11.011}{\emph{J. Geom. Phys.}
  {\bfseries 56} (2006) 2135}
  [\href{https://arxiv.org/abs/math/0505538}{{\ttfamily math/0505538}}].

\bibitem{Moncrief:1974am}
V.~Moncrief, \emph{{Gravitational perturbations of spherically symmetric
  systems. I. The exterior problem.}},
  \href{https://doi.org/10.1016/0003-4916(74)90173-0}{\emph{Annals Phys.}
  {\bfseries 88} (1974) 323}.

\bibitem{PhysRevD.12.1526}
V.~Moncrief, \emph{Gauge-invariant perturbations of reissner-nordstr\"om black
  holes}, \href{https://doi.org/10.1103/PhysRevD.12.1526}{\emph{Phys. Rev. D}
  {\bfseries 12} (1975) 1526}.

\bibitem{Gerlach:1979rw}
U.H.~Gerlach and U.K.~Sengupta, \emph{{Gauge invariant perturbations on most
  general spherically symmetric spacetimes}},
  \href{https://doi.org/10.1103/PhysRevD.19.2268}{\emph{Phys. Rev. D}
  {\bfseries 19} (1979) 2268}.

\bibitem{Gerlach:1980tx}
U.H.~Gerlach and U.K.~Sengupta, \emph{{Gauge invariant coupled gravitational,
  acoustical, and electromagnetic modes on most general spherical spacetimes}},
  \href{https://doi.org/10.1103/PhysRevD.22.1300}{\emph{Phys. Rev. D}
  {\bfseries 22} (1980) 1300}.

\bibitem{Gundlach:1999bt}
C.~Gundlach and J.M.~Martin-Garcia, \emph{{Gauge invariant and coordinate
  independent perturbations of stellar collapse. 1. The Interior}},
  \href{https://doi.org/10.1103/PhysRevD.61.084024}{\emph{Phys. Rev. D}
  {\bfseries 61} (2000) 084024}
  [\href{https://arxiv.org/abs/gr-qc/9906068}{{\ttfamily gr-qc/9906068}}].

\bibitem{Martin-Garcia:2000cgm}
J.M.~Martin-Garcia and C.~Gundlach, \emph{{Gauge invariant and coordinate
  independent perturbations of stellar collapse. 2. Matching to the exterior}},
  \href{https://doi.org/10.1103/PhysRevD.64.024012}{\emph{Phys. Rev. D}
  {\bfseries 64} (2001) 024012}
  [\href{https://arxiv.org/abs/gr-qc/0012056}{{\ttfamily gr-qc/0012056}}].

\bibitem{Pereniguez:2023wxf}
D.~Pere\~niguez, \emph{{Black hole perturbations and electric-magnetic
  duality}}, \href{https://doi.org/10.1103/PhysRevD.108.084046}{\emph{Phys.
  Rev. D} {\bfseries 108} (2023) 084046}
  [\href{https://arxiv.org/abs/2302.10942}{{\ttfamily 2302.10942}}].

\bibitem{Wald:1993nt}
R.M.~Wald, \emph{{Black hole entropy is the Noether charge}},
  \href{https://doi.org/10.1103/PhysRevD.48.R3427}{\emph{Phys. Rev. D}
  {\bfseries 48} (1993) R3427}
  [\href{https://arxiv.org/abs/gr-qc/9307038}{{\ttfamily gr-qc/9307038}}].

\bibitem{Lenzi:2021njy}
M.~Lenzi and C.F.~Sopuerta, \emph{{Darboux covariance: A hidden symmetry of
  perturbed Schwarzschild black holes}},
  \href{https://doi.org/10.1103/PhysRevD.104.124068}{\emph{Phys. Rev. D}
  {\bfseries 104} (2021) 124068}
  [\href{https://arxiv.org/abs/2109.00503}{{\ttfamily 2109.00503}}].

\bibitem{Solomon:2023ltn}
A.R.~Solomon, \emph{{Off-Shell Duality Invariance of Schwarzschild Perturbation
  Theory}}, \href{https://doi.org/10.3390/particles6040061}{\emph{Particles}
  {\bfseries 6} (2023) 943} [\href{https://arxiv.org/abs/2310.04502}{{\ttfamily
  2310.04502}}].

\bibitem{Jaramillo:2024qjz}
J.L.~Jaramillo, M.~Lenzi and C.F.~Sopuerta, \emph{{Integrability in Perturbed
  Black Holes: Background Hidden Structures}},
  \href{https://arxiv.org/abs/2407.14196}{{\ttfamily 2407.14196}}.

\bibitem{Lenzi:2025kqs}
M.~Lenzi, A.M.~Agudo and C.F.~Sopuerta, \emph{{Korteweg-de Vries integrals for
  modified black hole potentials: instabilities and other questions}},
  \href{https://doi.org/10.1088/1475-7516/2025/09/021}{\emph{JCAP} {\bfseries
  09} (2025) 021} [\href{https://arxiv.org/abs/2503.09918}{{\ttfamily
  2503.09918}}].

\bibitem{Lenzi:2025man}
M.~Lenzi, G.A.~Mena~Marug{\'a}n, A.~M{\'\i}nguez-S{\'a}nchez and C.F.~Sopuerta,
  \emph{{Master variables and Darboux symmetry for axial perturbations of the
  exterior and interior of black hole spacetimes}},
  \href{https://doi.org/10.1103/mcgc-tqmt}{\emph{Phys. Rev. D} {\bfseries 113}
  (2026) 064039} [\href{https://arxiv.org/abs/2512.10664}{{\ttfamily
  2512.10664}}].

\bibitem{DeLuca:2025zqr}
V.~De~Luca, B.~Khek, J.~Khoury and M.~Trodden, \emph{{Hidden symmetries for
  tidal Love numbers: Generalities and applications to analog black holes}},
  \href{https://doi.org/10.1103/5bdk-xclx}{\emph{Phys. Rev. D} {\bfseries 113}
  (2026) 044006} [\href{https://arxiv.org/abs/2512.06082}{{\ttfamily
  2512.06082}}].

\bibitem{Stephani:2003tm}
H.~Stephani, D.~Kramer, M.A.H.~MacCallum, C.~Hoenselaers and E.~Herlt,
  \emph{{Exact solutions of Einstein's field equations}}, Cambridge Monographs
  on Mathematical Physics, Cambridge Univ. Press, Cambridge (2003),
  \href{https://doi.org/10.1017/CBO9780511535185}{10.1017/CBO9780511535185}.

\bibitem{Martel:2003jj}
K.~Martel, \emph{{Gravitational wave forms from a point particle orbiting a
  Schwarzschild black hole}},
  \href{https://doi.org/10.1103/PhysRevD.69.044025}{\emph{Phys. Rev. D}
  {\bfseries 69} (2004) 044025}
  [\href{https://arxiv.org/abs/gr-qc/0311017}{{\ttfamily gr-qc/0311017}}].

\bibitem{Martel2003}
K.~Martel, \emph{Particles and Black Holes: Time-Domain Integration of the
  Equations of Black-Hole Perturbation Theory}, phd thesis, University of
  Guelph, Guelph, Ontario, Canada, 2003.

\bibitem{Kegeles:1979}
L.S.~Kegeles and J.M.~Cohen, \emph{Constructive procedure for perturbations of
  spacetimes}, \href{https://doi.org/10.1103/PhysRevD.19.1641}{\emph{Phys. Rev.
  D} {\bfseries 19} (1979) 1641}.

\bibitem{Wald:1978vm}
R.M.~Wald, \emph{{Construction of Solutions of Gravitational, Electromagnetic,
  Or Other Perturbation Equations from Solutions of Decoupled Equations}},
  \href{https://doi.org/10.1103/PhysRevLett.41.203}{\emph{Phys. Rev. Lett.}
  {\bfseries 41} (1978) 203}.

\bibitem{Green:2019nam}
S.R.~Green, S.~Hollands and P.~Zimmerman, \emph{{Teukolsky formalism for
  nonlinear Kerr perturbations}},
  \href{https://doi.org/10.1088/1361-6382/ab7075}{\emph{Class. Quant. Grav.}
  {\bfseries 37} (2020) 075001}
  [\href{https://arxiv.org/abs/1908.09095}{{\ttfamily 1908.09095}}].

\bibitem{Toomani:2021jlo}
V.~Toomani, P.~Zimmerman, A.~Spiers, S.~Hollands, A.~Pound and S.R.~Green,
  \emph{{New metric reconstruction scheme for gravitational self-force
  calculations}}, \href{https://doi.org/10.1088/1361-6382/ac37a5}{\emph{Class.
  Quant. Grav.} {\bfseries 39} (2022) 015019}
  [\href{https://arxiv.org/abs/2108.04273}{{\ttfamily 2108.04273}}].

\bibitem{Hollands:2024iqp}
S.~Hollands and V.~Toomani, \emph{{Metric Reconstruction in Kerr Spacetime}},
  \href{https://arxiv.org/abs/2405.18604}{{\ttfamily 2405.18604}}.

\bibitem{Aksteiner:2016pjt}
S.~Aksteiner, L.~Andersson and T.~B\"ackdahl, \emph{{New identities for
  linearized gravity on the Kerr spacetime}},
  \href{https://doi.org/10.1103/PhysRevD.99.044043}{\emph{Phys. Rev. D}
  {\bfseries 99} (2019) 044043}
  [\href{https://arxiv.org/abs/1601.06084}{{\ttfamily 1601.06084}}].

\bibitem{Wardell:2024yoi}
B.~Wardell, C.~Kavanagh and S.R.~Dolan, \emph{{Sourced metric perturbations of
  Kerr spacetime in Lorenz gauge}},
  \href{https://doi.org/10.1088/1361-6382/ae0918}{\emph{Class. Quant. Grav.}
  {\bfseries 42} (2025) 205007}
  [\href{https://arxiv.org/abs/2406.12510}{{\ttfamily 2406.12510}}].

\bibitem{Pitre:2026msx}
T.~Pitre, B.~Schneider and E.~Poisson, \emph{{Self-gravitating thin shells are
  dynamically unstable on all angular scales}},
  \href{https://arxiv.org/abs/2604.05980}{{\ttfamily 2604.05980}}.

\bibitem{Li:2023ulk}
D.~Li, A.~Hussain, P.~Wagle, Y.~Chen, N.~Yunes and A.~Zimmerman,
  \emph{{Isospectrality breaking in the Teukolsky formalism}},
  \href{https://doi.org/10.1103/PhysRevD.109.104026}{\emph{Phys. Rev. D}
  {\bfseries 109} (2024) 104026}
  [\href{https://arxiv.org/abs/2310.06033}{{\ttfamily 2310.06033}}].

\bibitem{Silva:2024ffz}
H.O.~Silva, G.~Tambalo, K.~Glampedakis, K.~Yagi and J.~Steinhoff,
  \emph{{Quasinormal modes and their excitation beyond general relativity}},
  \href{https://doi.org/10.1103/PhysRevD.110.024042}{\emph{Phys. Rev. D}
  {\bfseries 110} (2024) 024042}
  [\href{https://arxiv.org/abs/2404.11110}{{\ttfamily 2404.11110}}].

\bibitem{Silva:2026jih}
H.O.~Silva, G.~Tambalo, K.~Glampedakis and K.~Yagi, \emph{{Quasinormal modes
  and their excitation beyond general relativity. II. Isospectrality loss in
  gravitational waveforms}},
  \href{https://doi.org/10.1103/lgfg-lvjn}{\emph{Phys. Rev. D} {\bfseries 113}
  (2026) 084012} [\href{https://arxiv.org/abs/2601.13411}{{\ttfamily
  2601.13411}}].

\bibitem{Pope:2024ncb}
C.N.~Pope, D.O.~Rohrer and B.F.~Whiting, \emph{{Perturbations of Gibbons-Maeda
  black holes in Einstein-Maxwell-dilaton theories}},
  \href{https://doi.org/10.1103/PhysRevD.110.104036}{\emph{Phys. Rev. D}
  {\bfseries 110} (2024) 104036}
  [\href{https://arxiv.org/abs/2405.11042}{{\ttfamily 2405.11042}}].

\bibitem{Pope:2025jgz}
C.N.~Pope, D.O.~Rohrer and B.F.~Whiting, \emph{{Perturbations of black holes in
  Einstein-Maxwell-dilaton-axion theories}},
  \href{https://doi.org/10.1103/1b6k-f38p}{\emph{Phys. Rev. D} {\bfseries 112}
  (2025) 124064} [\href{https://arxiv.org/abs/2508.04589}{{\ttfamily
  2508.04589}}].

\bibitem{Cano:2024wzo}
P.A.~Cano and M.~David, \emph{{Isospectrality in Effective Field Theory
  Extensions of General Relativity}},
  \href{https://doi.org/10.1103/PhysRevLett.134.191401}{\emph{Phys. Rev. Lett.}
  {\bfseries 134} (2025) 191401}
  [\href{https://arxiv.org/abs/2407.12080}{{\ttfamily 2407.12080}}].

\bibitem{Cheung:2022rbm}
M.H.-Y.~Cheung et~al., \emph{{Nonlinear Effects in Black Hole Ringdown}},
  \href{https://doi.org/10.1103/PhysRevLett.130.081401}{\emph{Phys. Rev. Lett.}
  {\bfseries 130} (2023) 081401}
  [\href{https://arxiv.org/abs/2208.07374}{{\ttfamily 2208.07374}}].

\bibitem{Mitman:2022qdl}
K.~Mitman et~al., \emph{{Nonlinearities in Black Hole Ringdowns}},
  \href{https://doi.org/10.1103/PhysRevLett.130.081402}{\emph{Phys. Rev. Lett.}
  {\bfseries 130} (2023) 081402}
  [\href{https://arxiv.org/abs/2208.07380}{{\ttfamily 2208.07380}}].

\bibitem{Dyer:2025hdt}
R.~Dyer and C.J.~Moore, \emph{{The quasinormal mode content of binary black
  hole ringdown}},  \href{https://arxiv.org/abs/2510.13954}{{\ttfamily
  2510.13954}}.

\bibitem{Yi:2024elj}
S.~Yi, A.~Kuntz, E.~Barausse, E.~Berti, M.H.-Y.~Cheung, K.~Kritos et~al.,
  \emph{{Nonlinear quasinormal mode detectability with next-generation
  gravitational wave detectors}},
  \href{https://doi.org/10.1103/PhysRevD.109.124029}{\emph{Phys. Rev. D}
  {\bfseries 109} (2024) 124029}
  [\href{https://arxiv.org/abs/2403.09767}{{\ttfamily 2403.09767}}].

\bibitem{Lagos:2024ekd}
M.~Lagos, T.~Andrade, J.~Rafecas-Ventosa and L.~Hui, \emph{{Black hole
  spectroscopy with nonlinear quasinormal modes}},
  \href{https://doi.org/10.1103/PhysRevD.111.024018}{\emph{Phys. Rev. D}
  {\bfseries 111} (2025) 024018}
  [\href{https://arxiv.org/abs/2411.02264}{{\ttfamily 2411.02264}}].

\bibitem{Wang:2026rev}
Y.-F.~Wang, S.~Ma, N.~Khera and H.~Yang, \emph{{A nonlinear voice from GW250114
  ringdown}},  \href{https://arxiv.org/abs/2601.05734}{{\ttfamily 2601.05734}}.

\bibitem{Brizuela:2006ne}
D.~Brizuela, J.M.~Martin-Garcia and G.A.~Mena~Marugan, \emph{{Second and
  higher-order perturbations of a spherical spacetime}},
  \href{https://doi.org/10.1103/PhysRevD.74.044039}{\emph{Phys. Rev. D}
  {\bfseries 74} (2006) 044039}
  [\href{https://arxiv.org/abs/gr-qc/0607025}{{\ttfamily gr-qc/0607025}}].

\bibitem{Brizuela:2007zza}
D.~Brizuela, J.M.~Martin-Garcia and G.A.M.~Marugan, \emph{{High-order
  gauge-invariant perturbations of a spherical spacetime}},
  \href{https://doi.org/10.1103/PhysRevD.76.024004}{\emph{Phys. Rev. D}
  {\bfseries 76} (2007) 024004}
  [\href{https://arxiv.org/abs/gr-qc/0703069}{{\ttfamily gr-qc/0703069}}].

\bibitem{Brizuela:2009qd}
D.~Brizuela, J.M.~Martin-Garcia and M.~Tiglio, \emph{{A Complete
  gauge-invariant formalism for arbitrary second-order perturbations of a
  Schwarzschild black hole}},
  \href{https://doi.org/10.1103/PhysRevD.80.024021}{\emph{Phys. Rev. D}
  {\bfseries 80} (2009) 024021}
  [\href{https://arxiv.org/abs/0903.1134}{{\ttfamily 0903.1134}}].

\bibitem{Ioka:2007ak}
K.~Ioka and H.~Nakano, \emph{{Second and higher-order quasi-normal modes in
  binary black hole mergers}},
  \href{https://doi.org/10.1103/PhysRevD.76.061503}{\emph{Phys. Rev. D}
  {\bfseries 76} (2007) 061503}
  [\href{https://arxiv.org/abs/0704.3467}{{\ttfamily 0704.3467}}].

\bibitem{Bucciotti:2023ets}
B.~Bucciotti, A.~Kuntz, F.~Serra and E.~Trincherini, \emph{{Nonlinear
  quasi-normal modes: uniform approximation}},
  \href{https://doi.org/10.1007/JHEP12(2023)048}{\emph{JHEP} {\bfseries 12}
  (2023) 048} [\href{https://arxiv.org/abs/2309.08501}{{\ttfamily
  2309.08501}}].

\bibitem{Redondo-Yuste:2023seq}
J.~Redondo-Yuste, G.~Carullo, J.L.~Ripley, E.~Berti and V.~Cardoso, \emph{{Spin
  dependence of black hole ringdown nonlinearities}},
  \href{https://doi.org/10.1103/PhysRevD.109.L101503}{\emph{Phys. Rev. D}
  {\bfseries 109} (2024) L101503}
  [\href{https://arxiv.org/abs/2308.14796}{{\ttfamily 2308.14796}}].

\bibitem{Redondo-Yuste:2023ipg}
J.~Redondo-Yuste, D.~Pere{\~n}iguez and V.~Cardoso, \emph{{Ringdown of a
  dynamical spacetime}},
  \href{https://doi.org/10.1103/PhysRevD.109.044048}{\emph{Phys. Rev. D}
  {\bfseries 109} (2024) 044048}
  [\href{https://arxiv.org/abs/2312.04633}{{\ttfamily 2312.04633}}].

\bibitem{May:2024rrg}
T.~May, S.~Ma, J.L.~Ripley and W.E.~East, \emph{{Nonlinear effect of absorption
  on the ringdown of a spinning black hole}},
  \href{https://doi.org/10.1103/PhysRevD.110.084034}{\emph{Phys. Rev. D}
  {\bfseries 110} (2024) 084034}
  [\href{https://arxiv.org/abs/2405.18303}{{\ttfamily 2405.18303}}].

\bibitem{Bucciotti:2024zyp}
B.~Bucciotti, L.~Juliano, A.~Kuntz and E.~Trincherini, \emph{{Quadratic
  Quasi-Normal Modes of a Schwarzschild Black Hole}},
  \href{https://arxiv.org/abs/2405.06012}{{\ttfamily 2405.06012}}.

\bibitem{Bucciotti:2024jrv}
B.~Bucciotti, L.~Juliano, A.~Kuntz and E.~Trincherini, \emph{{Amplitudes and
  Polarizations of Quadratic Quasi-Normal Modes for a Schwarzschild Black
  Hole}},  \href{https://arxiv.org/abs/2406.14611}{{\ttfamily 2406.14611}}.

\bibitem{BenAchour:2024skv}
J.~Ben~Achour and H.~Roussille, \emph{{Quadratic perturbations of the
  Schwarzschild black hole: the algebraically special sector}},
  \href{https://doi.org/10.1088/1475-7516/2024/07/085}{\emph{JCAP} {\bfseries
  07} (2024) 085} [\href{https://arxiv.org/abs/2406.08159}{{\ttfamily
  2406.08159}}].

\bibitem{Ma:2024qcv}
S.~Ma and H.~Yang, \emph{{Excitation of quadratic quasinormal modes for Kerr
  black holes}}, \href{https://doi.org/10.1103/PhysRevD.109.104070}{\emph{Phys.
  Rev. D} {\bfseries 109} (2024) 104070}
  [\href{https://arxiv.org/abs/2401.15516}{{\ttfamily 2401.15516}}].

\bibitem{Bourg:2024jme}
P.~Bourg, R.~Panosso~Macedo, A.~Spiers, B.~Leather, B.~Bonga and A.~Pound,
  \emph{{Quadratic quasi-normal mode dependence on linear mode parity}},
  \href{https://arxiv.org/abs/2405.10270}{{\ttfamily 2405.10270}}.

\bibitem{Singh:2025xzd}
J.~Singh and V.~Suneeta, \emph{{Computing nonlinearity ratios using second
  order black hole perturbation theory}},
  \href{https://arxiv.org/abs/2512.00943}{{\ttfamily 2512.00943}}.

\bibitem{Cardoso:2026llh}
V.~Cardoso, J.~Redondo-Yuste, U.~Sperhake and F.~Tuncer, \emph{{Nonlinear
  Dynamics in General Relativity}},
  \href{https://arxiv.org/abs/2603.04501}{{\ttfamily 2603.04501}}.

\bibitem{Speeney:2024mas}
N.~Speeney, E.~Berti, V.~Cardoso and A.~Maselli, \emph{{Black holes surrounded
  by generic matter distributions: Polar perturbations and energy flux}},
  \href{https://doi.org/10.1103/PhysRevD.109.084068}{\emph{Phys. Rev. D}
  {\bfseries 109} (2024) 084068}
  [\href{https://arxiv.org/abs/2401.00932}{{\ttfamily 2401.00932}}].

\bibitem{Redondo-Yuste:2024vdb}
J.~Redondo-Yuste, \emph{{Perturbations of relativistic dissipative stars}},
  \href{https://doi.org/10.1088/1361-6382/adbfef}{\emph{Class. Quant. Grav.}
  {\bfseries 42} (2025) 075012}
  [\href{https://arxiv.org/abs/2411.16841}{{\ttfamily 2411.16841}}].

\bibitem{Franchini:2023xhd}
N.~Franchini, \emph{{Slow rotation black hole perturbation theory}},
  \href{https://doi.org/10.1103/PhysRevD.108.044079}{\emph{Phys. Rev. D}
  {\bfseries 108} (2023) 044079}
  [\href{https://arxiv.org/abs/2305.19313}{{\ttfamily 2305.19313}}].

\bibitem{Berti:2004ny}
E.~Berti, F.~White, A.~Maniopoulou and M.~Bruni, \emph{{Rotating neutron stars:
  An Invariant comparison of approximate and numerical spacetime models}},
  \href{https://doi.org/10.1111/j.1365-2966.2005.08812.x}{\emph{Mon. Not. Roy.
  Astron. Soc.} {\bfseries 358} (2005) 923}
  [\href{https://arxiv.org/abs/gr-qc/0405146}{{\ttfamily gr-qc/0405146}}].

\bibitem{Gerosa:2013laa}
D.~Gerosa, M.~Kesden, E.~Berti, R.~O'Shaughnessy and U.~Sperhake,
  \emph{{Resonant-plane locking and spin alignment in stellar-mass black-hole
  binaries: a diagnostic of compact-binary formation}},
  \href{https://doi.org/10.1103/PhysRevD.87.104028}{\emph{Phys. Rev. D}
  {\bfseries 87} (2013) 104028}
  [\href{https://arxiv.org/abs/1302.4442}{{\ttfamily 1302.4442}}].

\bibitem{xAct}
J.M.~Mart{\'i}n-Garc{\'i}a, ``xact: Efficient tensor computer algebra for the
  wolfram language.'' \url{https://www.xact.es/}, 2002--.

\end{thebibliography}\endgroup
\clearpage

\appendix

\section{Proofs about $\C$ and $\J$}\label{sec:Ap1}

\subsection{Proof of Eq.~\eqref{eq:algC}}
Here we show that $\C$ inherits the algebraic symmetries of the Riemann tensor and, in addition, satisfies Eqs.~\eqref{eq:algC}. This follows from the well-known fact that the Weyl tensor,
\begin{equation}\label{eq:Weyl_Def}
    \tensor{C}{_\mu_\nu^\alpha^\beta}\coloneqq \tensor{R}{_\mu_\nu^\alpha^\beta}+\frac{1}{3}R\delta^{\ [\alpha}_{ \mu}\delta^{\ \beta]}_{ \nu}-2\delta^{\ [\alpha}_{ [\mu} R^{\ \beta]}_{ \nu]}\, ,
\end{equation}
is, by construction, trace-free, $\tensor{C}{^\alpha_\mu_\alpha_\nu}=0$, inherits the algebraic symmetries of the Riemann tensor, and has equal left and right duals \cite{Penrose:1985bww},
\begin{equation}\label{eq:dsW}
     \tensor{\epsilon}{_\mu_\nu^\alpha^\beta}C_{\alpha\beta\rho\sigma}=\tensor{\epsilon}{_\rho_\sigma^\alpha^\beta}C_{\alpha\beta\mu\nu}\, .
\end{equation}
The algebraic properties $\C_{(\mu\nu)\rho\sigma}=\C_{\mu\nu(\rho\sigma)}=0$ and $\C_{\mu\nu\rho\sigma}=\C_{\rho\sigma\mu\nu}$ are immediate. To establish the cyclic property, first note that
\begin{equation}
\begin{aligned}
     \tensor{\epsilon}{^\mu^\alpha^\beta^\gamma}\C_{\nu\alpha\beta\gamma}&=\tensor{\epsilon}{^\mu^\alpha^\beta^\gamma}C_{\nu\alpha\beta\gamma}-\frac{1}{2}i\tensor{\epsilon}{^\mu^\alpha^\beta^\gamma}\tensor{\epsilon}{_\nu_\alpha_\delta_\eta}\tensor{C}{^\delta^\eta_\beta_\gamma}\\
     &=\tensor{\epsilon}{^\mu^\alpha^\beta^\gamma}C_{\nu[\alpha\beta\gamma]}-\frac{1}{2}i\tensor{\epsilon}{^\mu^\alpha^\beta^\gamma}\tensor{\epsilon}{_\beta_\gamma_\delta_\eta}\tensor{C}{^\delta^\eta_\nu_\alpha}\\
     &=2i\delta^{\mu}_{[\delta}\delta^{\alpha}_{\eta]}\tensor{C}{^\delta^\eta_\nu_\alpha}\\
     &=2i\tensor{C}{^\mu^\alpha_\nu_\alpha}\\
    &=0\, ,
\end{aligned}   
\end{equation}
where in the second step we used \eqref{eq:dsW}, and in the third the identity
\begin{equation}\label{eq:epsid}
    \epsilon^{\mu_{1}...\mu_{p}\mu_{p+1}...\mu_{4}}\epsilon_{\nu_{1}...\nu_{p}\mu_{p+1}...\mu_{4}}=-p!(4-p)!\delta^{\mu_{1}}_{[\nu_{1}}...\delta^{\mu_{p}}_{\nu_{p}]}\, ,\qquad \left(0\leq p\leq 4\right)\, .
\end{equation}
It then follows that $\C_{\mu[\nu\rho\sigma]}=0$, since
\begin{equation}\label{eq:cyclic_rel}
\begin{aligned}
        0&=\tensor{\epsilon}{_\mu_\xi_\lambda_\sigma}\tensor{\epsilon}{^\mu^\alpha^\beta^\gamma}\C_{\nu\alpha\beta\gamma}\\
        &=-3!\delta^{\alpha}_{[\xi}\delta^{\beta}_{\lambda}\delta^{\gamma}_{\sigma]}\C_{\nu\alpha\beta\gamma}\\
        &=-3!\C_{\nu[\xi\lambda\sigma]}\, .
\end{aligned}
\end{equation}
That $\C$ is trace-free follows from
\begin{equation}
     \tensor{\C}{^\gamma_\mu_\gamma_\nu}=-\frac{1}{2}i\tensor{\epsilon}{^\gamma_\mu^\alpha^\beta}C_{\alpha\beta\gamma\nu}=-\frac{1}{2}i\tensor{\epsilon}{^\gamma_\mu^\alpha^\beta}C_{[\alpha\beta\gamma]\nu}=0\, .
\end{equation}
Finally, the equality of the left and right duals, as well as self-duality, follow from
\begin{equation}
\begin{aligned}
    \tensor{\epsilon}{_\mu_\nu^\alpha^\beta}\C_{\alpha\beta\rho\sigma}&=\tensor{\epsilon}{_\mu_\nu^\alpha^\beta}\left(C_{\alpha\beta\rho\sigma}-\frac{1}{2}i\tensor{\epsilon}{_\alpha_\beta^\gamma^\delta}C_{\gamma\delta\rho\sigma}\right)\\
    &=\tensor{\epsilon}{_\mu_\nu^\alpha^\beta}C_{\alpha\beta\rho\sigma}+2i \delta^{\gamma}_{[\mu}\delta^{\delta}_{\nu]}C_{\gamma\delta\rho\sigma}\\
    &=\tensor{\epsilon}{_\mu_\nu^\alpha^\beta}C_{\alpha\beta\rho\sigma}+2i C_{\mu\nu\rho\sigma}\, .
\end{aligned}
\end{equation}
It is then immediate that $\tensor{\epsilon}{_\mu_\nu^\alpha^\beta}\C_{\alpha\beta\rho\sigma}=\tensor{\epsilon}{_\rho_\sigma^\alpha^\beta}\C_{\alpha\beta\mu\nu}$, and that $\tensor{\epsilon}{_\mu_\nu^\alpha^\beta}\C_{\alpha\beta\rho\sigma}=2i \C_{\mu\nu\rho\sigma}$.

\subsection{Proof of Eq.~\eqref{eq:difC}}

Here we prove \eqref{eq:difC}. First, we write
\begin{equation}\label{eq:div_C}
    \nabla^{\mu}\tensor{C}{_\mu_\nu^\alpha^\beta}=\nabla^{\mu}\tensor{R}{_\mu_\nu^\alpha^\beta}+\frac{1}{3}\delta^{\alpha}_{[\mu}\delta^{\beta}_{\nu]}\nabla^{\mu}R-2\delta^{[\alpha}_{[\mu}\nabla^{\vert \mu \rvert}\tensor{R}{^{\beta]}_{\nu]}}\,,
\end{equation}
where we have used the definition of the Weyl tensor \eqref{eq:Weyl_Def}. From the differential Bianchi identity for the Riemann tensor, $\nabla_{[\mu}R_{\nu\rho]\sigma\alpha}=0$, it follows that
\begin{equation}
    \nabla^{\mu}R_{\mu\beta\gamma\nu}=2\nabla_{[\gamma}R_{\nu]\beta}\, .
\end{equation}
Using $G_{\mu\nu}=T_{\mu\nu}$, the right-hand side of \eqref{eq:div_C} can therefore be expressed entirely in terms of $T_{\mu\nu}$, yielding
\begin{equation}\label{eq:div_C_To_T}
    \nabla^{\mu}\tensor{C}{_\mu_\nu^\alpha^\beta}=\nabla^{[\alpha}\tensor{\tilde{T}}{^{\beta]}_\nu}\,,
\end{equation}
where $\tilde{T}_{\mu\nu}$ was defined in \eqref{eq:Jcurrent}. It then follows immediately that
\begin{equation}
\begin{aligned}
    \nabla^{\mu}\tensor{\C}{_\mu_\nu^\alpha^\beta}
    &= \nabla^{\mu}\tensor{C}{_\mu_\nu^\alpha^\beta}-\frac{1}{2}i\tensor{\epsilon}{_\mu_\nu^\lambda^\rho}\nabla^{\mu}\tensor{C}{_\lambda_\rho^\alpha^\beta}\\
    &= \nabla^{\mu}\tensor{C}{_\mu_\nu^\alpha^\beta}-\frac{1}{2}i\tensor{\epsilon}{^\alpha^\beta^\lambda^\rho}\nabla^{\mu}\tensor{C}{_\lambda_\rho_\mu_\nu}\\
    &= \tensor{\J}{_\nu^\alpha^\beta}\, ,
\end{aligned}
\end{equation}
where in the second step we used \eqref{eq:algC} (together with $\nabla\epsilon=0$), and in the final step we used \eqref{eq:div_C_To_T} and the definition of $\J$ given in \eqref{eq:Jcurrent}.

\subsection{Proof of Eq.~\eqref{eq:algsdJ}}

First, write $\J$ as
\begin{equation}
    \tensor{\J}{_\mu^\alpha^\beta}= \tensor{Q}{_\mu^\alpha^\beta}+i \tensor{M}{_\mu^\alpha^\beta}\,,
\end{equation}
where 
\begin{equation}
 \tensor{Q}{_\mu^\alpha^\beta}\coloneqq\nabla^{[\alpha}\tilde{T}^{\beta]}_{\ \mu}\, , 
 \quad 
 \tensor{M}{_\mu^\alpha^\beta}\coloneqq-\frac{1}{2}\tensor{\epsilon}{_\lambda_\rho^\alpha^\beta}\tensor{Q}{_\mu^\lambda^\rho}\, .
\end{equation}
It is straightforward to verify that
\begin{equation}
    \tensor{Q}{_\mu^\mu^\beta}=\frac{1}{2}\nabla_{\mu}T^{\mu\beta}\, ,\quad Q_{[\mu\alpha\beta]}=0\, ,
\end{equation}
where the second identity follows from $\tilde{T}_{[\mu\nu]}=0$. Although we leave $T_{\mu\nu}$ unspecified, it must in any case be divergence-free, $\nabla_\mu T^{\mu\nu}=0$ (see Eq.~\eqref{eq:ConsT}), by consistency with the Einstein equation \eqref{eq:Einstein}. Hence $\tensor{Q}{_\mu^\mu^\beta}=0$.

For $\tensor{M}{_\mu^\alpha^\beta}$, one finds
\[
\tensor{M}{_\mu^\mu^\beta}= -\frac{1}{2}\epsilon^{\sigma\rho\mu\beta}Q_{\mu\sigma\rho}=0\, .
\]
Moreover, $\epsilon^{\mu\rho\sigma\nu}M_{\mu\rho\sigma}=\nabla^{\mu}T_{\mu}^{\ \nu}=0$, which implies $M_{[\mu\rho\sigma]}=0$ (see Eq.~\eqref{eq:cyclic_rel}). This completes the proof. 

\subsection{Proof of Eq.~\eqref{eq:triC2}}

A relatively direct proof of \eqref{eq:difC} follows by using the differential-form version of the curvature equations. We follow the notation and conventions of \cite{Mukkamala:2024dxf}, where the Hodge dual and the exterior covariant derivative $D$ of a tensor-valued differential $p$-form $X_{\mu_{1}...\mu_{p}\mathcal{A}}$ (with $\mathcal{A}$ an arbitrary index structure and $\mu_{1}...\mu_{p}$ fully antisymmetric) are
\begin{equation}
\begin{aligned}
    D_{\mu}X_{\mu_{1}...\mu_{p}\mathcal{A}}&\equiv(p+1)\nabla_{[\mu}X_{\mu_{1}...\mu_{p}]\mathcal{A}}\, ,\\ 
    (\star X)_{\mu_{1}...\mu_{4-p}\mathcal{A}}&\equiv\frac{1}{p!}\tensor{\epsilon}{_{\mu_{1}}_{...}_{\mu_{4-p}}^{\nu_{1}}^{...}^{\nu_{p}}}X_{\nu_{1}...\nu_{p}\mathcal{A}}\, .
\end{aligned}
\end{equation}
In this notation, Eq.~\eqref{eq:difC} becomes
\begin{equation}\label{eq:DCJ}
    \star D \star\C=\J\, .
\end{equation}
To derive \eqref{eq:triC2}, we compute $\triangle\C$ in two ways, using the self-duality of $\C$. First,
\begin{equation}
\begin{aligned}\label{eq:tri_C_1}
    \triangle\C&=\left(\star D\star D+D\star D\star\right)\C\\
    &=\left(-i\star D\star D\star+D\star D\star\right)\C\\
    &=\left(D-i\star D\right)\J\, ,
\end{aligned}
\end{equation}
where in the second step we used the self-duality of $\C$, and in the third step \eqref{eq:DCJ}. This expresses $\triangle\C$ in terms of the current $\J$.

To obtain a form with manifestly hyperbolic principal part, we rewrite
\begin{equation}
    \triangle\C=\left(\star D\star D+D\star D\star\right)\C\\
    =\left(\star D\star D+i D\star D\right)\C
\end{equation}
using again the self-duality of $\C$. It therefore suffices to compute $\star D\star D\C$, since $D\star D \C$ is just (minus) its dual. One finds
\begin{equation}\label{eq:sDsDC}
\begin{aligned}
    (\star D\star D\C)_{\mu\nu\rho\sigma}&=\nabla^{\alpha}D_{\mu}\C_{\nu\alpha\rho\sigma}\\
    &=3\nabla^{\alpha}\nabla_{[\mu}\C_{\nu\alpha]\rho\sigma}\\
    &=\square\C_{\mu\nu\rho\sigma}+2\nabla^{\alpha}\nabla_{[\mu}\C_{\nu]\alpha\rho\sigma}\\
    &=\square\C_{\mu\nu\rho\sigma}+2\nabla_{[\mu}\nabla^{\alpha}\C_{\nu]\alpha\rho\sigma}+2[\nabla^{\alpha},\nabla_{[\mu}]\C_{\nu]\alpha\rho\sigma}\\
    &=\square\C_{\mu\nu\rho\sigma}-D_{\mu}\J_{\nu\rho\sigma}+2[\nabla^{\alpha},\nabla_{[\mu}]\C_{\nu]\alpha\rho\sigma}\, ,
\end{aligned}
\end{equation}
where we used, respectively, a standard differential-form identity, the definition of $D$, the algebraic symmetries of $\C$, commutation of covariant derivatives, and finally the definitions of $D$ and $\J$.

From this, we obtain immediately
\begin{equation}\label{eq:DsDC}
\begin{aligned}
        ( iD\star D\C)_{\mu\nu\rho\sigma}&=-i\star( \star D\star D\C)_{\mu\nu\rho\sigma}\\
        &=\square\C_{\mu\nu\rho\sigma}+i\left(\star D\J\right)_{\mu\nu\rho\sigma}-i\tensor{\epsilon}{_\mu_\nu^\alpha^\beta}[\nabla^{\gamma},\nabla_{\alpha}]\C_{\beta\gamma\rho\sigma}\, .
    \end{aligned}
\end{equation}
Adding \eqref{eq:sDsDC} and \eqref{eq:DsDC} gives
\begin{equation}
\begin{aligned}\label{eq:tri_C_2}
    \triangle\C&=2\square\C_{\mu\nu\rho\sigma}-D_{\mu}\J_{\nu\rho\sigma}+i(\star D\J)_{\mu\nu\rho\sigma}\\
    &+2[\nabla^{\alpha},\nabla_{[\mu}]\C_{\nu]\alpha\rho\sigma}-i\tensor{\epsilon}{_\mu_\nu^\alpha^\beta}[\nabla^{\gamma},\nabla_{\alpha}]\C_{\beta\gamma\rho\sigma}\, .
\end{aligned}
\end{equation}
The commutators of covariant derivatives are rewritten using Ricci's identity. The Riemann tensor is then decomposed into Weyl and Ricci parts using \eqref{eq:Weyl_Def}, and the Ricci tensor is replaced by $T_{\mu\nu}$ through \eqref{eq:Einstein}. This yields contractions involving $C_{\mu\nu\rho\sigma}$ and $\C_{\mu\nu\rho\sigma}$, which must be expressed entirely in terms of $\C$.

This follows from the identities
\begin{equation}\label{eq:alg_Csq_1}
        C^{\mu\nu\alpha\beta}\C_{\rho\sigma\alpha\beta}=\frac{1}{2}\C^{\mu\nu\alpha\beta}\C_{\rho\sigma\alpha\beta}\,, 
\end{equation}
and 
\begin{equation}\label{eq:alg_Csq_2}
        \tensor{C}{_{[\mu}^\lambda^\rho^{[\alpha}}\tensor{\C}{^{\beta]}_{\lvert\lambda}_{\rho\rvert}_{\nu]}}=\tensor{\C}{_{[\mu}^\lambda^\rho^{[\alpha}}\tensor{\C}{^{\beta]}_{\lvert\lambda}_{\rho\rvert}_{\nu]}}+\frac{1}{4}\C^{\lambda\rho\delta[\alpha}\delta^{\beta]}_{[\mu}\C_{\nu]\delta\lambda\rho}-\frac{1}{8}\C_{\mu\nu\lambda\rho}\C^{\alpha\beta\lambda\rho}\, .
\end{equation}
The identity \eqref{eq:alg_Csq_1} follows from the self-duality of $\C$. To derive \eqref{eq:alg_Csq_2}, write
\begin{equation}
\begin{aligned}
    \tensor{\C}{_{[\mu}^\lambda^\rho^{[\alpha}}\tensor{\C}{^{\beta]}_{\lvert\lambda}_{\rho\rvert}_{\nu]}}&=\tensor{C}{_{[\mu}^\lambda^\rho^{[\alpha}}\tensor{\C}{^{\beta]}_{\lvert\lambda}_{\rho\rvert}_{\nu]}}-\frac{1}{2}i\tensor{\epsilon}{_{[\mu}^\lambda^{\eta}^\xi}\tensor{C}{_{\lvert\eta}_\xi^\rho^{[\alpha}}\tensor{\C}{^{\beta]}_{\lambda}_{\rho\rvert}_{\nu]}}\\
    &=\tensor{C}{_{[\mu}^\lambda^\rho^{[\alpha}}\tensor{\C}{^{\beta]}_{\lvert\lambda}_{\rho\rvert}_{\nu]}}-\frac{1}{4}\tensor{\epsilon}{_{[\mu}^\lambda^{\eta}^\xi}\tensor{C}{_{\lvert\eta}_\xi^\rho^{[\alpha}}\tensor{\epsilon}{^{\beta]}_{\lambda}^\delta^\sigma}\tensor{\C}{_\delta_\sigma_{\rho\rvert}_{\nu]}}\, ,
\end{aligned}
\end{equation}
where in the first line we expanded one $\C$ in terms of $C$, and in the second used the self-duality of $\C$. Applying the identity \eqref{eq:epsid} and the algebraic symmetries of $\C_{\mu\nu\rho\sigma}$ yields \eqref{eq:alg_Csq_2}.

With these results, the second line of \eqref{eq:tri_C_2} reduces to algebraic contractions of $T_{\mu\nu}$ with $\C_{\mu\nu\rho\sigma}$, and of $\C_{\mu\nu\rho\sigma}$ with itself. Combining \eqref{eq:tri_C_1} and \eqref{eq:tri_C_2} yields \eqref{eq:triC2}.

\section{Fundamental Equations}\label{sec:Ap3}

We present the fundamental equations explicitly for perturbations about a Schwarzschild black hole in GR, including the presence of matter sources. That is, we will impose the following background relations,
\begin{equation}
    r^{a}r_{a}=1-\frac{2M}{r}\, ,\quad T_{ab}=\mathcal{P}=0\, ,
\end{equation}
but keep general the fluctuations of the energy-momentum tensor. Under the same assumtions, we also provide the explicit expressions for the components of $\delta\J$ and $\delta\C$ in terms of $\delta T$ and $h$, respectively. The general expressions for an arbitrary background $T_{\mu\nu}$ and arbitrary fluctuations $\delta T_{\mu\nu}$ are provided in the accompanying \texttt{Mathematica} notebook. To denote the covariant derivative of $g_{ab}$ we will use $\nabla_{a}$ and ``$:$'', and we define $\square=\nabla_{a}\nabla^{a}$. 

\subsection{Einstein Equations}\label{sec:EinstEq}

Here we present the harmonic components of the linearised Einstein equation \eqref{eq:Einstein},
\begin{equation}
    \delta G_{\mu\nu}=\delta T_{\mu\nu}\, .
\end{equation}
\begin{itemize}
    \item $(\mu,\nu)=(a,b)$, $Y$ component: 
    \begin{align}\notag
        \Phi^{+}{}_{ab}&=\frac{2 M h^{+}{}_{ab}}{r^3} + \frac{\Lambda^2 \
h^{+}{}_{ab}}{2 r^2} -  \frac{2 M g_{ab} \
h^{+}{}^{c}{}_{c}}{r^3} -  \frac{\Lambda^2 g_{ab} \
h^{+}{}^{c}{}_{c}}{2 r^2} -  \frac{2 M g_{ab} \
h^{+}{}}{r^5} + \frac{g_{ab} h^{+}{}}{r^4} \\ \notag
&- \frac{\Lambda^2 g_{ab} h^{+}{}}{2 r^4}  -  \
\frac{2 h^{+}{} r_{a} r_{b}}{r^4} + \
\frac{h^{+}{}{}_{:a} r_{b}}{r^3} + \frac{r_{a} \
h^{+}{}{}_{:b}}{r^3} -  \
\frac{h^{+}{}{}_{:a}{}_{:b}}{r^2} \\ \notag
&+ \frac{g_{ab} h^{+}{}{}^{:c}{}_{:c}}{r^2} + \
\frac{h^{+}{}_{bc}{}_{:a} r^{c}}{r} + \
\frac{h^{+}{}_{ac}{}_{:b} r^{c}}{r} -  \
\frac{h^{+}{}_{ab}{}_{:c} r^{c}}{r} + \frac{g_{ab} \
h^{+}{}^{d}{}_{d}{}_{:c} r^{c}}{r} -  \frac{g_{ab} \
h^{+}{}{}_{:c} r^{c}}{r^3} \\ 
&-  \frac{2 g_{ab} r^{c} h^{+}{}_{c}{}^{d}{}_{:d}}{r} -  \frac{g_{ab} \
h^{+}{}_{cd} r^{c} r^{d}}{r^2} \, .
    \end{align}
Notice we present it such that no second derivatives of $h_{ab}$ are present. As discussed in  \cite{Pereniguez:2023wxf}, this is possible using the linearised Gauss-Bonnet identity of 2-dimensional manifolds (specialised to the Shcwarzschild background it is given in Eq.~\eqref{eq:GB_1st}).

    \item  $(\mu,\nu)=(a,A)$, $Z$ component:
    \begin{align}\label{eq:Einstein_Lorentz}
\Phi^{+}{}_{a}&= \frac{h^{+}{}^{b}{}_{b} r_{a}}{2 r} + \
\frac{h^{+}{} r_{a}}{r^3} -  \frac{1}{2} \
h^{+}{}^{b}{}_{b}{}_{:a} -  \frac{h^{+}{}{}_{:a}}{2 \
r^2} + \frac{1}{2} h^{+}{}_{a}{}^{b}{}_{:b}\, ,
    \end{align}

\item  $(\mu,\nu)=(a,A)$, $X$ component:
    \begin{align}\notag 
\Phi^{-}{}_{a}&=- \frac{M h^{-}{}_{a}}{r^3} + \frac{\Lambda^2 \
h^{-}{}_{a}}{2 r^2} -  \
\frac{h^{-}{}^{b} r_{a} r_{b}}{r^2} -  \frac{r_{a} \
h^{-}{}^{b}{}_{:b}}{r} + \frac{1}{2} \
h^{-}{}^{b}{}_{:a}{}_{:b} \\ 
&-  \frac{1}{2} h^{-}{}_{a}{}^{:b}{}_{:b} + \
\frac{h^{-}{}_{b}{}_{:a} r^{b}}{r} \, ,
    \end{align}
    
\item  $(\mu,\nu)=(A,B)$, $U$ component:
    \begin{align}\notag
\Phi^{+}&=- \frac{1}{4} \Lambda^2 h^{+}{}^{a}{}_{a} -  \frac{4 M \
h^{+}{}}{r^3} + \frac{h^{+}{}}{r^2} + \
\frac{1}{2} h^{+}{}{}^{:a}{}_{:a} + \frac{1}{2} r \
h^{+}{}^{b}{}_{b}{}_{:a} r^{a} -  \
\frac{h^{+}{}{}_{:a} r^{a}}{r} \\ 
&-  r r^{a} h^{+}{}_{a}{}^{b}{}_{:b} -  \frac{1}{2} r^2 \
h^{+}{}^{ab}{}_{:a}{}_{:b} + \frac{1}{2} r^2 \
h^{+}{}^{a}{}_{a}{}^{:b}{}_{:b} \, ,
    \end{align}

\item  $(\mu,\nu)=(A,B)$, $V$ component:
    \begin{align}\label{eq:Einstein_Trace}
\Phi^{\oplus}&=- \frac{1}{2} h^{+}{}^{a}{}_{a} \, ,
    \end{align}

\item  $(\mu,\nu)=(A,B)$, $W$ component:
    \begin{align}\label{eq:Einstein_Div}
\Phi^{-}{} &= h^{-}{}^{a}{}_{:a}\, .
    \end{align}
    
\end{itemize}

\subsection{Energy-momentum Conservation}\label{app:Tcons}

Here we present the harmonic components of the linearised conservation of the energy-momentum tensor \eqref{eq:ConsT},
\begin{equation}
    \delta \left(\nabla^{\nu}T_{\nu\mu}\right)=0\, .
\end{equation}
\begin{itemize}
    \item $\mu=a$, $Y$ component:
    \begin{align}
        \Phi^{+}{}^{a}{}_{b}{}_{:a}&=\frac{\Lambda^2 \Phi^{+}{}_{b}}{r^2} -  \frac{2 \Phi^{+}{}_{ba} \
r^{a}}{r} + \frac{2 \Phi^{+}{} r_{b}}{r^3}\, ,
    \end{align}

     \item $\mu=A$, $Z$ component:
    \begin{align}
        \Phi^{+}{}^{a}{}_{:a}&=- \frac{\Phi^{\oplus}{}}{r^2} + \frac{\Lambda^2 \Phi^{\oplus}{}}{2 \
r^2} -  \frac{\Phi^{+}{}}{r^2} -  \frac{2 \Phi^{+}{}^{a} r_{a}}{r}\, ,
    \end{align}

     \item $\mu=A$, $X$ component:
    \begin{align}
        \Phi^{-}{}^{a}{}_{:a}&=- \frac{\Phi^{-}{}}{r^2} + \frac{\Lambda^2 \Phi^{-}{}}{2 r^2} -  \
\frac{2 \Phi^{-}{}^{a} r_{a}}{r}\, .
    \end{align}
\end{itemize}

\subsection{Algebraic Equations of $\delta \C$}\label{app:algebraic_C}

From the traceless property in \eqref{eq:algC}, 
\begin{equation}
    \delta\left(\tensor{\C}{^\alpha_\mu_\alpha_\nu}\right)=0\, ,
\end{equation}
we find:
\begin{itemize}
    \item $(\mu,\nu)=(a,b)$, $Y$ component:
    \begin{align}\label{eq:Trace_ab}
        0&=\frac{2 M h^{+}{}_{ab}}{r^3} -  \frac{2 M g_{ab} \
h^{+}{}^{c}{}_{c}}{r^3} + \frac{2 M g_{ab} \
h^{+}{}}{r^5} + \frac{2 \Psi^{+}{}_{ab}}{r^2} -  g_{ab} \
\Psi^{+}{}\, ,
    \end{align}

    \item $(\mu,\nu)=(a,A)$, $Z$ component:
    \begin{align}\label{eq:Trace_aA_Z}
     0&=- \frac{3i M \varepsilon_{ab} h^{-}{}^{b}}{r^3} + \
\frac{\Xi^{+}{}_{a}}{r^2} + \varepsilon_{ab} \Psi^{+}{}^{b}\, ,
    \end{align}

    \item $(\mu,\nu)=(a,A)$, $X$ component:
    \begin{align}\label{eq:Trace_aA_X}
     0&=- \frac{M h^{-}{}_{a}}{r^3} -  \frac{\Xi^{-}{}_{a}}{r^2} + \
\varepsilon_{ab} \Psi^{-}{}^{b}\,,
    \end{align}

    \item $(\mu,\nu)=(A,B)$, $U$ component:
    \begin{align}
     0&=\frac{M \mathit{h}^{+}{}^{a}{}_{a}}{r} -  \frac{2 M \
\mathit{h}^{+}{}}{r^3} + \frac{\Xi^{+}{}}{r^2} + \Psi^{+}{}^{a}{}_{a}\, ,
    \end{align}

    \item $(\mu,\nu)=(A,B)$, $V$ component:
    \begin{align}\label{eq:Trace_AB_V}
     0&=\Psi^{\oplus}{}^{a}{}_{a}\, ,
    \end{align}

     \item $(\mu,\nu)=(A,B)$, $W$ component:
    \begin{align}
     0&=\Psi^{-}{}^{a}{}_{a}\, .
    \end{align}  
\end{itemize}
From the self-duality property in \eqref{eq:algC}, 
\begin{equation}
    \delta \left(\C_{\mu\nu\rho\sigma}+\frac{1}{2}i \tensor{\epsilon}{_\mu_\nu^\alpha^\beta}\tensor{\C}{_\alpha_\beta_\rho_\sigma}\right)=0\, ,
\end{equation}
one finds:
\begin{itemize}
    \item $(\mu,\nu,\rho,\sigma)=(a,b,c,d)$, $Y$ component: it is $\sim \varepsilon_{ab}\varepsilon_{cd}$, and contracting it with $\varepsilon^{ab}\varepsilon^{cd}$ yields
    \begin{align}
        0&=\frac{M \mathit{h}^{+}{}^{a}{}_{a}}{r^3} -  \frac{2 M \
\mathit{h}^{+}{}}{r^5} + \frac{i \Psi^{-}{}}{r^2} + \Psi^{+}{}\,,
    \end{align}

    \item $(\mu,\nu,\rho,\sigma)=(A,a,b,c)$, $Z$ component: it is $\sim \varepsilon_{bc}$ and contracting it with $\varepsilon^{bc}$ gives
    \begin{align}
        0&=- \frac{2i M \mathit{h}^{-}{}_{a}}{r^3} + i \varepsilon_{ab} \
\Psi^{-}{}^{b} + \Psi^{+}{}_{a}\, ,
    \end{align}

    \item $(\mu,\nu,\rho,\sigma)=(a,A,b,B)$, $U$ component: 
    \begin{align}\label{eq:SD_aAbB_U}
        0&=\frac{M \mathit{h}^{+}{}_{ab}}{r} -  \frac{M g_{ab} \
\mathit{h}^{+}{}^{c}{}_{c}}{2 r} + \Psi^{+}{}_{ab} + \frac{1}{2}i \
g_{ab} \Psi^{-}{}\,,
    \end{align}

     \item $(\mu,\nu,\rho,\sigma)=(a,A,b,B)$, $W$ component: 
    \begin{align}\label{eq:SD_aAbB_W}
       0&=\Psi^{-}{}_{ab} - i \varepsilon_{a}{}^{c} \Psi^{\oplus}{}_{bc}\, ,
    \end{align}

    \item $(\mu,\nu,\rho,\sigma)=(a,b,A,B)$, $Y \epsilon$ component: it is $\sim\varepsilon_{ab}$, so contracting it with $\varepsilon^{ab}$ gives
    \begin{align}\label{eq:SD_abAB}
       0&=\frac{i M \mathit{h}^{+}{}^{a}{}_{a}}{r} -  \frac{2i M \
\mathit{h}^{+}{}}{r^3} + \frac{i \Xi^{+}{}}{r^2} + \Psi^{-}{}\, ,
    \end{align}

     \item $(\mu,\nu,\rho,\sigma)=(a,A,B,C)$, $Z \otimes\epsilon$ component: 
    \begin{align}
       0&=- \frac{2 M \mathit{h}^{-}{}_{a}}{r} -  \Xi^{-}{}_{a} - i \varepsilon_{ab} \Xi^{+}{}^{b}\, .
    \end{align}
\end{itemize}
Some of the components where not shown because they yield equations that are manifestly the same as the ones above. Likewise, the equation between left and right duals in \eqref{eq:algC} does not provide new information. 

\subsection{First-derivative Equations of $\delta \C$}

Here we provide the equations following from the linearisation of \eqref{eq:difC},
\begin{equation}
    \delta \left(\nabla^{\alpha}\C_{\alpha\mu\nu\rho}-\J_{\mu\nu\rho}\right)=0\, .
\end{equation}
We provide all of them although it is clear not all are independent. 
\begin{itemize}
    \item $(\mu,\nu,\rho)=(a,b,c)$, $Y$ component: it is $\sim \varepsilon _{bc}$, so contracting with $\varepsilon^{bc}$ gives
    \begin{align}\notag
        \Pi^{+}{}_{a}&=\frac{3i \Lambda^2 M \mathit{h}^{-}{}_{a}}{r^5} -  \
\frac{\Lambda^2 \Psi^{+}{}_{a}}{r^2} + \frac{2 M \varepsilon_{a}{}^{c} \
\mathit{h}^{+}{}_{bc} r^{b}}{r^4} -  \frac{8 M \varepsilon_{ab} \
\mathit{h}^{+}{} r^{b}}{r^6} \\ \notag
&+ \frac{2 \varepsilon_{b}{}^{c} \
\Psi^{+}{}_{ac} r^{b}}{r^3} -  \frac{2 \varepsilon_{ab} \Psi^{+}{} \
r^{b}}{r} + \frac{3 M \varepsilon_{ab} \
\mathit{h}^{+}{}{}^{:b}}{r^5} -  \varepsilon_{ab} \Psi^{+}{}{}^{:b} \\
&-  \frac{2 M \varepsilon^{bc} \mathit{h}^{+}{}_{ab}{}_{:c}}{r^3} -  \
\frac{2 M \varepsilon_{a}{}^{b} \mathit{h}^{+}{}_{b}{}^{c}{}_{:c}}{r^3}\, ,
    \end{align}

    \item $(\mu,\nu,\rho)=(a,b,A)$, $Z$ component:
    \begin{align}\notag
     \Pi^{+}{}_{ab}&= - \frac{5 M \mathit{h}^{+}{}_{ab}}{2 r^3} + \frac{3 M g_{ab} \
\mathit{h}^{+}{}^{c}{}_{c}}{2 r^3} -  \frac{M g_{ab} \
\mathit{h}^{+}{}}{r^5}  -  \varepsilon_{ab} \Pi^{+}{} \
-  \frac{\Psi^{\oplus}{}_{ab}}{r^2} \\ \notag
&+ \frac{\Lambda^2 \
\Psi^{\oplus}{}_{ab}}{2 r^2} -  \frac{\Psi^{+}{}_{ab}}{r^2} + \
\frac{3i M \varepsilon_{ac} \mathit{h}^{-}{}^{c} r_{b}}{r^4} -  \
\frac{\Xi^{+}{}_{a} r_{b}}{r^3} -  \frac{3i M \varepsilon_{ac} \
\mathit{h}^{-}{}^{c}{}_{:b}}{2 r^3} \\ \notag
&+ \frac{3i M \varepsilon_{ab} \
\mathit{h}^{-}{}^{c} r_{c}}{r^4} -  \frac{i M \varepsilon_{ab} \
\mathit{h}^{-}{}^{c}{}_{:c}}{r^3} + \frac{\varepsilon_{bc} \
\Psi^{+}{}_{a} r^{c}}{r} + \frac{\varepsilon_{ac} \Psi^{+}{}_{b} \
r^{c}}{r}\\
&-  \frac{i M \varepsilon_{bc} \
\mathit{h}^{-}{}_{a}{}^{:c}}{r^3} -  \frac{i M \varepsilon_{ac} \
\mathit{h}^{-}{}_{b}{}^{:c}}{2 r^3} + \varepsilon_{ac} \
\Psi^{+}{}_{b}{}^{:c}\,,
    \end{align}

    \item $(\mu,\nu,\rho)=(a,b,A)$, $X$ component:
    \begin{align}\notag
   \Pi^{-}{}_{ab} &=- \frac{i M \varepsilon_{b}{}^{c} \mathit{h}^{+}{}_{ac}}{r^3} -  \
\frac{i M \varepsilon_{a}{}^{c} \mathit{h}^{+}{}_{bc}}{2 r^3} -  \
\frac{i M \varepsilon_{ab} \mathit{h}^{+}{}^{c}{}_{c}}{2 r^3} + \
\frac{i M \varepsilon_{ab} \mathit{h}^{+}{}}{r^5}  -  \
\varepsilon_{ab} \Pi^{-}{} \\\notag
&-  \frac{\Psi^{-}{}_{ab}}{r^2} + \
\frac{\Lambda^2 \Psi^{-}{}_{ab}}{2 r^2} + \frac{\varepsilon_{ab} \
\Psi^{-}{}}{2 r^2} -  \frac{2 M \mathit{h}^{-}{}_{b} r_{a}}{r^4} + \
\frac{3 M \mathit{h}^{-}{}_{b}{}_{:a}}{2 r^3} \\\notag
&+ \frac{3 M \
\mathit{h}^{-}{}_{a} r_{b}}{r^4} + \frac{\Xi^{-}{}_{a} \
r_{b}}{r^3} -  \frac{M \mathit{h}^{-}{}_{a}{}_{:b}}{2 r^3} + \
\frac{M g_{ab} \mathit{h}^{-}{}^{c} r_{c}}{r^4} -  \frac{M g_{ab} \
\mathit{h}^{-}{}^{c}{}_{:c}}{r^3} \\
&+ \frac{\varepsilon_{bc} \
\Psi^{-}{}_{a} r^{c}}{r} + \frac{\varepsilon_{ac} \Psi^{-}{}_{b} \
r^{c}}{r} + \varepsilon_{ac} \Psi^{-}{}_{b}{}^{:c}\,,
    \end{align}

     \item $(\mu,\nu,\rho)=(a,A,B)$, $Y \epsilon$ component:
    \begin{align}\notag
- 2 \Pi^{-}{}_{a}&=- \frac{2 \Lambda^2 M \mathit{h}^{-}{}_{a}}{r^3} -  \frac{\Lambda^2 \
\Xi^{-}{}_{a}}{r^2}  + \frac{i M \varepsilon_{a}{}^{b} \mathit{h}^{+}{}^{c}{}_{c}{}_{:b}}{r} -  \frac{2i M \
\varepsilon_{ab} \mathit{h}^{+}{} r^{b}}{r^4} -  \frac{\varepsilon_{ab} \Psi^{-}{} r^{b}}{r} \\\label{eq:FirstOrder_aAB_epsY}
&+ \frac{i M \varepsilon_{ab} \
\mathit{h}^{+}{}{}^{:b}}{r^3} -  \varepsilon_{ab} \Psi^{-}{}{}^{:b} -  \
\frac{2i M \varepsilon^{bc} \mathit{h}^{+}{}_{ab}{}_{:c}}{r} -  \
\frac{2i M \varepsilon_{a}{}^{b} \mathit{h}^{+}{}_{b}{}^{c}{}_{:c}}{r}\,,
    \end{align}

    \item $(\mu,\nu,\rho)=(A,a,b)$, $Z$ component: it is $\sim\varepsilon_{ab}$, so contracting with $\varepsilon^{ab}$ gives
    \begin{align}
-2 \Pi^{+}{}= -  \frac{9i M \mathit{h}^{-}{}^{a} r_{a}}{r^4} + \
\frac{3i M \mathit{h}^{-}{}^{a}{}_{:a}}{r^3} -  \Psi^{+}{}^{a}{}_{:a} \
-  \frac{\varepsilon_{ab} \Xi^{+}{}^{a} r^{b}}{r^3}\,,
    \end{align}

    \item $(\mu,\nu,\rho)=(A,a,b)$, $X$ component: it is $\sim\varepsilon_{ab}$, so contracting with $\varepsilon^{ab}$ gives
    \begin{align}\notag
-2 \Pi^{-}{}&=\frac{i M \mathit{h}^{+}{}^{a}{}_{a}}{2 r^3} -  \frac{2i M \
\mathit{h}^{+}{}}{r^5}  -  \frac{\Psi^{-}{}}{r^2} -  \
\Psi^{-}{}^{a}{}_{:a} + \frac{5 M \varepsilon_{ab} \
\mathit{h}^{-}{}^{a} r^{b}}{r^4} \\
&+ \frac{\varepsilon_{ab} \
\Xi^{-}{}^{a} r^{b}}{r^3} -  \frac{2 M \varepsilon_{ab} \
\mathit{h}^{-}{}^{a}{}^{:b}}{r^3}\,,
    \end{align}

    \item $(\mu,\nu,\rho)=(A,B,a)$, $U$ component: 
    \begin{align}\notag
 \Gamma^{+}{}_{a}&=  \frac{\Lambda^2 \Xi^{+}{}_{a}}{2 r^2} -  \frac{3 \
M \mathit{h}^{+}{} r_{a}}{r^4} + \frac{\Xi^{+}{} r_{a}}{r^3} + \
\frac{M \mathit{h}^{+}{}{}_{:a}}{2 r^3} -  \frac{M \
\mathit{h}^{+}{}_{a}{}^{b}{}_{:b}}{r}\\
&-  \Psi^{+}{}_{a}{}^{b}{}_{:b} \
+ \frac{2 M \mathit{h}^{+}{}_{ab} r^{b}}{r^2} + \
\frac{\Psi^{+}{}_{ab} r^{b}}{r}\,,
    \end{align}

     \item $(\mu,\nu,\rho)=(A,B,a)$, $V$ component: 
    \begin{align}\label{eq:FirstOrder_ABa_V}
 \Pi^{\oplus}{}_{a}= - \frac{3i M \varepsilon_{ab} \mathit{h}^{-}{}^{b}}{r^3} + \
\frac{\Xi^{+}{}_{a}}{r^2}  -  \
\Psi^{\oplus}{}_{a}{}^{b}{}_{:b} + \frac{\Psi^{\oplus}{}_{ab} \
r^{b}}{r}\,,
    \end{align}

    \item $(\mu,\nu,\rho)=(A,B,a)$, $W$ component: 
    \begin{align}
   \Gamma^{-}{}_{a}=\frac{M \mathit{h}^{-}{}_{a}}{r^3} -  \
\frac{\Xi^{-}{}_{a}}{r^2} -  \Psi^{-}{}_{a}{}^{b}{}_{:b} + \
\frac{\Psi^{-}{}_{ab} r^{b}}{r}\,,
    \end{align}

     \item $(\mu,\nu,\rho)=(A,B,a)$, $Y\epsilon$ component: 
    \begin{align}\notag
 \Pi^{-}{}_{a}&=\frac{\Lambda^2 M \mathit{h}^{-}{}_{a}}{r^3} + \frac{\Lambda^2 \
\Xi^{-}{}_{a}}{2 r^2}  -  \frac{i M \varepsilon_{a}{}^{b} \mathit{h}^{+}{}^{c}{}_{c}{}_{:b}}{2 r} + \frac{i M \
\varepsilon_{ab} \mathit{h}^{+}{} r^{b}}{r^4} + \frac{\varepsilon_{ab} \Psi^{-}{} r^{b}}{2 r} \\
&-  \frac{i M \varepsilon_{ab} \
\mathit{h}^{+}{}{}^{:b}}{2 r^3} + \frac{1}{2} \varepsilon_{ab} \
\Psi^{-}{}{}^{:b} + \frac{i M \varepsilon^{bc} \
\mathit{h}^{+}{}_{ab}{}_{:c}}{r} + \frac{i M \varepsilon_{a}{}^{b} \
\mathit{h}^{+}{}_{b}{}^{c}{}_{:c}}{r}\,,
    \end{align}

    \item $(\mu,\nu,\rho)=(A,B,C)$, $Z\otimes\epsilon$ component: 
    \begin{align}
  \Gamma^{-}{}=  \frac{4 M \mathit{h}^{-}{}^{a} r_{a}}{r^2} + \
\frac{\Xi^{-}{}^{a} r_{a}}{r} -  \frac{2 M \
\mathit{h}^{-}{}^{a}{}_{:a}}{r} -  \Xi^{-}{}^{a}{}_{:a}\,,
    \end{align}

    \item $(\mu,\nu,\rho)=(A,B,C)$, $X\otimes\epsilon$ component: 
    \begin{align}\notag
\Gamma^{+}{}&=- \frac{3 M \mathit{h}^{+}{}^{a}{}_{a}}{2 r} + \frac{4 M \
\mathit{h}^{+}{}}{r^3} -  \frac{\Xi^{+}{}}{r^2} + \
\frac{\Xi^{+}{}^{a} r_{a}}{r} -  \Xi^{+}{}^{a}{}_{:a} \\ 
&+ \frac{6i M \
\varepsilon_{ab} \mathit{h}^{-}{}^{a} r^{b}}{r^2} -  \frac{3i M \
\varepsilon_{ab} \mathit{h}^{-}{}^{a}{}^{:b}}{r}\,.
    \end{align}
\end{itemize}

\subsection{Second-derivative Equations of $\delta \C$}

Here we provide the linearisation of the curvature wave equation for $\delta \C$, in \eqref{eq:triC2}, with all free indices down. It is convenient to present them in the form
\begin{equation}
    \square \Psi=... \qquad \text{with $\Psi$ the components of $\delta \C$}\, ,
\end{equation}
so that they can be seen as wave equations for $\Psi$ with a well-behaved principal part. 
\begin{itemize}
    \item $(\mu,\nu,\rho,\sigma)=(a,b,c,d)$, $Y$ component: it is $\sim \varepsilon_{ab}\varepsilon_{cd}$, so contracting with $\varepsilon^{ab}\varepsilon^{cd}$ gives
    \begin{align}\notag
        \square \Psi^{+}{}{}&=\frac{6 M^2 \mathit{h}^{+}{}^{a}{}_{a}}{r^6} + \frac{2 \Lambda^2 M \
\mathit{h}^{+}{}^{a}{}_{a}}{r^5} -  \frac{104 M^2 \
\mathit{h}^{+}{}}{r^8} + \frac{40 M \mathit{h}^{+}{}}{r^7} + \frac{2i \
\Lambda^2 \Pi^{-}{}}{r^2} \\ \notag
&+ \frac{M \Phi^{+}{}^{a}{}_{a}}{r^3} + \
\frac{2 M \Phi^{+}{}}{r^5} -  \frac{14 M \Psi^{+}{}^{a}{}_{a}}{r^5} + \
\frac{4 \Psi^{+}{}^{a}{}_{a}}{r^4} + \frac{6i M \Psi^{-}{}}{r^5}-  \
\frac{8 M \Psi^{+}{}}{r^3} \\ \notag
& + \frac{4 \Psi^{+}{}}{r^2}+ \
\frac{\Lambda^2 \Psi^{+}{}}{r^2}  + \
\frac{5 M \mathit{h}^{+}{}^{b}{}_{b}{}_{:a} r^{a}}{r^4} -  \
\frac{18 M \mathit{h}^{+}{}{}_{:a} r^{a}}{r^6} -  \frac{2 \
\Psi^{+}{}{}_{:a} r^{a}}{r} \\ \notag
&+ \frac{6 M r^{a} \
\mathit{h}^{+}{}_{a}{}^{b}{}_{:b}}{r^4}-  \frac{2 M \
\mathit{h}^{+}{}^{a}{}_{a}{}^{:b}{}_{:b}}{r^3} + \frac{12i \Lambda^2 \
M \varepsilon_{ab} \mathit{h}^{-}{}^{a} r^{b}}{r^6} -  \frac{2i \
\varepsilon_{ab} \Pi^{-}{}^{a} r^{b}}{r^3}  \\ \label{eq:CWE_Psi_p}
&-  \frac{4 \Lambda^2 \
\varepsilon_{ab} \Psi^{+}{}^{a} r^{b}}{r^3} -  \frac{4 M \
\mathit{h}^{+}{}_{ab} r^{a} r^{b}}{r^5}-  \frac{4 \
\Psi^{+}{}_{ab} r^{a} r^{b}}{r^4} + \varepsilon_{ab} \
\Pi^{+}{}^{a}{}^{:b}\,,
    \end{align}

    \item $(\mu,\nu,\rho,\sigma)=(A,a,b,c)$, $Z$ component: it is $\sim \varepsilon_{bc}$, so contracting with $\varepsilon^{bc}$ gives
    \begin{align}\notag
\square \Psi^{+}{}_{a}&= - \frac{12i M^2 \mathit{h}^{-}{}_{a}}{r^6} -  \frac{3i \Lambda^2 M \
\mathit{h}^{-}{}_{a}}{2 r^5} + \frac{3i M \Xi^{-}{}_{a}}{2 r^5} + \
\frac{3 M \varepsilon_{ab} \Xi^{+}{}^{b}}{2 r^5} + \Pi^{+}{}_{a}\\ \notag
& + \
\frac{3i M \varepsilon_{ab} \Psi^{-}{}^{b}}{2 r^3} -  \frac{19 M \
\Psi^{+}{}_{a}}{2 r^3} + \frac{4 \Psi^{+}{}_{a}}{r^2} + \
\frac{\Lambda^2 \Psi^{+}{}_{a}}{r^2} -  \frac{i \Pi^{-}{}^{b}{}_{b} \
r_{a}}{r} \\ \notag
& + \frac{\Pi^{+}{} r_{a}}{r} + 2 \Pi^{+}{}{}_{:a} + \
\frac{27i M \mathit{h}^{-}{}^{b} r_{a} r_{b}}{r^5} -  \frac{2 \
\Psi^{+}{}^{b} r_{a} r_{b}}{r^2} -  \frac{3i M r_{a} \
\mathit{h}^{-}{}^{b}{}_{:b}}{r^4}\\ \notag
& + \frac{3i M \
\mathit{h}^{-}{}^{b}{}_{:a}{}_{:b}}{2 r^3} + \frac{3i M \
\mathit{h}^{-}{}_{a}{}^{:b}{}_{:b}}{2 r^3} -  \frac{6 M \varepsilon_{b}{}^{c} \
\mathit{h}^{+}{}_{ac} r^{b}}{r^4} -  \frac{2 M \varepsilon_{a}{}^{c} \mathit{h}^{+}{}_{bc} r^{b}}{r^4}\\ \notag
& -  \frac{3 M \
\varepsilon_{ab} \mathit{h}^{+}{}^{c}{}_{c} r^{b}}{2 r^4} + \frac{5 \
M \varepsilon_{ab} \mathit{h}^{+}{} r^{b}}{r^6} + \frac{i \
\Pi^{-}{}_{ab} r^{b}}{r} -  \frac{\varepsilon_{b}{}^{c} \
\Pi^{+}{}_{ac} r^{b}}{r} \\ \notag
&-  \frac{i \varepsilon_{ab} \Pi^{-}{} \
r^{b}}{r} -  \frac{2 \varepsilon_{b}{}^{c} \Psi^{\oplus}{}_{ac} \
r^{b}}{r^3} + \frac{\Lambda^2 \varepsilon_{b}{}^{c} \
\Psi^{\oplus}{}_{ac} r^{b}}{r^3} -  \frac{2 \varepsilon_{b}{}^{c} \
\Psi^{+}{}_{ac} r^{b}}{r^3}  \\ \notag
&+ \frac{2 \varepsilon_{ab} \Psi^{+}{} \
r^{b}}{r}-  \frac{9i M \mathit{h}^{-}{}_{b}{}_{:a} r^{b}}{r^4} \
-  \frac{6i M \mathit{h}^{-}{}_{a}{}_{:b} r^{b}}{r^4} -  \frac{3 M \
\varepsilon_{ab} \mathit{h}^{+}{}{}^{:b}}{2 r^5}  \\ 
&- 2i \varepsilon_{ab} \
\Pi^{-}{}{}^{:b}+ \frac{3 M \varepsilon_{a}{}^{b} \
\mathit{h}^{+}{}_{b}{}^{c}{}_{:c}}{2 r^3} + \frac{2 \varepsilon_{bc} \
\Xi^{+}{}^{b} r_{a} r^{c}}{r^4}\,,
    \end{align}

    \item $(\mu,\nu,\rho,\sigma)=(A,a,b,c)$, $X$ component: it is $\sim \varepsilon_{bc}$, so contracting with $\varepsilon^{bc}$ gives
    \begin{align}\notag
\square \Psi^{-}{}_{a}&=- \frac{24 M^2 \varepsilon_{ab} \mathit{h}^{-}{}^{b}}{r^6} + \frac{12 \
M \varepsilon_{ab} \mathit{h}^{-}{}^{b}}{r^5} -  \frac{\Lambda^2 M \
\varepsilon_{ab} \mathit{h}^{-}{}^{b}}{2 r^5} -  \frac{3 M \varepsilon_{ab} \Xi^{-}{}^{b}}{2 r^5} \\ \notag
&+ \frac{3i M \Xi^{+}{}_{a}}{2 r^5} + i \
\varepsilon_{ab} \Pi^{+}{}^{b} -  \frac{19 M \Psi^{-}{}_{a}}{2 r^3} + \
\frac{4 \Psi^{-}{}_{a}}{r^2} + \frac{\Lambda^2 \Psi^{-}{}_{a}}{r^2} \\ \notag
&- \
 \frac{3i M \varepsilon_{ab} \Psi^{+}{}^{b}}{2 r^3} + \frac{3i M \
\mathit{h}^{+}{}^{b}{}_{b} r_{a}}{2 r^4} + \frac{9i M \
\mathit{h}^{+}{} r_{a}}{r^6} + \frac{i \Pi^{+}{}^{b}{}_{b} \
r_{a}}{r} + \frac{\Pi^{-}{} r_{a}}{r} \\ \notag
&+ \frac{3 \Psi^{-}{} \
r_{a}}{r^3} -  \frac{3i M \mathit{h}^{+}{}{}_{:a}}{2 r^5} + 2 \
\Pi^{-}{}{}_{:a} -  \frac{2 \Psi^{-}{}^{b} r_{a} r_{b}}{r^2} + \
\frac{3i M \mathit{h}^{+}{}_{a}{}^{b}{}_{:b}}{2 r^3} \\ \notag
&-  \frac{3 M \varepsilon_{ac} \
\mathit{h}^{-}{}^{b}{}^{:c}{}_{:b}}{2 r^3} -  \frac{6i M \
\mathit{h}^{+}{}_{ab} r^{b}}{r^4} -  \frac{\varepsilon_{b}{}^{c} \
\Pi^{-}{}_{ac} r^{b}}{r} -  \frac{i \Pi^{+}{}_{ab} r^{b}}{r} + \
\frac{i \varepsilon_{ab} \Pi^{+}{} r^{b}}{r} \\ \notag
&-  \frac{2 \varepsilon_{b}{}^{c} \Psi^{-}{}_{ac} r^{b}}{r^3} + \frac{\Lambda^2 \varepsilon_{b}{}^{c} \Psi^{-}{}_{ac} r^{b}}{r^3} -  \frac{6 M \varepsilon_{ac} \mathit{h}^{-}{}^{c}{}_{:b} r^{b}}{r^4} + 2i \varepsilon_{ab} \Pi^{+}{}{}^{:b}  \\ \notag
&+ \frac{3 M \varepsilon_{ab} r^{b} \
\mathit{h}^{-}{}^{c}{}_{:c}}{r^4}+ \frac{M \varepsilon_{ab} \
\mathit{h}^{-}{}^{b}{}^{:c}{}_{:c}}{2 r^3} -  \frac{6 M \varepsilon_{bc} \mathit{h}^{-}{}^{b} r_{a} r^{c}}{r^5} -  \frac{2 \
\varepsilon_{bc} \Xi^{-}{}^{b} r_{a} r^{c}}{r^4} \\
& -  \frac{17 M \
\varepsilon_{ac} \mathit{h}^{-}{}^{b} r_{b} r^{c}}{r^5} + \
\frac{9 M \varepsilon_{ac} r^{b} \mathit{h}^{-}{}_{b}{}^{:c}}{r^4}\,,
    \end{align}

    \item $(\mu,\nu,\rho,\sigma)=(a,b,A,B)$, $Y\epsilon$ component: it is $\sim \varepsilon_{ab}$, so contracting with $\varepsilon^{ab}$ gives
    \begin{align}\notag
  \square \Psi^{-}&=\frac{i \Lambda^2 M \mathit{h}^{+}{}^{a}{}_{a}}{r^3} -  \frac{20i M^2 \
\mathit{h}^{+}{}}{r^6} + \frac{8i M \mathit{h}^{+}{}}{r^5} + \frac{2i \
\Lambda^2 M \mathit{h}^{+}{}}{r^5} -  \frac{i \Lambda^2 \
\Gamma^{+}{}}{r^2} + \frac{i M \Phi^{+}{}^{a}{}_{a}}{r} \\ \notag
&+ \frac{2i M \
\Phi^{+}{}}{r^3} -  \frac{6i M \Psi^{+}{}^{a}{}_{a}}{r^3} -  \frac{10 \
M \Psi^{-}{}}{r^3} + \frac{4 \Psi^{-}{}}{r^2} + \frac{\Lambda^2 \
\Psi^{-}{}}{r^2} + \frac{2i \Gamma^{+}{}^{a} r_{a}}{r} \\ \notag
&+ \frac{2 \
\Lambda^2 \Psi^{-}{}^{a} r_{a}}{r} -  \frac{2i M \
\mathit{h}^{+}{}{}^{:a}{}_{:a}}{r^3}  + \
\frac{i M \mathit{h}^{+}{}^{b}{}_{b}{}_{:a} r^{a}}{r^2} -  \
\frac{2i M \mathit{h}^{+}{}{}_{:a} r^{a}}{r^4} + \frac{2 \
\Psi^{-}{}{}_{:a} r^{a}}{r} \\ \notag
&+ \frac{6i M r^{a} \
\mathit{h}^{+}{}_{a}{}^{b}{}_{:b}}{r^2} -  \frac{i M \
\mathit{h}^{+}{}^{a}{}_{a}{}^{:b}{}_{:b}}{r} -  \frac{6 \Lambda^2 M \
\varepsilon_{ab} \mathit{h}^{-}{}^{a} r^{b}}{r^4} -  \frac{2 \
\Lambda^2 \varepsilon_{ab} \Xi^{-}{}^{a} r^{b}}{r^3} \\ \label{eq:CWE_Psi_m}
&+ \frac{4 \
\varepsilon_{ab} \Pi^{-}{}^{a} r^{b}}{r} - 2 \varepsilon_{ab} \
\Pi^{-}{}^{a}{}^{:b}\, ,
    \end{align}

    \item $(\mu,\nu,\rho,\sigma)=(A,a,B,b)$, $U$ component: 
    \begin{align}\notag
    \square \Psi^{+}{}_{ab}{}&=\frac{6 M^2 \mathit{h}^{+}{}_{ab}}{r^4} + \frac{\Lambda^2 M \
\mathit{h}^{+}{}_{ab}}{r^3} -  \frac{3 M^2 g_{ab} \
\mathit{h}^{+}{}^{c}{}_{c}}{r^4} -  \frac{2 M^2 g_{ab} \
\mathit{h}^{+}{}}{r^6} + \frac{\Lambda^2 M g_{ab} \
\mathit{h}^{+}{}}{r^5} \\ \notag
&+ \frac{1}{2}i \Lambda^2 \varepsilon_{a}{}^{c} \
\Pi^{-}{}_{bc} + \frac{1}{2} \Lambda^2 \Pi^{+}{}_{ab} + \frac{1}{2}i \Lambda^2 g_{ab} \Pi^{-}{} + \frac{1}{2} \Lambda^2 \
\varepsilon_{ab} \Pi^{+}{} + \frac{M g_{ab} \Phi^{+}{}^{c}{}_{c}}{2 r}\\ \notag
&+ \frac{M g_{ab} \Phi^{+}{}}{r^3} + \frac{4 M \Psi^{+}{}_{ab}}{r^3} + \
\frac{\Lambda^2 \Psi^{+}{}_{ab}}{r^2} -  \frac{3 M g_{ab} \
\Psi^{+}{}^{c}{}_{c}}{r^3} + \frac{3i M g_{ab} \Psi^{-}{}}{r^3} + 2 \
g_{ab} \Psi^{+}{} \\ \notag
&-  \frac{4 M g_{ab} \Psi^{+}{}}{r} + \
\frac{\Gamma^{+}{}_{b} r_{a}}{r} -  \frac{\Lambda^2 \Xi^{+}{}_{b} \
r_{a}}{r^3} -  \Gamma^{+}{}_{b}{}_{:a} -  \frac{\Lambda^2 \
\Xi^{+}{}_{a} r_{b}}{r^3} + \frac{2i \varepsilon_{ac} \Pi^{-}{}^{c} \
r_{b}}{r} \\ \notag
&+ \frac{8 M \mathit{h}^{+}{} r_{a} r_{b}}{r^5} -  \
\frac{2 \Xi^{+}{} r_{a} r_{b}}{r^4} - 2 \Psi^{+}{} r_{a} \
r_{b} -  \frac{M \mathit{h}^{+}{}_{ab}{}^{:c}{}_{:c}}{r} -  \
\frac{M g_{ab} \mathit{h}^{+}{}{}^{:c}{}_{:c}}{r^3} \\ \notag
&+ \frac{3i \Lambda^2 M \varepsilon_{bc} \
\mathit{h}^{-}{}_{a} r^{c}}{r^4} + \frac{3i \Lambda^2 M \varepsilon_{ac} \mathit{h}^{-}{}_{b} r^{c}}{r^4} -  \frac{i \varepsilon_{ac} \
\Pi^{-}{}_{b} r^{c}}{r} -  \varepsilon_{bc} r \Pi^{+}{}_{a} \
r^{c} \\ \notag
&-  \frac{\Lambda^2 \varepsilon_{bc} \Psi^{+}{}_{a} \
r^{c}}{r} -  \frac{\Lambda^2 \varepsilon_{ac} \Psi^{+}{}_{b} \
r^{c}}{r} -  \frac{4 M \mathit{h}^{+}{}_{bc} r_{a} \
r^{c}}{r^3} -  \frac{4 M \mathit{h}^{+}{}_{ac} r_{b} \
r^{c}}{r^3}\\ \notag
& + \frac{4 M \mathit{h}^{+}{}_{ab}{}_{:c} \
r^{c}}{r^2} -  \frac{3 M g_{ab} \mathit{h}^{+}{}^{d}{}_{d}{}_{:c} \
r^{c}}{2 r^2} -  \frac{M g_{ab} \mathit{h}^{+}{}{}_{:c} \
r^{c}}{r^4} + \frac{2 \Psi^{+}{}_{ab}{}_{:c} r^{c}}{r} \\ 
&+ i \
\varepsilon_{ac} \Pi^{-}{}_{b}{}^{:c} + \frac{3 M g_{ab} r^{c} \
\mathit{h}^{+}{}_{c}{}^{d}{}_{:d}}{r^2} + \frac{4 M g_{ab} \
\mathit{h}^{+}{}_{cd} r^{c} r^{d}}{r^3}\, ,
\end{align}

    \item $(\mu,\nu,\rho,\sigma)=(A,a,B,b)$, $V$ component: 
    \begin{align}\notag
   \square \Psi^{\oplus}{}_{ab}{}&=i \varepsilon_{a}{}^{c} \Pi^{-}{}_{bc} -  \Pi^{+}{}_{ab} + i g_{ab} \
\Pi^{-}{} -  \varepsilon_{ab} \Pi^{+}{} -  \frac{3i M \varepsilon_{b}{}^{c} \Psi^{-}{}_{ac}}{r^3} -  \frac{3i M \varepsilon_{a}{}^{c} \
\Psi^{-}{}_{bc}}{r^3}\\ \notag
& + \frac{10 M \Psi^{\oplus}{}_{ab}}{r^3} -  \
\frac{4 \Psi^{\oplus}{}_{ab}}{r^2} + \frac{\Lambda^2 \
\Psi^{\oplus}{}_{ab}}{r^2} -  \frac{3 M g_{ab} \
\Psi^{\oplus}{}^{c}{}_{c}}{r^3} + \frac{6i M \varepsilon_{bc} \
\mathit{h}^{-}{}^{c} r_{a}}{r^4} \\ \notag
&-  \frac{2 \Xi^{+}{}_{b} \
r_{a}}{r^3} + \frac{\Pi^{\oplus}{}_{b} r_{a}}{r} -  \
\Pi^{\oplus}{}_{b}{}_{:a} + \frac{6i M \varepsilon_{ac} \
\mathit{h}^{-}{}^{c} r_{b}}{r^4} -  \frac{2 \Xi^{+}{}_{a} \
r_{b}}{r^3}  -  \frac{i \
\varepsilon_{ac} \Gamma^{-}{}_{b} r^{c}}{r} \\ \label{eq:CWE_Psi_o}
&+ \frac{2 \varepsilon_{bc} \Psi^{+}{}_{a} r^{c}}{r} + \frac{2 \varepsilon_{ac} \
\Psi^{+}{}_{b} r^{c}}{r} + \frac{2 \Psi^{\oplus}{}_{ab}{}_{:c} \
r^{c}}{r} + i \varepsilon_{ac} \Gamma^{-}{}_{b}{}^{:c}\,,
    \end{align}

    \item $(\mu,\nu,\rho,\sigma)=(A,a,B,b)$, $W$ component: 
    \begin{align}\notag
    \square \Psi^{-}{}_{ab}{}&= - \Pi^{-}{}_{ab} - i \varepsilon_{a}{}^{c} \Pi^{+}{}_{bc} -  \varepsilon_{ab} \Pi^{-}{} - i g_{ab} \Pi^{+}{} + \frac{10 M \
\Psi^{-}{}_{ab}}{r^3} -  \frac{4 \Psi^{-}{}_{ab}}{r^2} + \
\frac{\Lambda^2 \Psi^{-}{}_{ab}}{r^2}\\ \notag
& -  \frac{3 M g_{ab} \
\Psi^{-}{}^{c}{}_{c}}{r^3} + \frac{3i M \varepsilon_{b}{}^{c} \
\Psi^{\oplus}{}_{ac}}{r^3} + \frac{3i M \varepsilon_{a}{}^{c} \
\Psi^{\oplus}{}_{bc}}{r^3} -  \frac{6 M \mathit{h}^{-}{}_{b} \
r_{a}}{r^4} + \frac{\Gamma^{-}{}_{b} r_{a}}{r} \\ \notag
&+ \frac{2 \
\Xi^{-}{}_{b} r_{a}}{r^3} -  \Gamma^{-}{}_{b}{}_{:a} -  \frac{6 M \
\mathit{h}^{-}{}_{a} r_{b}}{r^4} + \frac{2 \Xi^{-}{}_{a} \
r_{b}}{r^3} + \frac{8 M g_{ab} \mathit{h}^{-}{}^{c} r_{c}}{r^4} + \frac{i \varepsilon_{ac} \
\Pi^{\oplus}{}_{b} r^{c}}{r} \\
&+ \frac{2 \varepsilon_{bc} \
\Psi^{-}{}_{a} r^{c}}{r} + \frac{2 \varepsilon_{ac} \Psi^{-}{}_{b} \
r^{c}}{r} + \frac{2 \Psi^{-}{}_{ab}{}_{:c} r^{c}}{r} - i \
\varepsilon_{ac} \Pi^{\oplus}{}_{b}{}^{:c}\,,
    \end{align}

    \item $(\mu,\nu,\rho,\sigma)=(A,a,B,b)$, $Y\epsilon$ component: contracting with $\varepsilon^{ab}$,
    \begin{align}\notag
    \square\Psi^{-}{}{}&=\frac{i \Lambda^2 M \mathit{h}^{+}{}^{a}{}_{a}}{r^3} -  \frac{20i M^2 \
\mathit{h}^{+}{}}{r^6} + \frac{8i M \mathit{h}^{+}{}}{r^5} + \frac{2i \
\Lambda^2 M \mathit{h}^{+}{}}{r^5} + \frac{1}{2}i \Lambda^2 \
\Pi^{+}{}^{a}{}_{a} -  \Lambda^2 \Pi^{-}{}\\ \notag
& + \frac{i M \
\Phi^{+}{}^{a}{}_{a}}{r} + \frac{2i M \Phi^{+}{}}{r^3} -  \frac{6i M \
\Psi^{+}{}^{a}{}_{a}}{r^3} -  \frac{10 M \Psi^{-}{}}{r^3} + \frac{4 \
\Psi^{-}{}}{r^2} + \frac{\Lambda^2 \Psi^{-}{}}{r^2} + \frac{i \
\Gamma^{+}{}^{a} r_{a}}{r} \\ \notag
&+ \frac{2 \Lambda^2 \Psi^{-}{}^{a} \
r_{a}}{r} - i \Gamma^{+}{}^{a}{}_{:a} -  \frac{2i M \
\mathit{h}^{+}{}{}^{:a}{}_{:a}}{r^3}  + \
\frac{i M \mathit{h}^{+}{}^{b}{}_{b}{}_{:a} r^{a}}{r^2} -  \
\frac{2i M \mathit{h}^{+}{}{}_{:a} r^{a}}{r^4} \\ \notag
&+ \frac{2 \
\Psi^{-}{}{}_{:a} r^{a}}{r} + \frac{6i M r^{a} \
\mathit{h}^{+}{}_{a}{}^{b}{}_{:b}}{r^2} -  \frac{i M \
\mathit{h}^{+}{}^{a}{}_{a}{}^{:b}{}_{:b}}{r} -  \frac{6 \Lambda^2 M \
\varepsilon_{ab} \mathit{h}^{-}{}^{a} r^{b}}{r^4}  \\ 
&-  \frac{2 \
\Lambda^2 \varepsilon_{ab} \Xi^{-}{}^{a} r^{b}}{r^3}+ \frac{3 \
\varepsilon_{ab} \Pi^{-}{}^{a} r^{b}}{r} - i \varepsilon_{ab} r \
\Pi^{+}{}^{a} r^{b} -  \varepsilon_{ab} \Pi^{-}{}^{a}{}^{:b}\,,
    \end{align}

\item $(\mu,\nu,\rho,\sigma)=(A,B,C,a)$, $\epsilon\otimes Z$ component: 
    \begin{align}\notag
    \square \Xi^{-}{}_{a}{}&=- \frac{6 M^2 \mathit{h}^{-}{}_{a}}{r^4} + \frac{6 M \
\mathit{h}^{-}{}_{a}}{r^3} + \frac{\Lambda^2 M \
\mathit{h}^{-}{}_{a}}{2 r^3} -  \Gamma^{-}{}_{a} + \frac{1}{2} \
\Lambda^2 \Gamma^{-}{}_{a} + \frac{21 M \Xi^{-}{}_{a}}{2 r^3} -  \
\frac{4 \Xi^{-}{}_{a}}{r^2} \\ \notag
&+ \frac{\Lambda^2 \Xi^{-}{}_{a}}{r^2} + \
\frac{3i M \varepsilon_{ab} \Xi^{+}{}^{b}}{2 r^3} -  \Pi^{-}{}_{a} -  \
\frac{3 M \varepsilon_{ab} \Psi^{-}{}^{b}}{2 r} -  \frac{3i M \
\Psi^{+}{}_{a}}{2 r} + \frac{\Gamma^{-}{} r_{a}}{r}  \\ \notag
&- i r \
\Pi^{+}{} r_{a} + i r^2 \Pi^{+}{}{}_{:a}-  \frac{19 M \
\mathit{h}^{-}{}^{b} r_{a} r_{b}}{r^3} + \frac{2 \Xi^{-}{}^{b} \
r_{a} r_{b}}{r^2} + \frac{3 M r_{a} \
\mathit{h}^{-}{}^{b}{}_{:b}}{r^2}   \\ \notag
&-  \frac{3 M \
\mathit{h}^{-}{}^{b}{}_{:a}{}_{:b}}{2 r} -  \frac{M \
\mathit{h}^{-}{}_{a}{}^{:b}{}_{:b}}{2 r}-  \frac{3i M \varepsilon_{b}{}^{c} \
\mathit{h}^{+}{}_{ac} r^{b}}{r^2} -  \frac{3i M \varepsilon_{a}{}^{c} \mathit{h}^{+}{}_{bc} r^{b}}{r^2} \\ \notag
&-  \frac{3i M \
\varepsilon_{ab} \mathit{h}^{+}{}^{c}{}_{c} r^{b}}{2 r^2} + \
\frac{9i M \varepsilon_{ab} \mathit{h}^{+}{} r^{b}}{r^4} - i \
\varepsilon_{b}{}^{c} r \Pi^{+}{}_{ac} r^{b} - 2 \varepsilon_{ab} r \
\Pi^{-}{} r^{b} \\ \notag
&-  \frac{2 \Psi^{-}{}_{ab} r^{b}}{r} + \
\frac{\Lambda^2 \Psi^{-}{}_{ab} r^{b}}{r} + \frac{3 \varepsilon_{ab} \Psi^{-}{} r^{b}}{r} + \frac{9 M \
\mathit{h}^{-}{}_{b}{}_{:a} r^{b}}{r^2} + \frac{4 \
\Xi^{-}{}_{a}{}_{:b} r^{b}}{r} \\ 
&-  \frac{3i M \varepsilon_{ab} \
\mathit{h}^{+}{}{}^{:b}}{2 r^3} + \frac{3i M \varepsilon_{a}{}^{b} \
\mathit{h}^{+}{}_{b}{}^{c}{}_{:c}}{2 r} - i \varepsilon^{bc} r^2 \
\Pi^{+}{}_{ab}{}_{:c} + 2 \varepsilon_{ac} \Psi^{-}{}^{b} r_{b} \
r^{c}\, ,
\end{align}

    \item $(\mu,\nu,\rho,\sigma)=(A,B,C,a)$, $\epsilon\otimes X$ component: 
    \begin{align}\notag
\square \Xi^{+}{}_{a}{}&= - \frac{42i M^2 \varepsilon_{ab} \mathit{h}^{-}{}^{b}}{r^4} + \
\frac{18i M \varepsilon_{ab} \mathit{h}^{-}{}^{b}}{r^3} -  \frac{3i \
\Lambda^2 M \varepsilon_{ab} \mathit{h}^{-}{}^{b}}{2 r^3} -  \
\Gamma^{+}{}_{a} -  \frac{3i M \varepsilon_{ab} \Xi^{-}{}^{b}}{2 r^3} \\ \notag
&+ \frac{21 M \Xi^{+}{}_{a}}{2 r^3} -  \frac{4 \Xi^{+}{}_{a}}{r^2} + \
\frac{\Lambda^2 \Xi^{+}{}_{a}}{r^2} + \Pi^{\oplus}{}_{a} -  \frac{1}{2} \Lambda^2 \Pi^{\oplus}{}_{a} -  \frac{3i M \
\Psi^{-}{}_{a}}{2 r}  \\ \notag
&+ \frac{3 M \varepsilon_{ab} \Psi^{+}{}^{b}}{2 r} \
+ \frac{3 M \mathit{h}^{+}{}^{b}{}_{b} r_{a}}{2 r^2} -  \frac{13 M \
\mathit{h}^{+}{} r_{a}}{r^4} + \frac{\Gamma^{+}{} r_{a}}{r} + \
\frac{2 \Xi^{+}{} r_{a}}{r^3} - i r \Pi^{-}{} r_{a}  \\ \notag
&+ \frac{3 M \
\mathit{h}^{+}{}{}_{:a}}{2 r^3} + i r^2 \Pi^{-}{}{}_{:a} + \frac{2 \
\Xi^{+}{}^{b} r_{a} r_{b}}{r^2} -  \frac{3 M \
\mathit{h}^{+}{}_{a}{}^{b}{}_{:b}}{2 r} -  \frac{3i M \varepsilon_{ac} \
\mathit{h}^{-}{}^{b}{}^{:c}{}_{:b}}{2 r}  \\ \notag
&+ \frac{4 M \
\mathit{h}^{+}{}_{ab} r^{b}}{r^2} - i \varepsilon_{b}{}^{c} r \
\Pi^{-}{}_{ac} r^{b} + 2 \varepsilon_{ab} r \Pi^{+}{} r^{b} + \
\frac{2 \Psi^{\oplus}{}_{ab} r^{b}}{r} -  \frac{\Lambda^2 \
\Psi^{\oplus}{}_{ab} r^{b}}{r}  \\ \notag
&-  \frac{2 \Psi^{+}{}_{ab} \
r^{b}}{r} -  \frac{3i M \varepsilon_{bc} \
\mathit{h}^{-}{}^{c}{}_{:a} r^{b}}{r^2} -  \frac{9i M \varepsilon_{ac} \mathit{h}^{-}{}^{c}{}_{:b} r^{b}}{r^2} + \frac{4 \
\Xi^{+}{}_{a}{}_{:b} r^{b}}{r}  \\ \notag
&+ \frac{3i M \varepsilon_{ab} \
r^{b} \mathit{h}^{-}{}^{c}{}_{:c}}{r^2} - i \varepsilon^{bc} r^2 \
\Pi^{-}{}_{ab}{}_{:c} + \frac{3i M \varepsilon_{ab} \
\mathit{h}^{-}{}^{b}{}^{:c}{}_{:c}}{2 r} -  \frac{6i M \varepsilon_{bc} \mathit{h}^{-}{}^{b} r_{a} r^{c}}{r^3}  \\
&-  \frac{9i M \
\varepsilon_{ac} \mathit{h}^{-}{}^{b} r_{b} r^{c}}{r^3} - 2 \
\varepsilon_{ac} \Psi^{+}{}^{b} r_{b} r^{c} + \frac{3i M \
\varepsilon_{bc} r^{b} \mathit{h}^{-}{}_{a}{}^{:c}}{r^2} + \frac{6i \
M \varepsilon_{ac} r^{b} \mathit{h}^{-}{}_{b}{}^{:c}}{r^2}\, ,
\end{align}

\item $(\mu,\nu,\rho,\sigma)=(A,B,C,D)$, $Y \epsilon\otimes \epsilon$ component: 
    \begin{align}\notag
    \square \Xi^{+} &=\frac{6 M^2 \mathit{h}^{+}{}^{a}{}_{a}}{r^2} -  \frac{64 M^2 \
\mathit{h}^{+}{}}{r^4} + \frac{24 M \mathit{h}^{+}{}}{r^3} -  \frac{4 \
\Lambda^2 M \mathit{h}^{+}{}}{r^3} + \Lambda^2 \Gamma^{+}{} + \
\frac{24 M \Xi^{+}{}}{r^3} -  \frac{8 \Xi^{+}{}}{r^2} \\ \notag
&+ \
\frac{\Lambda^2 \Xi^{+}{}}{r^2} -  M r \Phi^{+}{}^{a}{}_{a} -  \
\frac{2 M \Phi^{+}{}}{r} + \frac{6 M \Psi^{+}{}^{a}{}_{a}}{r} -  \
\frac{6i M \Psi^{-}{}}{r} - 2 r \Gamma^{+}{}^{a} r_{a} + \frac{4 \
\Lambda^2 \Xi^{+}{}^{a} r_{a}}{r} \\ \notag
&+ \frac{4 M \
\mathit{h}^{+}{}{}^{:a}{}_{:a}}{r} + 3 M \
\mathit{h}^{+}{}^{b}{}_{b}{}_{:a} r^{a} -  \frac{14 M \
\mathit{h}^{+}{}{}_{:a} r^{a}}{r^2} + \frac{6 \Xi^{+}{}{}_{:a} \
r^{a}}{r} - 6 M r^{a} \mathit{h}^{+}{}_{a}{}^{b}{}_{:b}\\ \label{eq:CWE_Xi_p}
& + 4i \
\varepsilon_{ab} r \Pi^{-}{}^{a} r^{b} -  \frac{4 M \
\mathit{h}^{+}{}_{ab} r^{a} r^{b}}{r} - 4 \Psi^{+}{}_{ab} \
r^{a} r^{b} - 2i \varepsilon_{ab} r^2 \Pi^{-}{}^{a}{}^{:b}\, .
\end{align}
\end{itemize}

\subsection{$\delta \C$ in terms of $h$}\label{app:sdC_To_h}

Here we provide the components of $\delta \C$ in terms of $h$. To simplify the expressions, we automatically use \eqref{eq:Einstein_Rep} as well as \eqref{eq:DDh} to replace some derivatives of the metric perturbation. In the expression for $\Psi^{+}{}_{ab}$, we also used the linearised Gauss-Bonnet identity, which on the Schwarzschild background reads \footnote{We recall this identity follows from the fact that the Einstein tensor vanishes identically in two dimensions. Hence on any background 2$D$ geometry one has $\delta G_{ab}(h)=0$.}
\begin{equation}\label{eq:GB_1st}
\begin{aligned}
    \square h^{+}{}_{ab}&=\frac{4 M \mathit{h}^{+}{}_{ab}}{r^3} + \
\mathit{h}^{+}{}_{b}{}^{c}{}_{:c}{}_{:a} -  \
\mathit{h}^{+}{}^{c}{}_{c}{}_{:a}{}_{:b} + \
\mathit{h}^{+}{}_{a}{}^{c}{}_{:c}{}_{:b} \\
&+ g_{ab} (- \frac{2 M \
\mathit{h}^{+}{}^{c}{}_{c}}{r^3} -  \
\mathit{h}^{+}{}^{cd}{}_{:c}{}_{:d} + \
\mathit{h}^{+}{}^{c}{}_{c}{}^{:d}{}_{:d})\, ,
\end{aligned}
\end{equation}
which allows one to eliminate all second derivatives of $h^{+}_{ab}$ from the expression.
\begin{align}\notag
    \Psi^{+}&=\left(\frac{4 M}{r^5} -  r^{-4} -  \frac{\Lambda^2}{2 r^4}\right) \
\mathit{h}^{+}{} -  \frac{1}{3} \Phi^{+}{}^{a}{}_{a} + \left(\frac{4 \
M}{r^3} -  \frac{1}{3 r^2} + \frac{\Lambda^2}{6 r^2}\right) \Phi^{\oplus}{} \
-  \frac{2 \Phi^{+}{}^{a} r_{a}}{3 r} \\
&-  \frac{1}{3} \
\Phi^{+}{}^{a}{}_{:a} + \frac{\mathit{h}^{+}{}{}_{:a} r^{a}}{r^3} \
-  \frac{i \Lambda^2 \varepsilon_{ab} \mathit{h}^{-}{}^{a} \
r^{b}}{r^3} -  \frac{\mathit{h}^{+}{}_{ab} r^{a} r^{b}}{r^2} \
+ \frac{i \Lambda^2 \varepsilon_{ab} \mathit{h}^{-}{}^{a}{}^{:b}}{2 \
r^2}\,,\\ \notag \\ \notag
    \Psi^{-}&=\left(\frac{2i M}{r^3} -  \frac{i}{r^2} -  \frac{i \Lambda^2}{2 r^2}\right) \
\mathit{h}^{+}{} -  \frac{1}{3}i r^2 \Phi^{+}{}^{a}{}_{a} + \left(- \frac{1}{3}i + \frac{1}{6}i \Lambda^2 + \frac{2i M}{r}\right) \
\Phi^{\oplus}{} \\\notag
&-  \frac{2}{3}i r \Phi^{+}{}^{a} r_{a} -  \frac{1}{3}i r^2 \Phi^{+}{}^{a}{}_{:a} + \frac{i \
\mathit{h}^{+}{}{}_{:a} r^{a}}{r} + \frac{\Lambda^2 \varepsilon_{ab} \mathit{h}^{-}{}^{a} r^{b}}{r} \\
&- i \mathit{h}^{+}{}_{ab} \
r^{a} r^{b} -  \frac{1}{2} \Lambda^2 \varepsilon_{ab} \
\mathit{h}^{-}{}^{a}{}^{:b}\,,\\ \notag \\ \notag
\Psi^{+}_{a}&=\left(\frac{i M}{r^3} + \frac{i \Lambda^2}{4 r^2}\right) \mathit{h}^{-}{}_{a} -  \
\frac{1}{2} \varepsilon_{ab} \Phi^{+}{}^{b} + \frac{i \Phi^{-}{} \
r_{a}}{2 r} -  \frac{1}{4}i \Phi^{-}{}{}_{:a} -  \frac{i \
\mathit{h}^{-}{}^{b} r_{a} r_{b}}{2 r^2} + \frac{1}{4}i \
\mathit{h}^{-}{}_{a}{}^{:b}{}_{:b} \\ 
&-  \frac{\varepsilon_{b}{}^{c} \
\mathit{h}^{+}{}_{ac} r^{b}}{2 r} -  \frac{i \
\mathit{h}^{-}{}_{a}{}_{:b} r^{b}}{2 r} -  \frac{1}{2} \varepsilon^{bc} \mathit{h}^{+}{}_{ab}{}_{:c}\,,\\ \notag \\ \notag
\Psi^{-}_{a}&=\varepsilon_{ab} \left(\frac{M}{r^3} -  \frac{\Lambda^2}{4 r^2}\right) \
\mathit{h}^{-}{}^{b} + \frac{1}{2}i \Phi^{+}{}_{a} -  \frac{i \
\mathit{h}^{+}{} r_{a}}{r^3} + \frac{i \mathit{h}^{+}{}{}_{:a}}{2 \
r^2} -  \frac{i \mathit{h}^{+}{}_{ab} r^{b}}{2 r} -  \
\frac{\varepsilon_{ab} \Phi^{-}{} r^{b}}{2 r}\\
&+ \frac{\varepsilon_{af} \mathit{h}^{-}{}^{f}{}_{:b} r^{b}}{2 r} + \frac{1}{4} \
\varepsilon_{ab} \Phi^{-}{}{}^{:b} -  \frac{1}{4} \varepsilon_{ab} \
\mathit{h}^{-}{}^{b}{}^{:f}{}_{:f} + \frac{\varepsilon_{af} \
\mathit{h}^{-}{}^{b} r_{b} r^{f}}{2 r^2}\,,\\ \notag \\ \notag
    \Psi^{+\oplus}_{ab}&=- \frac{1}{2} h^{+}{}_{ab} - \frac{1}{2} g_{ab} \Phi^{\oplus}
    -\frac{1}{2}i\left(h^{-}{}^{c}{}_{:(b}\varepsilon_{a)c} + h^{-}{}_{(b}{}^{:c}\varepsilon_{a)c}\right)\,,\\ \notag \\ \notag  
    \Psi^{-}_{ab}&=h^{-}{}_{(a}{}_{:b)}-\frac{1}{2}g_{ab}h^{-}{}^{c}{}_{:c}
    -\frac{1}{2}i\varepsilon_{(a}{}^{c} h^{+}{}_{b)c}\,,\\ \notag \\ \notag
     \Psi^{+}_{ab}&=- \frac{M \mathit{h}^{+}{}_{ab}}{r} + g_{ab} \left(\frac{M}{r^3} -  \
\frac{1}{2 r^2} -  \frac{\Lambda^2}{4 r^2}\right) \mathit{h}^{+}{} + \left(- \frac{1}{6} + \frac{1}{12} \Lambda^2\right) \Phi^{\oplus}{} g_{ab}  \\ \notag
& + \frac{i \Lambda^2  \mathit{h}^{-}{}^{c} r_{(a}\varepsilon_{b)c}}{2 r}-  \frac{1}{4}i \Lambda^2 \
 \mathit{h}^{-}{}^{c}{}_{:(a}\varepsilon_{b)c}-  \frac{i \Lambda^2 \
 \mathit{h}^{-}{}_{(a} r^{c}\varepsilon_{b)c}}{2 r}+ \frac{1}{4}i \Lambda^2  \mathit{h}^{-}{}_{(a}{}^{:c}\varepsilon_{b)c} \\
&+ g_{ab} \left(- \frac{1}{6} r^2 \Phi^{+}{}^{c}{}_{c} -  \frac{1}{3} r \
\Phi^{+}{}^{c} r_{c} -  \frac{1}{6} r^2 \Phi^{+}{}^{c}{}_{:c} + \
\frac{\mathit{h}^{+}{}{}_{:c} r^{c}}{2 r} -  \frac{1}{2} \
\mathit{h}^{+}{}_{cd} r^{c} r^{d}\right)\, .
\end{align}

\subsection{$\delta \J$ in terms of $\delta T$} \label{app:dJ_dT}

Here we provide the components of $\delta \J$ in terms of those of $\delta T$:

\begin{align}\notag
    \Pi^{+}{}_{ab}&=- \frac{1}{2} \Phi^{+}{}_{ab} + \frac{1}{6} g_{ab} \
\Phi^{+}{}^{c}{}_{c} + \frac{g_{ab} \Phi^{+}{}}{3 r^2} -  \frac{i \
\varepsilon_{bc} \Phi^{-}{}^{c} r_{a}}{4 r} + \frac{\Phi^{+}{}_{b} \
r_{a}}{4 r} + \frac{1}{4} \Phi^{+}{}_{b}{}_{:a}\\
&-  \frac{i \
\varepsilon_{ac} \Phi^{-}{}^{c} r_{b}}{4 r} + \frac{\Phi^{+}{}_{a} \
r_{b}}{4 r} + \frac{1}{4} \Phi^{+}{}_{a}{}_{:b} -  \frac{1}{4}i \
\varepsilon_{bc} \Phi^{-}{}_{a}{}^{:c} -  \frac{1}{4}i \varepsilon_{ac} \Phi^{-}{}_{b}{}^{:c}\, ,\\ \notag \\ \notag
\Pi^{-}{}_{ab}&=- \frac{1}{4}i \varepsilon_{b}{}^{c} \Phi^{+}{}_{ac} -  \frac{1}{4}i \
\varepsilon_{a}{}^{c} \Phi^{+}{}_{bc} + \frac{\Phi^{-}{}_{b} \
r_{a}}{4 r} + \frac{i \varepsilon_{bc} \Phi^{+}{}^{c} r_{a}}{4 \
r} + \frac{1}{4} \Phi^{-}{}_{b}{}_{:a} + \frac{\Phi^{-}{}_{a} \
r_{b}}{4 r} \\ 
&+ \frac{i \varepsilon_{ac} \Phi^{+}{}^{c} r_{b}}{4 \
r} + \frac{1}{4} \Phi^{-}{}_{a}{}_{:b} + \frac{1}{4}i \varepsilon_{bc} \Phi^{+}{}_{a}{}^{:c} + \frac{1}{4}i \varepsilon_{ac} \
\Phi^{+}{}_{b}{}^{:c} \, ,\\ \notag \\ 
\Pi^{+}{}_{a}&=- \frac{i \Lambda^2 \Phi^{-}{}_{a}}{2 r^2} -  \frac{1}{6} \varepsilon_{a}{}^{b} \Phi^{+}{}^{c}{}_{c}{}_{:b} + \frac{2 \varepsilon_{ab} \
\Phi^{+}{} r^{b}}{3 r^3} -  \frac{\varepsilon_{ab} \
\Phi^{+}{}{}^{:b}}{3 r^2} + \frac{1}{2} \varepsilon^{bc} \
\Phi^{+}{}_{ab}{}_{:c}\, ,\\ \notag \\ \notag
\Pi^{-}{}_{a}&=- \frac{1}{4} \Lambda^2 \Phi^{-}{}_{a} + \frac{1}{4}i \Lambda^2 \
\varepsilon_{ab} \Phi^{+}{}^{b} -  \frac{1}{6}i \varepsilon_{a}{}^{b} \
r^2 \Phi^{+}{}^{c}{}_{c}{}_{:b} -  \frac{1}{2}i \varepsilon_{a}{}^{c} \
r \Phi^{+}{}_{bc} r^{b} \\ 
&+ \frac{i \varepsilon_{ab} \Phi^{+}{} \
r^{b}}{6 r} + \frac{1}{6}i \varepsilon_{ab} \Phi^{+}{}{}^{:b}\, ,\\ \notag \\ 
\Pi^{\oplus}{}_{a}&=- \frac{1}{2}i \varepsilon_{ab} \Phi^{-}{}^{b} + \frac{1}{2} \
\Phi^{+}{}_{a} + \frac{\Phi^{\oplus}{} r_{a}}{2 r} -  \frac{1}{2} \
\Phi^{\oplus}{}{}_{:a} -  \frac{i \varepsilon_{ab} \Phi^{-}{} \
r^{b}}{2 r} + \frac{1}{2}i \varepsilon_{ab} \Phi^{-}{}{}^{:b}\, ,\\ \notag \\ 
\Pi^{+}&=\frac{i \Phi^{-}{}^{a} r_{a}}{4 r} + \frac{1}{4}i \
\Phi^{-}{}^{a}{}_{:a} + \frac{\varepsilon_{ab} \Phi^{+}{}^{a} \
r^{b}}{4 r} -  \frac{1}{4} \varepsilon_{ab} \Phi^{+}{}^{a}{}^{:b}\, ,\\ \notag \\ 
\Pi^{-}&=\frac{1}{12}i \Phi^{+}{}^{a}{}_{a} -  \frac{i \Phi^{+}{}}{3 r^2} -  \
\frac{i \Phi^{+}{}^{a} r_{a}}{4 r} -  \frac{1}{4}i \
\Phi^{+}{}^{a}{}_{:a} + \frac{\varepsilon_{ab} \Phi^{-}{}^{a} \
r^{b}}{4 r} -  \frac{1}{4} \varepsilon_{ab} \Phi^{-}{}^{a}{}^{:b}\, ,\\ \notag \\ 
\Gamma^{+}{}_{a}&=- \frac{1}{4}i \Lambda^2 \varepsilon_{ab} \Phi^{-}{}^{b} -  \frac{1}{4} \Lambda^2 \Phi^{+}{}_{a} -  \frac{\Phi^{+}{} r_{a}}{6 \
r} + \frac{1}{6} r^2 \Phi^{+}{}^{b}{}_{b}{}_{:a} -  \frac{1}{6} \
\Phi^{+}{}{}_{:a} + \frac{1}{2} r \Phi^{+}{}_{ab} r^{b}\, ,\\ \notag \\ 
\Gamma^{-}{}_{a}&=\frac{1}{2} \Phi^{-}{}_{a} + \frac{1}{2}i \varepsilon_{ab} \
\Phi^{+}{}^{b} + \frac{\Phi^{-}{} r_{a}}{2 r} -  \frac{1}{2} \
\Phi^{-}{}{}_{:a} + \frac{i \varepsilon_{ab} \Phi^{\oplus}{} \
r^{b}}{2 r} -  \frac{1}{2}i \varepsilon_{ab} \Phi^{\oplus}{}{}^{:b}\, ,\\ \notag \\ \notag
\Gamma^{+}&=\frac{1}{6} r^2 \Phi^{+}{}^{a}{}_{a} + \frac{1}{2} \Phi^{\oplus}{} \
-  \frac{1}{4} \Lambda^2 \Phi^{\oplus}{} -  \frac{1}{6} \Phi^{+}{} \
+ \frac{1}{2} r \Phi^{+}{}^{a} r_{a}\\
&-  \frac{1}{2}i \varepsilon_{ab} r \Phi^{-}{}^{a} r^{b} + \frac{1}{2}i \varepsilon_{ab} r^2 \
\Phi^{-}{}^{a}{}^{:b}\, ,\\ \notag \\ 
\Gamma^{-}&=- \frac{1}{2} \Phi^{-}{} + \frac{1}{4} \Lambda^2 \Phi^{-}{} -  \frac{1}{2} r \Phi^{-}{}^{a} r_{a} -  \frac{1}{2}i \varepsilon_{ab} \
r \Phi^{+}{}^{a} r^{b} + \frac{1}{2}i \varepsilon_{ab} r^2 \
\Phi^{+}{}^{a}{}^{:b}\, .
\end{align}

\end{document}